\documentclass[twocolumn]{aastex62}

\usepackage{amsmath, mathrsfs, amssymb}
\usepackage{graphicx}
\usepackage{rotating}
\usepackage{morefloats}
\usepackage{color}
\usepackage{enumitem}
\usepackage{verbatim}
\usepackage{hyperref}
\usepackage{booktabs}

\newcommand{\fsec}{\mbox{\ensuremath{.\!\!^{\rm s}}}}

\newcommand{\ihat}{\hat{\mathbf{i}}}
\newcommand{\jhat}{\hat{\mathbf{j}}}
\newcommand{\khat}{\hat{\mathbf{k}}}

\setlist[itemize]{leftmargin=*}

\received{November 05, 2019}
\accepted{April 22, 2020}
\published{June 2020}
\submitjournal{\textit{Monthly Notices of the Royal Astronomical Society}}

\begin{document}
\label{firstpage}

%===========================================================================
%===========================================================================
% TITLE & AUTHORS
\shorttitle{Spin-Orbit Misalignment in V4641 Sgr}
\shortauthors{Salvesen \& Pokawanvit (2020)}

\title{\LARGE{Origin of Spin-Orbit Misalignments: The Microblazar V4641 Sgr}}

\correspondingauthor{Greg Salvesen}
\email{gregsalvesen@gmail.com}

\author[0000-0002-9535-4914]{Greg Salvesen}
\thanks{NSF Astronomy \& Astrophysics Postdoctoral Fellow.}
\affil{CCS-2, Los Alamos National Laboratory, P.O. Box 1663, Los Alamos, NM 87545, USA.}
\affil{Center for Theoretical Astrophysics, Los Alamos National Laboratory, Los Alamos, NM 87545, USA.}
\affil{Department of Physics, University of California, Santa Barbara, CA 93106, USA.}

\author{Supavit Pokawanvit}
\affil{Department of Physics, University of California, Santa Barbara, CA 93106, USA.}

%===========================================================================
%===========================================================================
% ABSTRACT
\begin{abstract}
Of the known microquasars, V4641 Sgr boasts the most severe lower limit ($> 52^{\circ}$) on the misalignment angle between the relativistic jet axis and the binary orbital angular momentum. Assuming the jet and black hole spin axes coincide, we attempt to explain the origin of this extreme spin-orbit misalignment with a natal kick model, whereby an aligned binary system becomes misaligned by a supernova kick imparted to the newborn black hole. The model inputs are the kick velocity distribution, which we measure customized to V4641 Sgr, and the immediate pre/post-supernova binary system parameters. Using a grid of binary stellar evolution models, we determine post-supernova configurations that evolve to become consistent with V4641 Sgr today and obtain the corresponding pre-supernova configurations by using standard prescriptions for common envelope evolution. Using each of these potential progenitor system parameter sets as inputs, we find that a natal kick struggles to explain the origin of the V4641 Sgr spin-orbit misalignment. Consequently, we conclude that evolutionary pathways involving a standard common envelope phase followed by a supernova kick are highly unlikely for V4641 Sgr. An alternative interpretation is that the jet axis does not reliably trace the black hole spin axis. Our results raise concerns about compact object merger statistics gleaned from binary population synthesis models, which rely on unverified prescriptions for common envelope evolution and natal kicks. We also challenge the spin-orbit alignment assumption routinely invoked to measure black hole spin magnitudes.
\end{abstract}

%===========================================================================
%===========================================================================
% KEYWORDS

%===========================================================================
%===========================================================================
% INTRODUCTION
\section{Introduction}
\label{sec:intro}
In a black hole X-ray binary system, the angular momentum of the black hole need not be aligned with that of the binary orbit. Measuring such a ``spin-orbit'' misalignment is challenging, in part because the orientation of a spinning black hole must be inferred indirectly. Conventional theory predicts the long-axis of a relativistic jet to be parallel to the black hole spin vector \citep{BlandfordZnajek1977, BlandfordPayne1982}. If true, then the jet inclination $i_{\mathrm{jet}}$ offers an observational tracer of the black hole spin inclination $i_{\bullet}$. The binary orbital inclination $i_{\mathrm{orb}}$ is more directly accessible through ellipsoidal modeling of the infrared/optical light curve, which changes with orbital phase due to the tidally distorted companion star \citep{AvniBahcall1975}. However, the physically interesting quantity is \textit{not} the inclination difference between the black hole spin and binary orbit, but rather the angle between their \textit{angular momenta}.

% TABLE 1:
\setlength{\tabcolsep}{3pt}
\begin{table*}[!t]
\centering
\begin{tabular}{l l l l l l}
\toprule
Source & $i_{\mathrm{orb}}$ & $i_{\mathrm{jet}}$ & $\theta_{\mathrm{pro}}$ & $\theta_{\mathrm{ret}}$ & Comments About the $i_{\mathrm{jet}}$ Measurement \\
\midrule
V4641 Sgr & $72\fdg3 \pm 4\fdg1$ (5) & $< 16^{\circ}$ (\S\ref{sec:ijet}) & $52^{\circ}$--$92^{\circ}$ & $88^{\circ}$--$180^{\circ}$ & Transverse Doppler effect of superluminal jet \\
XTE J1118+480 & $68$--$79^{\circ}$ (4) & $< 30^{\circ}$ (6) & $38^{\circ}$--$109^{\circ}$ & $71^{\circ}$--$180^{\circ}$ & Modeling the flat/inverted radio/IR spectrum \\
GRO J1655--40 & $70\fdg2 \pm 1\fdg0$ (1) & $85^{\circ} \pm 2^{\circ}$ (2) & $12^{\circ}$--$158^{\circ}$ & $22^{\circ}$--$168^{\circ}$ & Multiple ejections, result not reproduced since? \\
XTE J1550--564 & $74\fdg7 \pm 3\fdg8$ (12) & $70\fdg8^{+7.3}_{-4.5}$ (14) & $0^{\circ}$--$157^{\circ}$ & $23^{\circ}$--$180^{\circ}$ & Multiple ejections/outbursts, kinematic model \\
\hline
H1743--322 & $<75^{\circ}$ (9) & $75^{\circ} \pm 3^{\circ}$ (13) & $0^{\circ}$--$153^{\circ}$ & $27^{\circ}$--$180^{\circ}$ & Symmetric (two-sided) ejection, kinematic model \\
4U 1543--47 & $20\fdg7 \pm 1\fdg5$ (10,11) & $32^{\circ}\phantom{}^{+3}_{-4}$ (8) & $6^{\circ}$--$57^{\circ}$ & $123^{\circ}$--$174^{\circ}$ & Disk inclination from X-ray reflection (not $i_{\mathrm{jet}}$) \\
V404 Cyg & $67^{\circ}\phantom{}^{+3}_{-1}$ (3) & $14\fdg0$--$40\fdg6$ (7) & $39^{\circ}$--$97^{\circ}$ & $83^{\circ}$--$180^{\circ}$ & Three resolved approaching/receding ejecta pairs \\
\midrule
\bottomrule
\end{tabular}
\caption{From \textit{left} to \textit{right}, the columns give the black hole X-ray binary source name, binary orbital inclination $i_{\mathrm{orb}}$, jet axis inclination $i_{\mathrm{jet}}$, and the allowable spin-orbit misalignment angles for prograde ($\theta_{\mathrm{pro}}$) and retrograde ($\theta_{\mathrm{ret}}$) spin-orbit scenarios, given no information about the position angle of the binary orbital angular momentum, and incorporating the quoted uncertainties on $i_{\mathrm{orb}}$ and $i_{\mathrm{jet}}$. The last column provides comments about the $i_{\mathrm{jet}}$ measurement for each source. We section off H1743--322 because of its weak $i_{\mathrm{orb}}$ constraint, 4U 1543--47 because the value in the $i_{\mathrm{jet}}$ column is an inner disk inclination measurement, and V404 Cyg because the large variations in $i_{\mathrm{jet}}$ complicate estimating the spin-orbit misalignment. To obtain $\theta_{\mathrm{pro}}$ and $\theta_{\mathrm{ret}}$ for V404 Cyg, we assume (questionably) that the jet precesses about an axis inclined $27^{\circ}$ from our line-of-sight, which is half-way between the $i_{\mathrm{jet}}$ extremes. The numbers in parentheses map to the following references: (1) \citet{Greene2001}; (2) \citet{HjellmingRupen1995}; (3) \citet{Khargharia2010}; (4) \citet{Khargharia2013}; (5) \citet{MacDonald2014}; (6) \citet{Maitra2009}; (7) \citet{MillerJones2019}; (8) \citet{MorningstarMiller2014}; (9) \citet{Motta2010}; (10) \citet{Orosz1998}; (11) \citet{Orosz2003}; (12) \citet{Orosz2011b}; (13) \citet{Steiner2012a}; (14) \citet{SteinerMcClintock2012}.}
\vspace{-0mm}
\label{tab:misalign}
\end{table*}

% FIGURE 1:
%\begin{comment}
\begin{figure*}[!t]
  \begin{center}
    \includegraphics[width=0.495\textwidth]{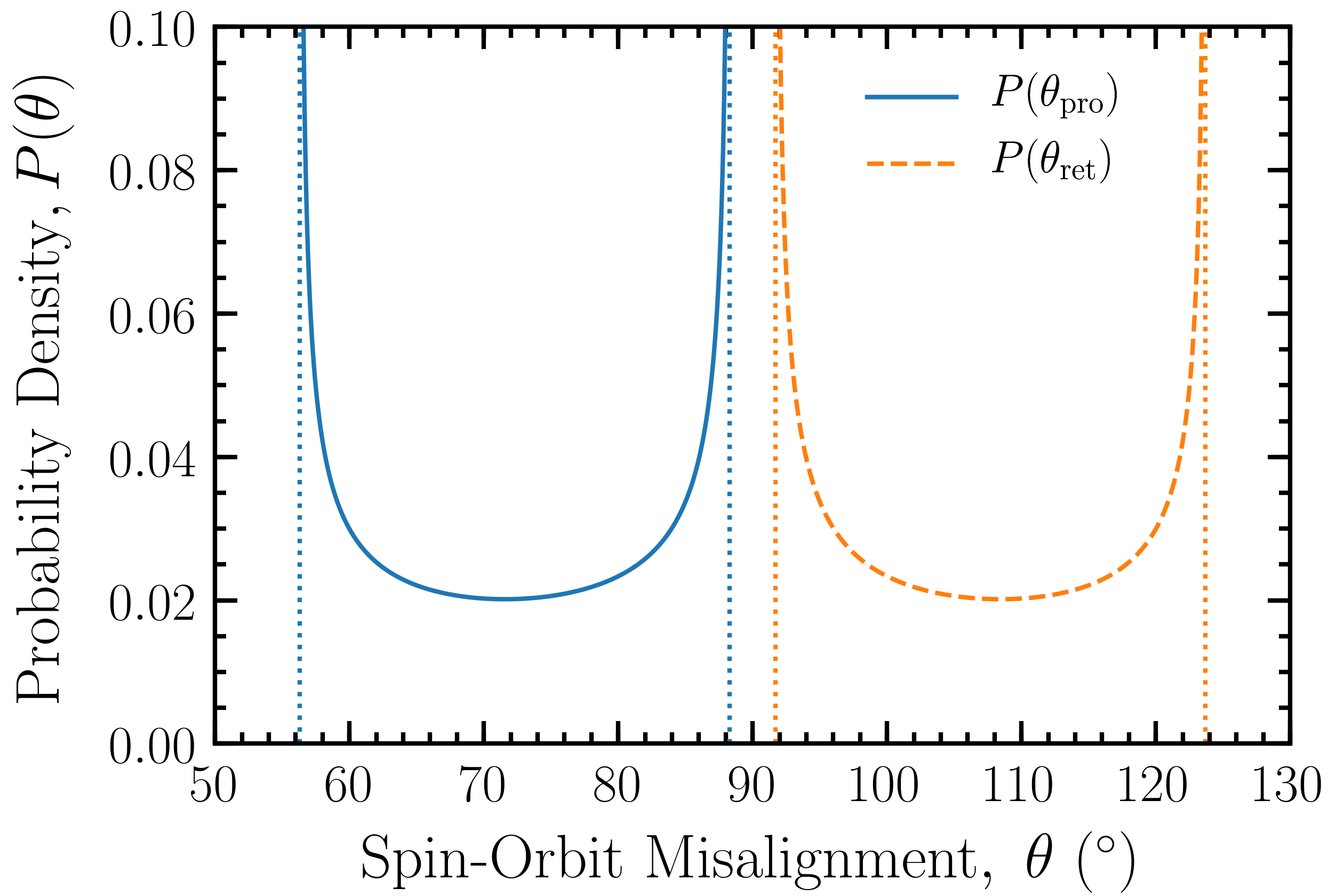}
    \hfill
    \includegraphics[width=0.495\textwidth]{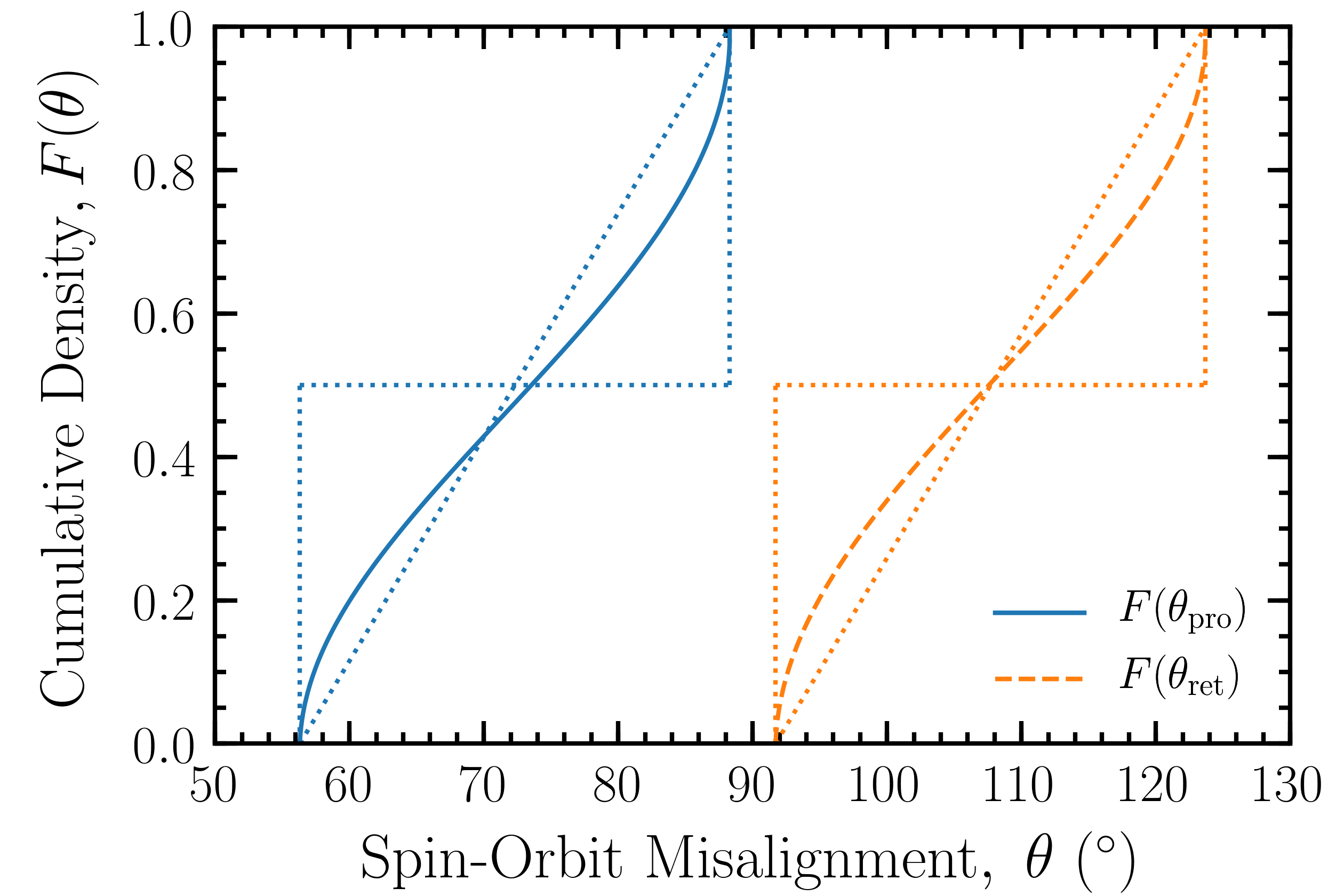}    
  \caption{\textit{Left}: Probability density functions (PDFs) of the spin-orbit misalignment angle $P(\theta)$ in V4641 Sgr, for prograde (\textit{solid blue line}) and retrograde (\textit{dashed orange line}) spin-orbit configurations. The \textit{vertical lines} mark the lower/upper $\theta$ limits for the PDFs. The PDFs assume $i_{\bullet} = i_{\mathrm{jet}} < 16^{\circ}$, adopt $i_{\mathrm{orb}} = 72^{\circ}$, and use a uniform distribution for the position angle of the binary orbital angular momentum following \citet{Martin2008a} (see Table \ref{tab:misalign}). \textit{Right}: The corresponding cumulative density functions (CDFs) are similar to those expected if $P(\theta)$ were uniformly distributed between the $\theta$ extremes (\textit{dotted diagonal line}), but deviate from those expected if $P(\theta)$ were isolated to the $\theta$ extremes (\textit{dotted step line}). This means there is significant likelihood for spin-orbit misalignments intermediate between the extremes and justifies treating $\min( \theta_{\mathrm{pro}} )$ as a lower limit.}
  \vspace{-0mm}
  \label{fig:pdfcdf}
  \end{center}
\end{figure*}
%\end{comment}

In practice, this spin-orbit misalignment angle $\theta$ can only be confined to within a broad range because the position angle of the binary orbital angular momentum is unconstrained. Consequently, the binary orbital angular momentum vector can lie anywhere on the surface of a double cone, with an axis along our line-of-sight and a half-opening angle $i_{\mathrm{orb}}$ \citep[for diagrams, see][]{Fragile2001, Martin2008a}. Furthermore, given only $i_{\bullet}$ (taken to be $i_{\mathrm{jet}}$) and $i_{\mathrm{orb}}$, the spin-orbit misalignment range depends on whether the black hole spin and binary orbit are in the prograde or retrograde sense. In X-ray binaries, the black hole spin parameter $a_{\ast}$ is consistently measured to be positive, favoring the prograde spin-orbit scenario \citep[e.g.,][]{Reynolds2014, McClintock2014}.

In Table \ref{tab:misalign}, we compiled the current spin-orbit misalignment constraints for microquasars, as \textit{inferred} from their jet axis inclinations. The most extreme case to-date is V4641 Sgr and the subject of this paper, boasting $\theta > 52^{\circ}$ as determined from the apparent superluminal motion of its radio jet (see \S\ref{sec:ijet}). Figure \ref{fig:pdfcdf} shows the probability density function (PDF) and cumulative density function (CDF) of the spin-orbit misalignment angle in V4641 Sgr, assuming a uniform distribution for the unknown position angle of the binary orbital angular momentum \citep{Martin2008a}. Although the PDF peaks at the two extremes of $\theta$, the CDF shows that the intermediate $\theta$ values contain the majority of the integrated probability. This justifies treating $52^{\circ}$ as the lower limit, rather than the actual value, of the spin-orbit misalignment in V4641 Sgr today.

From Table \ref{tab:misalign}, the microquasars GRO J1655--40, XTE J1118+480, and V404 Cyg also have significant spin-orbit misalignments, again under the assumption that the jet axis traces the black hole spin axis. Absent knowledge of $i_{\mathrm{orb}}$ (e.g., H1743--322), a non-precessing jet aligned to the black hole spin (but misaligned to the binary orbit) would be observationally indistinguishable from the jet expected in a spin-orbit aligned system. Therefore, deducing an unambiguous spin-orbit misalignment requires measurements of both $i_{\mathrm{jet}}$ and $i_{\mathrm{orb}}$. In XTE J1550--564, the close agreement along the inclination dimension (i.e., $i_{\mathrm{jet}} \simeq i_{\mathrm{orb}}$) supports the prospect of spin-orbit alignment; however, the observational inaccessibility of the position angle dimension of the binary orbital angular momentum makes $\theta$ formally unconstrained \citep{SteinerMcClintock2012}.

The spin-orbit misalignment today $\theta$, provides a lower limit on the initial misalignment at black hole birth $\theta_{0}$. This is because the reaction force of a misaligned accretion flow to the Lense-Thirring torque acts to align the black hole spin to the total angular momentum of the binary system \citep{BardeenPetterson1975, LenseThirring1918}. The binary orbital angular momentum is the dominant component, so the system evolves toward spin-orbit alignment over time. The timescale for alignment $t_{\mathrm{align}}$ depends on several things, such as the internal viscous stresses of the disk and the mass accretion rate \citep[e.g.,][]{Martin2007, Martin2008a}. Adopting reasonable parameter ranges for V4641 Sgr,\footnote{Using Equation \eqref{eqn:talign} with $M_{1} = 6.4~M_{\odot}$, $\alpha_{1} = 0.2$, $\alpha_{2} = 2$, $\beta = 3 / 4$, and either $a_{\ast} = 0.1$, $\dot{M} = 10^{-7}~M_{\odot}/\mathrm{yr}$ for fast alignment, or $a_{\ast} = 1.0$, $\dot{M} = 10^{-9}~M_{\odot}/\mathrm{yr}$ for slow alignment.} we find that $t_{\mathrm{align}} \simeq 0.6$--$200~\mathrm{Myr}$ can easily be an appreciable fraction of (or exceed) the lifetime of the system, so we do not necessarily expect complete alignment today. This establishes that spin-orbit misalignments are likely to persist, as observed in Table \ref{tab:misalign}, but what is their origin?

The generally accepted spin-orbit misalignment production mechanism appeals to imparting the newborn black hole with a momentum impulse, or ``kick'', from an asymmetric core-collapse supernova \citep[e.g.,][]{FlanneryvandenHeuvel1975}, combined with mass loss from the explosion \citep[e.g.,][]{Blaauw1961, Boersma1961}. Indeed, supportive evidence for supernova kicks comes from the population of pulsars and X-ray binaries with high Galactic latitudes and peculiar velocities of several hundred km/s \citep[e.g.,][]{Hobbs2005, JonkerNelemans2004}. In principle, a natal kick might alter the orientation of the binary orbital axis \textit{and} the rotational axis of the compact object from that of its progenitor. However, supernova kick models used in modern binary population synthesis studies do not permit an angular momentum kick to the remnant \citep[e.g.,][]{Belczynski2016, Stevenson2017}, which is appropriate for black holes (see \S\ref{sec:collapse}). Under this restriction and the assumption that the pre-supernova binary system is aligned, spin-orbit misalignments exceeding a few tens of degrees in black hole X-ray binaries are predicted to be rare \citep{Fragos2010}, which is at odds with the observational constraints in Table \ref{tab:misalign}.

In this paper, we test the ``natal kick model'' as the mechanism for producing the extreme spin-orbit misalignment in the microblazar V4641 Sgr. Applications of the kick model to black hole X-ray binaries have a long history: GRO J1655--40 \citep{Willems2005}, XTE J1118+480 \citep{Fragos2009}, M33 X--7 \citep{Valsecchi2010}, Cygnus X--1 \citep{Wong2012}, IC 10 X--1 \citep{Wong2014}, LMC X--3 \citep{Sorensen2017}. Aspects of our analysis closely follow these works; namely, integrating the past trajectory of the system through the Galaxy to obtain velocity constraints at the time of black hole birth (\S\ref{sec:vpec}) and using binary stellar evolution models to constrain system parameters at different epochs (\S\ref{sec:bse}). Our analysis is closest in spirit to \citet{Martin2010}, who derived spin-orbit misalignment constraints from the natal kick model and applied these to GRO J1655--40. We emphasize that the overarching theme of our analysis is to stack the deck in favor of a natal kick producing the extreme spin-orbit misalignment in V4641 Sgr. This approach ultimately strengthens our main result that the kick model fails when paired with a common envelope evolutionary history.

%----------------------------------------------------------------------------------------------------
% ROADMAP
\subsection{Roadmap}
\label{sec:roadmap}
To help guide the reader through our lengthy paper, we provide the following section-by-section roadmap:
\begin{itemize}
\setlength\itemsep{0mm}
\item \S\ref{sec:V4641} meticulously reviews all of the observational constraints for V4641 Sgr and Table \ref{tab:V4641} lists the constraints used in our subsequent analysis.
\item \S\ref{sec:model} describes the natal kick model, whose inputs are the kick velocity distribution and the pre/post-supernova system parameters. The output is a probability density for the spin-orbit misalignment angle, incorporating several velocity constraints.
\item \S\ref{sec:vpec} derives the kick velocity distribution specific to V4641 Sgr used in all applications of the kick model going forward. This is an improvement over other works that focused on a specific system, but adopted a kick distribution appropriate for populations of neutron stars or black holes.
\item \S\ref{sec:constraints} provides representative applications of the kick model to help develop an intuition for its behavior and to correct several mistakes in the literature.
\item \S\ref{sec:bse} determines sets of post-supernova system parameters that can evolve to match those of V4641 Sgr today (within their uncertainties).
\item \S\ref{sec:align} calculates the minimum spin-orbit misalignment at the time of black hole birth, based on the accretion history of each of the ``matching'' post-supernova system parameter sets from \S\ref{sec:bse}.
\item \S\ref{sec:CEE} uses prescriptions for common envelope evolution to determine pre-supernova system parameter sets that are consistent with each of the ``matching'' post-supernova system parameter sets from \S\ref{sec:bse}. Using each of these pre/post-supernova parameter sets as inputs, along with the kick distribution from \S\ref{sec:vpec}, we show that the natal kick model struggles to produce the requisite spin-orbit misalignment from \S\ref{sec:align}.
\item \S\ref{sec:disc} validates the main assumptions of the kick model, gives counterarguments to the jet being a black hole spin tracer, and disfavors a binary evolution origin for the V4641 Sgr misalignment.
\item \S\ref{sec:sumconc} concludes that a natal kick applied to a binary system that experienced a common envelope evolutionary pathway is a highly unlikely explanation for the spin-orbit misalignment in V4641 Sgr. This result has important implications for compact object merger studies and black hole spin measurements.
\end{itemize}

%===========================================================================
%===========================================================================
% THE MICROBLAZAR V4641 Sgr
\section{The Microblazar V4641 Sgr}
\label{sec:V4641}
V4641 Sgr is a fascinating source with a convoluted history. First discovered in June 1978 after brightening by two B-band magnitudes above quiescence \citep{Goranskij1978}, V4641 Sgr was incorrectly branded in the \textit{General Catalog of Variable Stars} \citep{GCVS4} as the distinct variable star GM Sgr \citep{Luyten1927}. This confusion arose from the $\sim1^{\prime}$ proximity of V4641 Sgr to GM Sgr in a crowded star field and the lack of a published finding chart for GM Sgr, but was resolved following the September 1999 major outburst of V4641 Sgr \citep{Williams1999, Samus1999, Hazen2000}.

%----------------------------------------------------------------------------------------------------
% THE SEPTEMBER 1999 MAJOR OUTBURST
\subsection{The September 1999 Major Outburst}
\label{sec:sep1999}
In February 1999, V4641 Sgr was independently discovered by \textit{BeppoSAX} \citep[SAX J1819.3-2525;][]{intZand1999, intZand2000} and \textit{RXTE} \citep[XTE J1819-254;][]{Markwardt1999a} as a faint X-ray transient with 2--10 keV flux varying from $<$0.001--0.08 Crab. On 1999 September 14.89 UT and preceding its most dramatic outburst observed to-date, V4641 Sgr flared for less than three hours with an X-ray flux reaching 4.5 Crab in a soft band \citep[2--12 keV;][]{Hjellming2000} and 5 Crab in a hard band \citep[20--100 keV;][]{McCollough1999}. Twelve hours later on Sep 15.40 UT, amateur astronomer \citet{Stubbings1999} \textit{visually} observed a major optical outburst reaching peak magnitude $m_{\mathrm{V}} = 8.8$ and sent a VSNET alert (\#3477). During the next 7 hours, \textit{RXTE} observed the 2--12 keV X-ray flux rise from 1.3 Crab to reach a whopping 12.2 Crab on Sep 15.70 UT \citep{Smith1999} while showcasing rapid variability \citep{WijnandsvanderKlis2000}, followed by a sharp decline to X-ray quiescence within 5 hours \citep{Markwardt1999b}. The optical brightness promptly declined during the X-ray rise, reaching quiescent levels after two days \citep{Kato1999}.

Following the X-ray and optical decay, on Sep 16.027 UT and 30 minutes later on Sep 16.048 UT, the \textit{VLA} imaged a one-sided, elongated (10:1 axis ratio), and extended ($\sim0.25^{\prime\prime}$ long) radio source in the vicinity of V4641 Sgr with a flux density at 4.9 GHz that decayed from $S_{5} = 420 \pm 20~\mathrm{mJy}$ to $S_{5} = 400 \pm 20~\mathrm{mJy}$ \citep{Hjellming1999a, Hjellming1999b, Hjellming2000}. Less than one day later on Sep 16.94 UT, the \textit{VLA} observed the radio source rapidly decay tenfold to $S_{5} = 45 \pm 4~\mathrm{mJy}$, which combined with inclement weather prevented reliable imaging. On Sep 17.94 UT, \textit{VLA} imaging showed that the extended radio morphology was gone and just the southern tip remained as a faint stationary core, with a flux density of $S_{5} = 19 \pm 5~\mathrm{mJy}$ that decayed to $S_{5} = 0.4 \pm 0.2~\mathrm{mJy}$ on Oct 6.04 UT. Unfortunately, the onset of the radio event was not caught, nor were moving components spatially resolved. Interpreting the extended radio source as a jet and associating its moment of ejection with either the early X-ray flare, the onset of the major X-ray outburst, or its initial quenching, \citet{Hjellming2000} estimated the apparent proper motion of the jet to be $\mu_{\mathrm{app}} = 0\farcs22 / \mathrm{day}$, $0\farcs36 / \mathrm{day}$, and $1\farcs1 / \mathrm{day}$, respectively.

Since the September 1999 event, many comparatively modest outbursts of V4641 Sgr occurred: Jul 2000 \citep{Hjellming2000ATel}, May 2002 \citep{Uemura2004a}, Aug 2003 \citep{Buxton2003, Bailyn2003, Rupen2003b}, Jul 2004 \citep{Swank2004, Rupen2004}, Jun 2005 \citep{Swank2005}, May 2007 \citep{Goranskij2007, CackettMiller2007}, Oct 2008 \citep{Yamaoka2008}, Aug 2010 \citep{Yamaoka2010}, Jan 2014 \citep{Tachibana2014, Uemura2014}, Jul 2015 \citep{Yoshii2015}, Aug 2018 \citep{Negoro2018, Kong2018}, and Jan 2020 \citep{Shaw2020, Imazato2020}. Short-lived optical/X-ray flares lasting $\sim$hours--days characterize most of these outbursts. This behavior is atypical of black hole X-ray transients, which usually display $\sim$months--year-long outburst cycles. A radio source was present during the 2000, 2002, 2003, and 2004 outbursts, but was not spatially extended. Interestingly, P Cygni profiles in the optical spectra from the 1999, 2002, and 2004 outbursts revealed a hard state accretion disk wind simultaneous with the unresolved radio jet \citep{MunozDarias2018}.

%----------------------------------------------------------------------------------------------------
% JET AXIS INCLINATION ANGLE
\subsection{Jet Axis Inclination Angle}
\label{sec:ijet}
The distance to V4641 Sgr is $d = 6.2 \pm 0.7~{\mathrm{kpc}}$ \citep{MacDonald2014}, which is derived from the extinction, apparent $V$ magnitude, and calculated absolute $V$ magnitude of the companion star, and is consistent with the parallax of $0.15 \pm 0.04~\mathrm{mas/yr}$ \citep{Gaia2018}. With this distance, the estimated apparent proper motions for the jet of $\mu_{\mathrm{app}} = 0\farcs22 / \mathrm{day}$, $0\farcs36 / \mathrm{day}$, $1\farcs1 / \mathrm{day}$ correspond to the highly super-luminal apparent speeds for the jet of $v_{\mathrm{app}} = 7.9 c$, $13 c$, $39 c$, respectively. 

% FIGURE 2:
%\begin{comment}
\begin{figure}[!t]
  \begin{center}
  \includegraphics[width=0.495\textwidth]{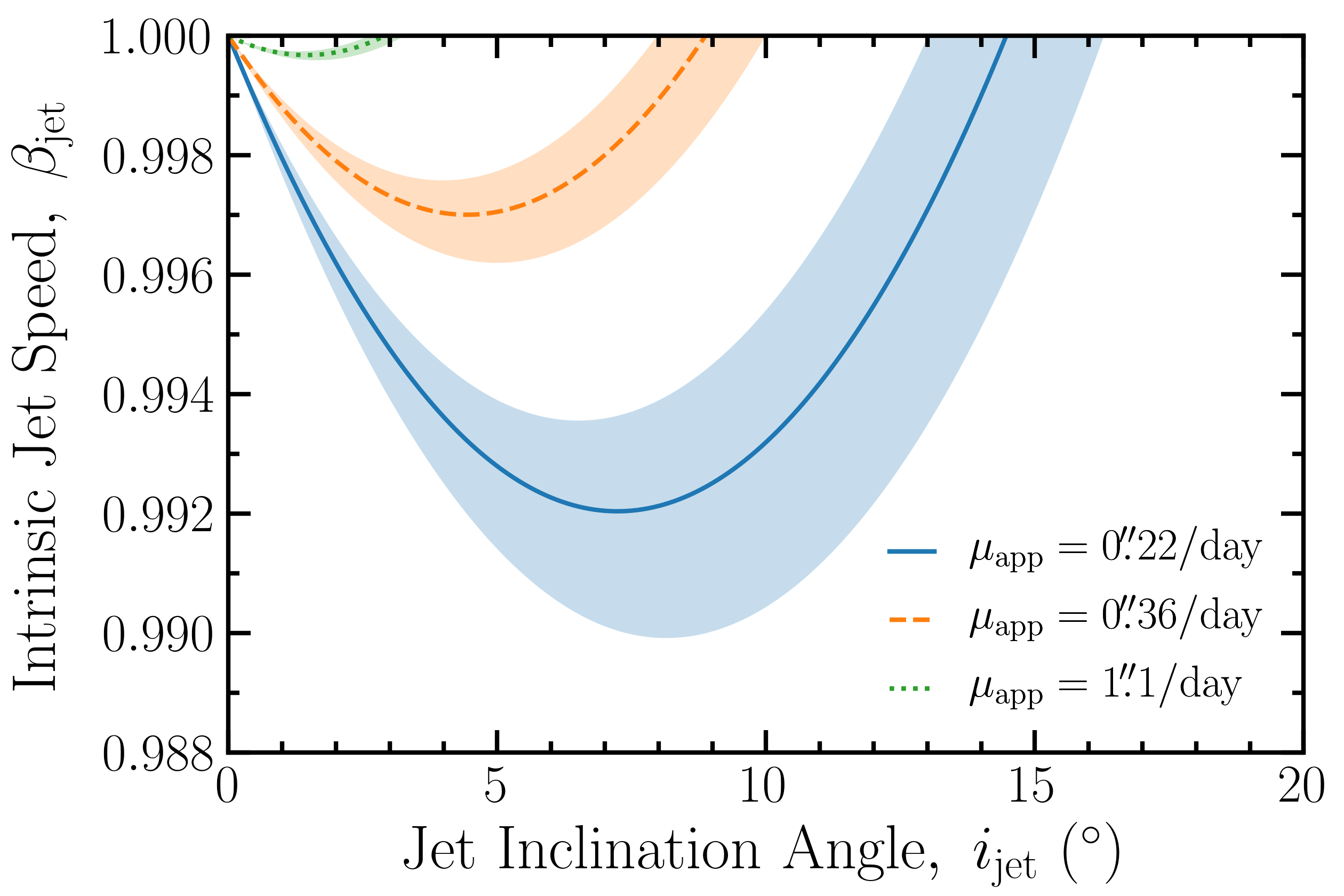}
  \vspace{-5mm}
  \caption{Intrinsic speed of the V4641 Sgr jet relative to the speed of light, $\beta_{\mathrm{jet}} = v_{\mathrm{jet}} / c$, as a function of the jet axis inclination angle, $i_{\mathrm{jet}}$ (see Equation \ref{eqn:muapp}). The different curves adopt the plausible apparent proper motions of the radio-emitting ejecta from the September 1999 outburst: $\mu_{\mathrm{app}} = 0\farcs22 / \mathrm{day}$ (\textit{blue solid line}), $\mu_{\mathrm{app}} = 0\farcs36 / \mathrm{day}$ (\textit{orange dashed line}), $\mu_{\mathrm{app}} = 1\farcs1 / \mathrm{day}$ (\textit{green dotted line}). Error bands reflect uncertainty in the distance $d = 6.2 \pm 0.7~{\mathrm{kpc}}$. The jet traveled faster than 99.0\% light-speed along an axis inclined slightly to the line of sight with $i_{\mathrm{jet}} < 16^{\circ}$.}
  \vspace{-4mm}
  \label{fig:ijet}
  \end{center}
\end{figure}
%\end{comment}

Associating the apparent proper motion with a jet approaching the observer allows constraints to be placed on the jet inclination $i_{\mathrm{jet}}$ and on $\beta_{\mathrm{jet}} = v_{\mathrm{jet}} / c$, the intrinsic jet speed $v_{\mathrm{jet}}$ relative to the speed of light $c$, using the relation \citep{Rees1966, MirabelRodriguez1999},
\begin{equation}
\mu_{\mathrm{app}}  = \frac{\beta_{\mathrm{jet}} \sin\left( i_{\mathrm{jet}} \right)}{1 - \beta_{\mathrm{jet}} \cos\left( i_{\mathrm{jet}} \right)} \frac{c}{d}. \label{eqn:muapp}
\end{equation}
Following \citet{Orosz2001}, Figure \ref{fig:ijet} shows the allowable $\left( i_{\mathrm{jet}}, \beta_{\mathrm{jet}} \right)$-space from Equation \eqref{eqn:muapp} for each of the three plausible $\mu_{\mathrm{app}}$ values above. Conservatively, the intrinsic jet speed is $v_{\mathrm{jet}} > 0.990 c$ and the jet inclination is $i_{\mathrm{jet}} < 16^{\circ}$ relative to our line-of-sight, which earns V4641 Sgr a ``microblazar'' distinction.

At the time of its September 1999 outburst, \citet{Williams1999} measured the J2000 position of V4641 Sgr to be $\alpha = 18^{\mathrm{h}} 19^{\mathrm{m}} 21\fsec61 \pm 0\fsec07$ and $\delta = -25^{\circ} 24^{\prime} 26\farcs3 \pm 1\farcs0$. Let us define $\alpha^{\mathrm{s}}$ and $\delta^{\prime\prime}$ as the seconds and arcseconds parts of $\alpha$ and $\delta$ for the J2000 epoch. The position of the centroid of the short-lived, radio extension detected on 1999 Sep 16.02 UT was $\alpha^{\mathrm{s}} = 21\fsec637 \pm 0\fsec007$ and $\delta^{\prime\prime} = 25\farcs60 \pm 0\farcs1$ \citep[90\% confidence;][]{Hjellming2000}. This radio extension disappeared by Sep 17.94 UT, but its southern tip persisted as a faint, stationary radio core at the position $\alpha^{\mathrm{s}} = 21\fsec634$ and $\delta^{\prime\prime} = 25\farcs85$, gradually decaying until Oct 7.95 UT when \textit{VLA} monitoring ceased. Therefore, the centroid of the short-lived, bright radio extension was offset from the long-lived, faint radio core by $\simeq 0\farcs25$, which corresponds to a separation of $\simeq 1600 \pm 200~\mathrm{AU}$ for the distance to V4641 Sgr.

\citet{Hjellming2000} associated the long-lived radio core with residual emission co-spatial with V4641 Sgr. However, the $\pm1\farcs0$ positional uncertainty for V4641 Sgr at the time raised the possibility of interpreting this long-lived radio core as an interaction of the jet with distant material surrounding V4641 Sgr \citep{Chaty2003}. Placing V4641 Sgr at the centroid of the short-lived radio extension, a sub-luminal jet ejected $> 9$ days before the major outburst would have time to travel $\simeq 0\farcs25$ ($\simeq 1600~\mathrm{AU}$) to the location of the long-lived radio core. This scenario might seem plausible given the pre-outburst optical activity \citep[e.g.,][]{Kato1999} and would nullify jet inclination constraints based on super-luminal motion. However, much improved optical astrometry now places V4641 Sgr at the position $\alpha^{\mathrm{s}} = 21\fsec63427 \pm 0\fsec00008$ and $\delta^{\prime\prime} = 25\farcs8493 \pm 0\farcs0009$ \citep[see Table \ref{tab:V4641};][]{Gaia2018}, which is precisely the position of the long-lived radio core. This firmly rules out the interpretation of the long-lived radio core as arising from sub-luminal jet ejecta interacting with distant material, and supports interpreting the short-lived radio extension as a moving jet ejection.

The proper motion of this jet ejection was not measured, due to its abrupt decay and unresolved structure. Consequently, the claim of apparent super-luminal motion relied on taking the jet to be launched around the time of the major outburst. To dismiss the enormous spin-orbit misalignment implied by such a super-luminal jet (see \S\ref{sec:sep1999}), previous works appeal to a sub-luminal jet launched weeks prior to the major outburst that interacts with surrounding material to produce the short-lived, extended radio emission \citep[e.g.,][]{NarayanMcClintock2005, SteinerMcClintock2012}. We disfavor this suggestion because it requires embracing the coincidence that a sub-luminal jet traveling for $\gtrsim 10$ days just happened to reach and interact with surrounding material, then promptly decay, contemporaneously with the major outburst that only lasts for a few hours itself.

Reasonably associating the moment of jet ejection around the time of the major outburst then makes a compelling case for super-luminal motion, which places a conservative upper limit on the jet axis inclination of $i_{\mathrm{jet}} < 16^{\circ}$ (see Figure \ref{fig:ijet}). Further support for low-inclination, apparent super-luminal jet motion comes from the lack of Doppler-shifted lines in the optical spectra during the days following the major outburst, as H$\alpha$ emission from approaching/receding ejecta with intrinsic speed $\gtrsim 0.95 c$ and inclination $\lesssim 10^{\circ}$ would be blue/redshifted into the UV/near-IR \citep{Chaty2003}.

%----------------------------------------------------------------------------------------------------
% BINARY SYSTEM PARAMETERS AND PROPERTIES
\subsection{Binary System Parameters and Properties}
\label{sec:binpars}
The orbital parameters of V4641 Sgr are accessible from modeling spectroscopic radial velocity curves in quiescence, yielding a binary orbital period $P = 2.817 \pm 0.002~{\mathrm{days}}$ and a mass function $f\left( M \right) = 2.74 \pm 0.04~M_{\odot}$ \citep{Lindstrom2005, Orosz2001}. The systemic radial velocity measurements of $\gamma = 72.7 \pm 3.3~\mathrm{km/s}$ \citep{Lindstrom2005} and $\gamma = 107.4 \pm 2.9~\mathrm{km/s}$ \citep{Orosz2001} are discrepant, but attributable to a systematic error in the \citet{Orosz2001} data reduction. We therefore favor the \citet{Lindstrom2005} radial velocity, but our conclusions are qualitatively unaffected by adopting the \citet{Orosz2001} value (see end of \S \ref{sec:vpec}).

\citet{Orosz2001} placed initial constraints on the V4641 Sgr binary orbital axis inclination of $60^{\circ} \lesssim i_{\mathrm{orb}} \le 70\fdg7$, derived from modeling an optical light curve with large uncertainties and the lack of observed X-ray eclipses. This inclination can be measured from optical light curve variations, caused by orbital modulation of the projected area of a distorted, Roche lobe-filling companion star. Applying this ellipsoidal variations technique to epochs of passive optical/infrared quiescence, \citet{MacDonald2014} confirmed that the companion star likely fills its Roche lobe and measured $i_{\mathrm{orb}} = 72\fdg3 \pm 4\fdg1$.

Therefore, the binary orbital axis ($i_{\mathrm{orb}} = 72\fdg3 \pm 4\fdg1$) and the approaching jet axis ($i_{\mathrm{jet}} < 16^{\circ}$) are largely misaligned by $\eta = | i_{\mathrm{jet}} - i_{\mathrm{orb}} | > 52^{\circ}$, as first noted by \citet{Orosz2001}. The orbital compactness of X-ray binaries precludes their visual separation, leaving the position angle of the binary orbital angular momentum unconstrained. Including the quoted uncertainties on $i_{\mathrm{orb}}$, the range of possible jet-orbit misalignment angles is then $52^{\circ} < \eta_{\odot} < 92^{\circ}$ or $88^{\circ} < \eta_{\otimes} < 180^{\circ}$ if the binary orbital angular momentum points toward or away from us, respectively \citep[see e.g., Figure 1 of][]{Martin2008a}. Interpreting the jet axis as the black hole spin axis, these ranges in $\eta_{\odot}$ and $\eta_{\otimes}$ translate respectively to ranges in the spin-orbit misalignment angle $\theta$ for prograde ($\theta_{\mathrm{pro}}$) and retrograde ($\theta_{\mathrm{ret}}$) spin-orbit scenarios.

% TABLE 2:
\setlength{\tabcolsep}{2pt}
\begin{table}[!tb]
\centering
\begin{tabular}{l c c c}
\toprule
Parameter [Units]\dotfill & Sym. & Value & Ref. \\
\midrule
Right Ascension (J2000)\dotfill & $\alpha$ & $18^{\mathrm{h}} 19^{\mathrm{m}} 21\fsec63427$ & 1 \\ %274\fdg8401428(3)\\ 
& & $\pm 0\fsec00008$ & \\
Declination (J2000)\dotfill & $\delta$ & $-25^{\circ} 24^{\prime} 25\farcs8493$ & 1 \\ %-25\fdg4071804(3)
& & $\pm 0\farcs0009$ & \\
Proper motion $\alpha$ [mas/yr]\dotfill & $\mu_{\alpha\ast}$ & $-0.734 \pm 0.070$ & 1 \\
Proper motion $\delta$ [mas/yr]\dotfill & $\mu_{\delta}$ & $0.418 \pm 0.056$ & 1 \\
Radial velocity [km/s]\dotfill & $\gamma$ & $72.7 \pm 3.3$ & 3 \\
Distance [kpc]\dotfill & $d$ & $6.2 \pm 0.7$ & 4 \\
Black hole mass $\left[ M_{\odot} \right]$\dotfill & $M_{\bullet}$ & $6.4 \pm 0.6$ & 4 \\
Companion star mass $\left[ M_{\odot} \right]$\dotfill & $M_{\star}$ & $2.9 \pm 0.4$ & 4 \\
Orbital period [days]\dotfill & $P$ & $2.817 \pm 0.002$ & 4 \\
Orbital separation $\left[ R_{\odot} \right]$\dotfill & $a$ & $17.5 \pm 1.0$ & 4 \\ 
Orbital axis inclination\dotfill & $i_{\mathrm{orb}}$ & $72\fdg3 \pm 4\fdg1$ & 4 \\
Jet axis inclination\dotfill & $i_{\mathrm{jet}}$ & $< 16^{\circ}$ & 2, 5, \S\ref{sec:ijet} \\
Spin-orbit misalignment\dotfill & $\theta$ & $> 52^{\circ}$ & 5, \S\ref{sec:binpars} \\
\midrule
\bottomrule
\end{tabular}
\caption{V4641 Sgr parameters with 68\%-level uncertainties. We transformed the \textit{Gaia} DR2 J2015.5 epoch to J2000 using the \texttt{Astropy} method \texttt{apply\_space\_motion}, using measured proper motions to account for source motion. The proper motion in right ascension includes a declination correction. \textsc{References:} (1) \citet{Gaia2018}; (2) \citet{Hjellming2000}; (3) \citet{Lindstrom2005}; (4) \citet{MacDonald2014}; (5) \citet{Orosz2001}.}
\vspace{-2mm}
\label{tab:V4641}
\end{table}

% FIGURE 3:
%\begin{comment}
\begin{figure*}[!t]
  \begin{center}
  \includegraphics[width=1.0\textwidth]{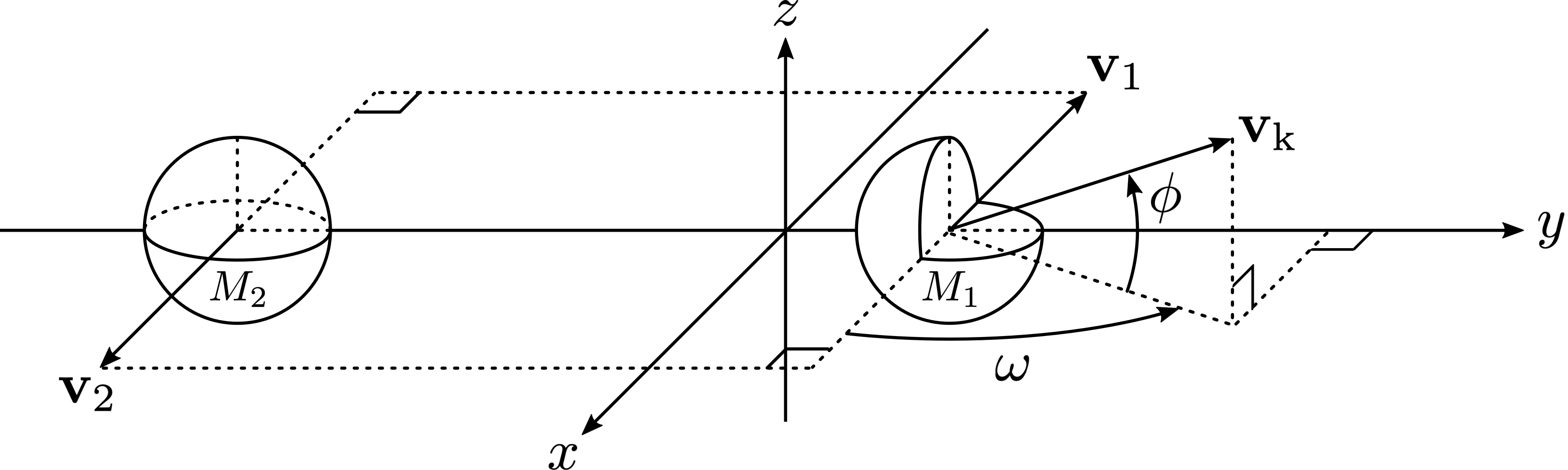}
  \caption{Schematic diagram in the pre-supernova center of mass frame, showing the moment when the black hole progenitor star of mass $M_{1}$ moving in the \textit{negative} $x$-direction with velocity $\mathbf{v}_{1}$ receives a supernova kick of velocity $\mathbf{v}_\mathrm{k}$ in the direction described by the angles $\omega$ and $\phi$. Instantaneously, a black hole of mass $M_{1}^{\prime}$ replaces its progenitor star and the system loses mass $M_{1} - M_{1}^{\prime}$. The companion star has mass $M_{2}$ and velocity $\mathbf{v}_{2}$, neither of which change immediately after the supernova.}
  \label{fig:kickdiagram}
  \end{center}
\end{figure*}
%\end{comment}

The projected rotational velocity of the companion star $v_{\mathrm{rot}} \sin( i_{\star} ) = 100.9 \pm 0.8~\mathrm{km/s}$ implies a primary-to-secondary mass ratio $Q = M_{\bullet} / M_{\star} = 2.2 \pm 0.2$, if one equates the inclination of the stellar rotational axis $i_{\star}$ to $i_{\mathrm{orb}}$. Combining this with the mass function and the binary orbital axis inclination gives the component masses $M_{\bullet} = 6.4 \pm 0.6~M_{\odot}$ and $M_{\star} = 2.9 \pm 0.4~M_{\odot}$ for the black hole and companion star, respectively \citep{MacDonald2014}. The primary is robustly a black hole, dynamically confirmed to exceed the $\simeq3~M_{\odot}$ neutron star stability threshold \citep{RhoadesRuffini1974}.

The companion star has a B9III classification, making it among the brightest, bluest, biggest, Roche lobe-filling companions of all known black hole X-ray binaries \citep{Orosz2001}. \citet{MacDonald2014} confirmed the B9III spectral type by spectroscopic comparison to three other stars (of types B8III, B9III, A0III), and found consistency with a generic B9III star reddened by $E (B - V) = 0.37 \pm 0.19$ from a photometric study during epochs of quiescence. Placing V4641 Sgr on a color-magnitude diagram \citep[Figure 6 of][]{Chaty2003} with color $(V - K) \simeq -0.15$ and absolute magnitude $M_{K} \simeq -1.3$ \citep[Figure 8 of][]{MacDonald2014} further establishes the companion star as a subgiant that is crossing the Hertzsprung gap (i.e., post-main sequence). 

Notably, \citet{Orosz2001} obtained component masses $M_{\bullet} = 9.61^{+2.08}_{-0.88}~M_{\odot}$ and $M_{\star} = 6.53^{+1.6}_{-1.03}~M_{\odot}$, which are discrepant with those quoted above. \citet{MacDonald2014} attributes this disagreement to the inferior spectral resolving power of \citet{Orosz2001} affecting the projected rotational velocity of the companion star, which is a proxy for the mass ratio $Q$. Consequently, this affects the \citet{Orosz2001} distance $d = 9.59^{+2.72}_{-2.19}~{\mathrm{kpc}}$, as does their underestimate of the binary orbital axis inclination. We therefore favor the \citet{MacDonald2014} masses and distance.

Table \ref{tab:V4641} lists the various parameters we adopt for V4641 Sgr in our subsequent analysis.

%===========================================================================
%===========================================================================
% SPIN-ORBIT MISALIGNMENT MODEL
\section{Spin-Orbit Misalignment Model}
\label{sec:model}
\begin{figure}[!b]
  \begin{center}
  \includegraphics[width=0.495\textwidth]{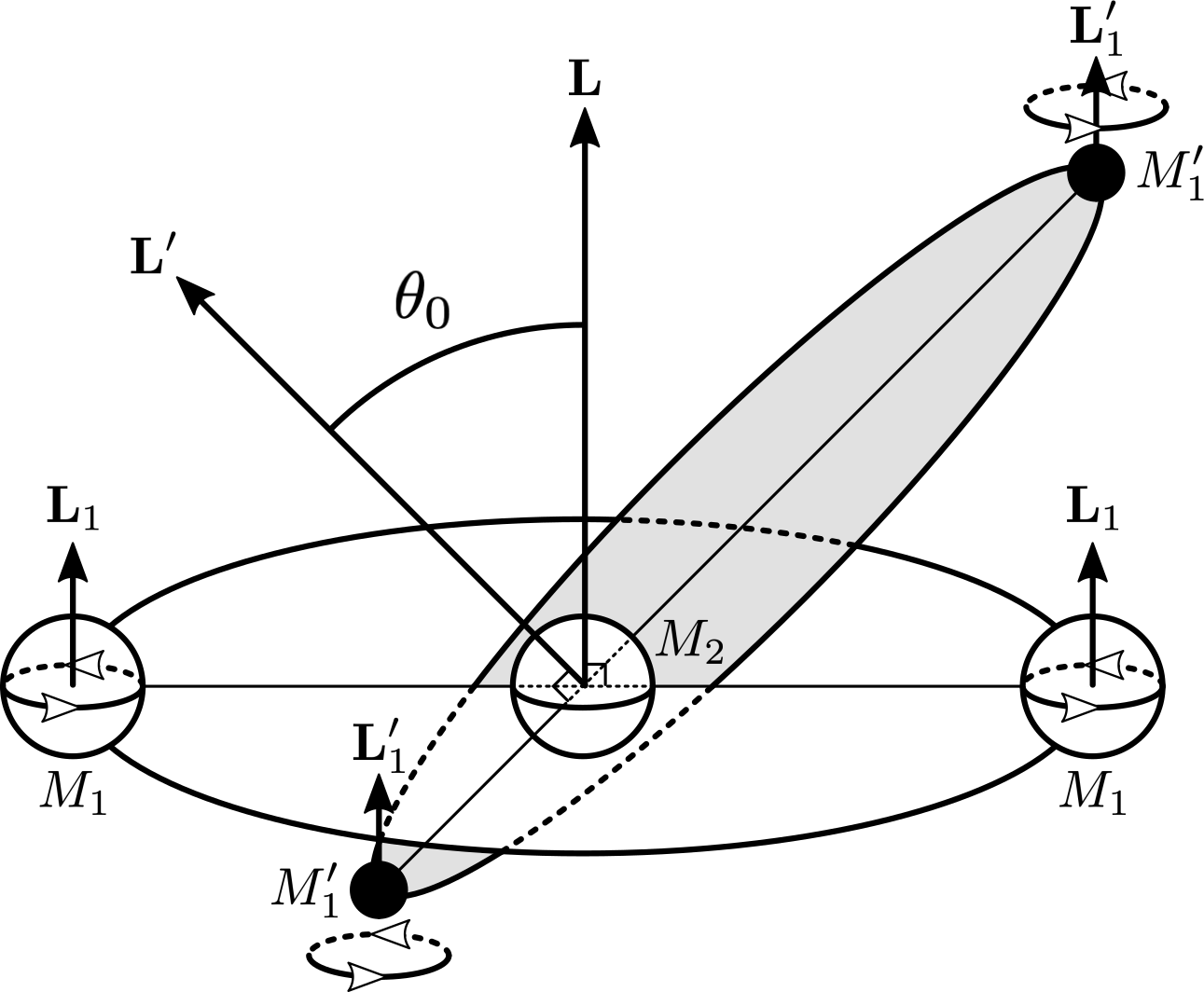}
  \caption{Diagram showing the supernova kick production mechanism for a spin-orbit misalignment angle $\theta_{0}$, as seen in the reference frame of the companion star of mass $M_{2}$. The pre-supernova binary orbit is circular with angular momentum $\mathbf{L}$ aligned to that of the black hole progenitor star $\mathbf{L}_{1}$ of mass $M_{1}$. The supernova delivers a linear momentum kick to the newly formed black hole of mass $M_{1}^{\prime}$, preserving the spin angular momentum direction to $\mathbf{L}_{1}^{\prime}$ but changing the orbital angular momentum direction to $\mathbf{L}^{\prime}$. This new eccentric orbit can circularize over time through secular processes (e.g., tidal heating, mass transfer, gravitational radiation).}
  \label{fig:orbitdiagram}
  \vspace{-0mm}
  \end{center}
\end{figure}

Our main objective is to calculate the conditional density $P\left( \theta_{0} | v_{\mathrm{orb}} \right)$ of the \textit{initial} spin-orbit misalignment angle $\theta_{0}$ to test whether a natal kick can produce the large spin-orbit misalignment in V4641 Sgr observed \textit{today} of $\theta > 52^{\circ}$, as inferred from the jet-orbit misalignment $\eta$ (see \S\ref{sec:binpars}). Shown schematically in Figures \ref{fig:kickdiagram} and \ref{fig:orbitdiagram}, the natal kick model solves the two-body problem including instantaneous mass loss and an arbitrarily-directed, linear momentum impulse (or ``kick'') imparted to a newborn black hole, presumably from an asymmetric core-collapse supernova. Together, the mass loss and the kick alter the binary orbit to produce a spin-orbit misalignment $\theta_{0}$ \textit{and} to give the system a translational space velocity. Our analysis closely follows \citet{Martin2009} and \citet{Martin2010} to constrain $\theta_{0}$ from a natal kick, but we acknowledge the many important developments to this model over the years: \citet{Blaauw1961, Boersma1961, FlanneryvandenHeuvel1975, Sutantyo1978, Hills1983, Wijers1992, BrandtPodsiadlowski1995, Kalogera1996, Kalogera2000, Hurley2002}.

% FIGURE 4:
%\begin{comment}
%\end{comment}

The chosen reference frame is the center of mass of the system immediately pre-supernova, which consists of a black hole progenitor star of mass $M_{1}$ and a companion star of mass $M_{2}$.\footnote{\citet{Martin2009, Martin2010} chose the opposite notation of subscripts 1 and 2 for the companion star and the black hole/progenitor star, respectively.} Just prior to the supernova, we assume each star follows a circular orbit around the center of mass with its spin aligned to the binary orbital angular momentum. At the moment when the supernova occurs, the progenitor star is traveling along the \textit{negative} $x$-axis with orbital velocity $\mathbf{v}_{\mathrm{orb}} = \mathbf{v}_{1} - \mathbf{v}_{2}$ relative to the companion star. Instantaneously, the progenitor loses mass $\Delta M = M_{1} - M_{1}^{\prime}$, while the mass of the companion remains unchanged. Simultaneously, a black hole of mass $M_{1}^{\prime}$ forms and receives a linear velocity kick of magnitude $v_{\mathrm{k}} \in [0, \infty)$ and direction specified by two angles: the angle $\phi \in [-\pi / 2, \pi / 2]$ out of the pre-supernova binary orbital plane, and the angle $\omega \in [0, 2 \pi)$ between the \textit{positive} $x$-axis and the projection of $\mathbf{v}_{\mathrm{k}}$ onto the pre-supernova binary orbital plane (see Figure \ref{fig:kickdiagram}).

This linear momentum kick causes a misalignment $\theta_{0}$ between the pre- and post-supernova binary orbital angular momentum vectors, while the rotational axes of the individual binary components remain unchanged (see Figure \ref{fig:orbitdiagram}). That is, the model assumes the supernova imparts no angular momentum to the natal black hole. Therefore, $\theta_{0}$ is equivalent to the spin-orbit misalignment angle at black hole birth, which we refer to as the \textit{initial} spin-orbit misalignment angle.

%----------------------------------------------------------------------------------------------------
% CONDITIONAL DENSITY OF THE INITIAL SPIN-ORBIT MISALIGNMENT ANGLE
\subsection{Conditional Density of the Initial Spin-Orbit Misalignment Angle}
\label{sec:Pi}
For a particular value of the pre-supernova relative orbital speed $v_{\mathrm{orb}}$ and a given natal kick velocity distribution $P( v_{\mathrm{k}} )$ (see Figure \ref{fig:Pvpec} and \S\ref{sec:vk}), the conditional density of $\theta_{0}$ is (see Appendix \ref{sec:appB}),
\begin{equation}
P\left( \theta_{0} | v_{\mathrm{orb}} \right) = \frac{1}{2 \pi} \iint_{R} P\left( v_{\mathrm{k}} \right) \frac{\left| \sin\left( \phi \right) \right|}{\left| \sin\left( \omega \right) \right| \sin^{2}\left( \theta_{0} \right)} dv_{\mathrm{k}} d\phi, \label{eqn:Pi}
\end{equation}
where $\omega = \omega\left( v_{\mathrm{k}}, \phi, \theta_{0} \right)$ from the relation \citep{Martin2009},
\begin{equation}
\cos\left( \omega \right) = \frac{v_{\mathrm{orb}}}{v_{\mathrm{k}}} \frac{1}{\cos\left( \phi \right)} - \frac{| \tan\left( \phi \right) |}{\tan\left( \theta_{0} \right)}.
\end{equation}

Several constraints derived in Appendices \ref{sec:appA} and \ref{sec:appB} combine to whittle down the integration region $R$ in $\left( v_{\mathrm{k}}, \phi \right)$-space of Equation \eqref{eqn:Pi}. The first constraint on $R$ is the mathematical consideration that $\cos\left( \omega \right)$ be real-valued, which restricts $v_{\mathrm{k}}$ to lie between $v_{+}$ and $v_{-}$, where \citep{Martin2009},
\begin{equation}
v_{\pm} = \frac{v_{\mathrm{orb}}}{\cos\left( \phi \right)} \left[ \frac{\left| \tan\left( \phi \right) \right|}{\tan\left( \theta_{0} \right)} \pm 1 \right]^{-1}. \label{eqn:vpm}
\end{equation}

Requiring the binary to remain intact implies a second constraint on $R$ that $v_{\mathrm{k}}$ be less than that required to unbind the system \citep{BrandtPodsiadlowski1995, Martin2009},
\begin{equation}
v_{\mathrm{bound}} = v_{\mathrm{orb}} \left[ \sqrt{1 + 2 \frac{M^{\prime}}{M} + \frac{\sin^{2}\left( \phi \right)}{\tan^{2} \left( \theta_{0} \right)}} - \frac{\left| \sin\left( \phi \right) \right|}{\tan\left( \theta_{0} \right)} \right], \label{eqn:vbound}
\end{equation}
where the total binary system mass pre- and post-supernova is $M = M_{1} + M_{2}$ and $M^{\prime} = M_{1}^{\prime} + M_{2}^{\prime}$.

The energy of the post-supernova system must be greater than the effective potential energy, which results in a third constraint on $R$ that $v_{\mathrm{k}}$ must exceed,
\begin{align}
&v_{\mathrm{eff}} = \nonumber \\
&v_{\mathrm{orb}} \left[ \sqrt{1 + \frac{M^{\prime}}{M} \frac{\left( 1 - e^{\prime 2} \right) v_{\mathrm{orb}}^{2}}{G M / a^{\prime}} + \frac{\sin^{2}\left( \phi \right)}{\tan^{2} \left( \theta_{0} \right)}} - \frac{\left| \sin\left( \phi \right) \right|}{\tan\left( \theta_{0} \right)} \right], \label{eqn:veff}
\end{align}
where $e^{\prime}$ and $a^{\prime}$ are the eccentricity and mean separation of the binary orbit immediately post-supernova.

The supernova kick gives the system as a whole a velocity $\mathbf{v}_{\mathrm{sys}}$ relative to the pre-supernova center of mass frame. A fourth constraint on $R$ comes from knowledge of this systemic velocity immediately post-supernova \citep{BrandtPodsiadlowski1995, Martin2010},\footnote{The convention that the black hole progenitor moves in the \textit{negative} $x$-direction requires a ``$+$'' sign for the second term on the right-hand side of Equation (10) in \citet{Martin2010}, whose inconsistent $x$-direction conventions for their $\cos(\omega)$ and $v_{\mathrm{sys}}^{2}$ equations led to incorrect constraints on the allowable ($v_{\mathrm{k}}$, $\phi$)-space.}
\begin{align}
v_{\mathrm{sys}}^{2} &= \frac{M_{1}^{\prime 2}}{M^{\prime 2}} v_{\mathrm{k}}^{2} - 2 f \frac{M_{1}^{\prime} M_{2}^{\prime}}{M^{\prime 2}} \frac{\left| \sin\left( \phi \right) \right|}{\tan\left( \theta_{0} \right)} v_{\mathrm{orb}} v_{\mathrm{k}} \nonumber \\
&+ f \frac{M_{2}^{\prime}}{M^{\prime 2}} \left( 2 M_{1}^{\prime} + f M_{2}^{\prime} \right) v_{\mathrm{orb}}^{2}, \label{eqn:vsys}
\end{align}
where $f = 1 - M^{\prime} / M$ is the fractional mass loss from the binary system due to the supernova.

Combining all of these constraints determines the integration region $R$ in $\left( v_{\mathrm{k}}, \phi \right)$-space for a given misalignment angle $\theta_{0}$, pre-supernova relative orbital speed $v_{\mathrm{orb}}$, black hole progenitor mass $M_{1}$, and post-supernova system parameters $\left\{ \right.$$M_{1}^{\prime}$, $M_{2}^{\prime}$, $e^{\prime}$, $a^{\prime}$$\left. \right\}$.

The pre-supernova relative orbital speed $v_{\mathrm{orb}}$ appears in the expressions for $v_{\pm}$, $v_{\mathrm{bound}}$, $v_{\mathrm{eff}}$, and $v_{\mathrm{sys}}$, which collectively determine the allowable integration region $R$ when calculating the spin-orbit misalignment angle conditional density $P\left( \theta_{0} | v_{\mathrm{orb}} \right)$. Therefore, $v_{\mathrm{orb}}$ possesses constraining power over $P\left( \theta_{0} | v_{\mathrm{orb}} \right)$. Indeed, we will see in \S\ref{sec:constraints} that whether the supernova kick model is deemed acceptable or rejectable hinges on the value of $v_{\mathrm{orb}}$.

%===========================================================================
%===========================================================================
% VELOCITY CONSTRAINTS
\section{Velocity Constraints}
\label{sec:vpec}
Here, we reasonably assume that the pre-supernova binary system participated in local Galactic rotation within the Galactic plane. This means that the systemic velocity $\mathbf{v}_{\mathrm{sys}}$ of Equation \ref{eqn:vsys} is equivalent to the peculiar velocity $\mathbf{v}_{\mathrm{pec}}$ of the immediate post-supernova system. The peculiar velocity of a source is defined by differencing its local Galactic circular rotational velocity $\mathbf{v}_{\mathrm{circ}}(R)$ from its Galactocentric velocity $\mathbf{v}$,\footnote{We caution that incorrect peculiar velocities for X-ray binaries exist in the literature \citep[e.g.,][]{Willems2005, Fragos2009}.}
\begin{equation}
\begin{bmatrix}
v_{\mathrm{pec}, X} \\
v_{\mathrm{pec}, Y}  \\
v_{\mathrm{pec}, Z}  \\
\end{bmatrix}
=
\begin{bmatrix}
v_{X} \\
v_{Y} \\
v_{Z} \\
\end{bmatrix}
-
\begin{bmatrix}
v_{\mathrm{circ}, X}(R) \\
v_{\mathrm{circ}, Y}(R) \\
0 \\
\end{bmatrix}
,
\end{equation}
where $R = \sqrt{X^{2} + Y^{2}}$ is the radial distance of the source from the $Z$-axis in a right-handed Galactocentric reference frame with coordinates $( X, Y, Z )$ and $XY$-plane coinciding with the Galactic plane. Axes directions are: $+X$ toward the Galactic center and along the projection of the Sun's position onto the Galactic plane; $+Y$ along the Galactic rotational velocity at the position of the Sun; and $+Z$ toward the north Galactic pole \citep{Blaauw1960, JohnsonSoderblom1987, ReidBrunthaler2004}.

% TABLE 3:
\setlength{\tabcolsep}{0pt}
\begin{table}[!t]
\centering
\begin{tabular}{l c c c}
\toprule
Parameter [Units]\dotfill & Sym. & Value & Ref. \\
\midrule
Sun's distance to Galactic center $[\mathrm{kpc}]$\dotfill & $R_{0}$ & $8.178 \pm 0.035$ & 1 \\
Sun's height above Galactic plane $[\mathrm{pc}]$\dotfill & $Z_{0}$ & $20.8 \pm 0.3$ & 2 \\
Sun's peculiar radial velocity $[\mathrm{km/s}]$\dotfill & $U_{\odot}$ & $10.0 \pm 1.0$ & 3 \\
Sun's peculiar rotation velocity $[\mathrm{km/s}]$\dotfill & $V_{\odot}$ & $12.0 \pm 2.0$ & 3 \\
Sun's peculiar vertical velocity $[\mathrm{km/s}]$\dotfill & $W_{\odot}$ & $7.3 \pm 0.4$ & 4 \\
Galactic circular speed at $R_{0}$ $[\mathrm{km/s}]$\dotfill & $\Theta_{0}$ & $236.9 \pm 4.2$ & 1 \\
\midrule
\bottomrule
\end{tabular}
\caption{Parameter choices for solar position ($R_{0}$, $Z_{0}$), solar motion ($U_{\odot}$, $V_{\odot}$, $W_{\odot}$), and Galactic motion $\Theta_{0}$ needed for the equatorial-to-Galactocentric coordinate transformation. \textsc{References:} (1) \citet{GRAVITY2019}; (2) \citet{BennettBovy2019}; (3) \citet{BlandHawthornGerhard2016}; (4) \citet{Schonrich2010}.}
\label{tab:SunGalPars}
\end{table}

% FIGURE 5:
%\begin{comment}
\begin{figure*}[!th]
  \begin{center}
  \includegraphics[width=1.0\textwidth]{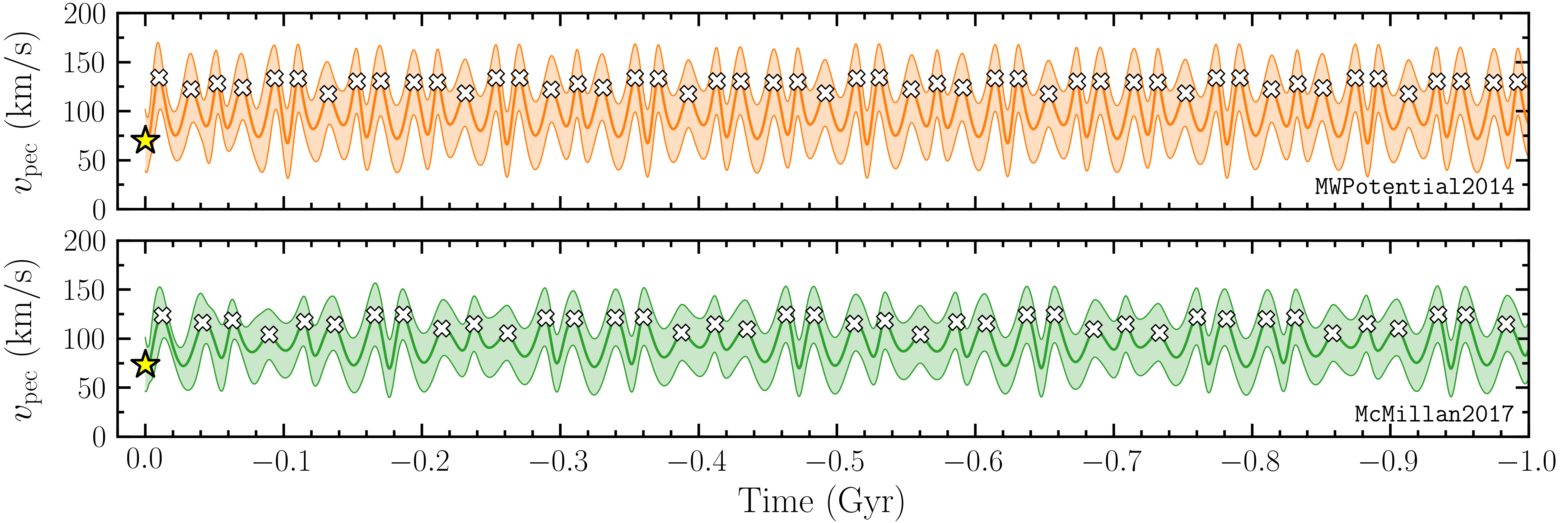}
  \vspace{-4mm}
  \caption{Peculiar velocity $v_{\mathrm{pec}}$ of V4641 Sgr traced backwards in time from today (\textit{yellow star}) out to 1 Gyr in the past using the Galactic potential models \texttt{MWPotential2014} (\textit{top panel}) and \texttt{McMillan2017} (\textit{bottom panel}) in \texttt{galpy}. The \textit{shaded region} shows the $v_{\mathrm{pec}}$ standard deviation from Monte Carlo sampling. The formation of V4641 Sgr as a black hole system, whether through a binary or dynamical channel, likely occurred within the Galactic plane, with each \textit{white $\times$} marking a Galactic plane crossing.}
  \label{fig:vpec}
  \end{center}
\end{figure*}
%\end{comment}

To constrain $v_{\mathrm{pec}}$ immediately post-supernova (and therefore $v_{\mathrm{sys}}$), the possible Galactic locations of V4641 Sgr at the moment of black hole birth must be known. To this end, we begin by calculating the present day Galactocentric position and velocity of V4641 Sgr by transforming the J2000 ICRS astrometric quantities $\left\{\right.$$\alpha$, $\delta$, $d$, $\mu_{\alpha \ast}$, $\mu_{\delta}$, $\gamma$$\left.\right\}$ in Table \ref{tab:V4641} to a Galactocentric frame using the \texttt{Astropy} core \texttt{Python} package \citep{Astropy2013, Astropy2018}. Table \ref{tab:SunGalPars} lists our parameter choices for the equatorial-to-Galactocentric coordinate transformation: solar position ($R_{0}$, $Z_{0}$), peculiar solar motion ($U_{\odot}, V_{\odot}, W_{\odot}$), and circular rotation speed $\Theta_{0}$ of the local standard of rest around the Galactic center. Today, V4641 Sgr has Galactocentric position $(X, Y, Z) = (-2.0\pm0.7, 0.73\pm0.08, -0.51 \pm 0.06)~\mathrm{kpc}$, and Galactocentric speed $v = 272 \pm 5~\mathrm{km/s}$. All quoted uncertainties in this section follow from Monte Carlo sampling of 10,000 combinations of \textit{all} astrometric and coordinate transformation parameters. We randomly sample from a normal distribution for each parameter, using its estimated value and 68\%-level uncertainty.

Despite not knowing the age of the black hole in V4641 Sgr, the relatively high mass of the companion star ($M_{\star} \simeq 3~M_{\odot}$) implies a system lifetime $< 1~\mathrm{Gyr}$. Therefore, we can place limits on the black hole birth location and the peculiar velocity of the system by tracing the orbital trajectory of V4641 Sgr backwards in time through the Galaxy for 1 Gyr, using its current position and velocity vectors as initial conditions and a timestep of 0.5 Myr.\footnote{We implicitly assume that any secular mass loss is symmetric (e.g., a stellar wind), so does not give the system a velocity boost.} Orbital integration is done using the \texttt{galpy} galactic dynamics \texttt{Python} package \citep{Bovy2015}, accounting for its left-handed coordinate convention. We try two different realistic Galactic potential models: \texttt{MWPotential2014} \citep{Bovy2012, Bovy2015} and \texttt{McMillan2017} \citep{McMillan2017}. Importantly, for every Monte Carlo iteration we always calibrate the Galactic potential model to the sampled parameter set $\left\{\right.$$R_{0}$, $Z_{0}$, $U_{\odot}$, $V_{\odot}$, $W_{\odot}$, $\Theta_{0}$$\left.\right\}$. This ensures that the model produces a Galactic rotation curve $\Theta(R)$ consistent with our parameter choices; that is, $\Theta(R_{0}) = \Theta_{0}$. This rescaling will cause a slight tension with the observational fits on which the models are based, but should still be reasonable and is necessary to calculate peculiar velocities self-consistently.

Figure \ref{fig:vpec} shows the 1-Gyr peculiar velocity history of V4641 Sgr, adopting either the Galactic potential model \texttt{MWPotential2014} (\textit{top panel}) or \texttt{McMillan2017} (\textit{bottom panel}), which constrain the peculiar velocity magnitude today to be $70^{+16}_{-19}~\mathrm{km/s}$ and $74^{+11}_{-12}~\mathrm{km/s}$, respectively. Notably, the positional history of V4641 Sgr cannot be accurately traced back beyond a few Myr when the trajectories begin to diverge between two plausible Galactic potential models. This means that we cannot reliably calculate the position or peculiar velocity of V4641 Sgr at any specific time in the past. Even though the orbital trajectories are not exact, they should do a reasonable job of sampling a representative swath of past locations for V4641 Sgr. In this sense, the distributions of positions and peculiar velocities are useful, despite the systematic uncertainties inherent to the Galactic models.

The black hole likely formed when V4641 Sgr was within the Galactic plane, but its moment of birth cannot be determined because of uncertainties in the Galactic model and multiple Galactic plane crossings (\textit{white $\times$'s} in Figure \ref{fig:vpec}). However, we can estimate the minimum black hole age as the time elapsed since the most recent Galactic plane crossing: $t_{\times} = 10.0^{+1.3}_{-1.1}~\mathrm{Myr}$ (\texttt{MWPotential2014}); $t_{\times} = 12.5^{+2.2}_{-2.1}~\mathrm{Myr}$ (\texttt{McMillan2017}). We can also estimate $v_{\mathrm{pec}}$ at black hole birth by only considering the times when the system crossed the Galactic plane \citep[e.g.,][]{Atri2019}.

% FIGURE 6:
%\begin{comment}
\begin{figure}[!th]
  \begin{center}
  \includegraphics[width=0.495\textwidth]{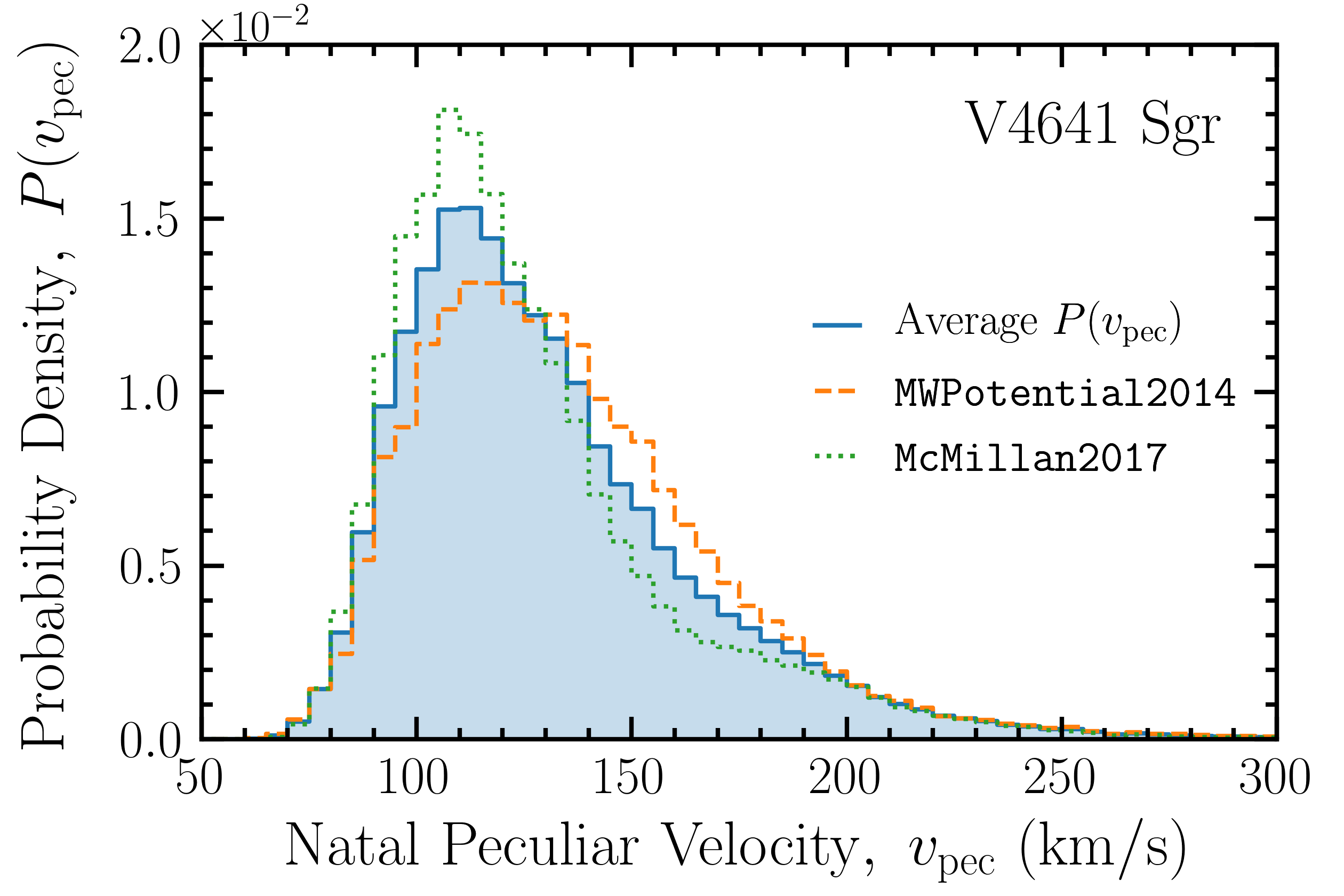}
  \vspace{-4mm}
  \caption{Natal peculiar velocity distribution $P( v_{\mathrm{pec}} )$ for V4641 Sgr (\textit{blue solid line}), derived from averaging the results of choosing either the \texttt{MWPotential2014} (\textit{orange dashed line}) or \texttt{McMillan2017} (\textit{green dotted line}) Galactic model in \texttt{galpy}. The construction of $P( v_{\mathrm{pec}} )$ followed from collecting the $v_{\mathrm{pec}}$ values at every Galactic plane crossing over the last 1 Gyr (\textit{white $\times$'s} in Figure \ref{fig:vpec}) for each of the 10,000 Monte Carlo iterations. Going forward, we interpret $P( v_{\mathrm{pec}} )$ as the natal kick velocity distribution $P( v_{\mathrm{k}} )$ specific to V4641 Sgr (see \S\ref{sec:vk}).}
  \vspace{-4mm}
  \label{fig:Pvpec}
  \end{center}
\end{figure}
%\end{comment}

Figure \ref{fig:Pvpec} shows this natal peculiar velocity distribution $P( v_{\mathrm{pec}} )$ specific to V4641 Sgr, generated by treating each Galactic plane crossing over the last 1 Gyr as the location and moment of black hole birth with equal likelihood. Encouragingly, two different Galactic potential models produce similar natal peculiar velocity distributions, so we average them together to arrive at our favored $P( v_{\mathrm{pec}} )$, shown by the \textit{blue solid line}. The natal $v_{\mathrm{pec}}$ median value is $123~\mathrm{km/s}$ and we treat the inter-95\% range of $P( v_{\mathrm{pec}} )$ as the extrema for the natal peculiar velocity (see Equation \ref{eqn:vkphi_vsys}): $v_{\mathrm{pec}}^{\mathrm{min}} = 78~\mathrm{km/s}$ and $v_{\mathrm{pec}}^{\mathrm{max}} = 202~\mathrm{km/s}$. These $v_{\mathrm{pec}}$ limits will be crucial for constraining the degree of spin-orbit misalignment that the natal kick model can produce (see \S\ref{sec:constraints}).

In \S\ref{sec:vk}, we will justify treating $P( v_{\mathrm{pec}} )$ as the natal kick distribution $P( v_{\mathrm{k}} )$ for V4641 Sgr, as needed by the natal kick model. Following \citet{Salvesen2019}, we tried fitting analytic distributions to $P( v_{\mathrm{pec}} )$ in Figure \ref{fig:Pvpec}, but neither a Maxwellian nor a normal distribution give an acceptable fit. Going forward then, we elect to use the binned $P( v_{\mathrm{pec}} )$ distribution shown in Figure \ref{fig:Pvpec} for $P( v_{\mathrm{k}} )$ in the natal kick model. The bin widths are $5~\mathrm{km/s}$ and we linearly interpolate between bins to calculate the probability density for an arbitrary $v_{\mathrm{k}}$.

Results for the most recent Galactic plane crossing time $t_{\times}$ and the inter-95\% natal peculiar velocity range from $P( v_{\mathrm{pec}} )$ are similar if we instead use the \citet{Orosz2001} radial systemic velocity $\gamma = 107.4 \pm 2.9~\mathrm{km/s}$ (see \S\ref{sec:binpars}). For $t_{\times}$, \texttt{MWPotential2014} gives $10.0^{+1.3}_{-1.1}~\mathrm{Myr}$ and \texttt{McMillan2017} gives $12.5\pm2.1~\mathrm{Myr}$. The averaged $P( v_{\mathrm{pec}} )$ distribution gives $v_{\mathrm{pec}}^{\mathrm{min}} =  93~\mathrm{km/s}$ and $ v_{\mathrm{pec}}^{\mathrm{max}} = 205~\mathrm{km/s}$. Ruling out large spin-orbit misalignments comes down to the upper limit $v_{\mathrm{pec}}^{\mathrm{max}}$ (see \S \ref{sec:constraints}), which is practically the same for either choice of $\gamma$.

%----------------------------------------------------------------------------------------------------
% NATAL KICK VELOCITY DISTRIBUTION
\subsection{Natal Kick Velocity Distribution}
\label{sec:vk}
Predicting the likelihood of a spin-orbit misalignment with the natal kick model requires knowledge of the natal kick velocity distribution $P( v_\mathrm{k} )$, which is not well-constrained. Black hole and neutron star X-ray binaries have similar vertical displacements out of the Galactic plane \citep[e.g.,][]{JonkerNelemans2004}, suggesting they both receive sizable velocity kicks upon birth. Recent work revealed a wide distribution of pulsar birth velocities that can be described by a double Maxwellian, with one population having an average speed of $120~\mathrm{km/s}$ and the other $540~\mathrm{km/s}$ \citep{Verbunt2017}, while older work found consistency with a single Maxwellian having an average speed of $400 \pm 40~{\mathrm{km/s}}$ \citep{Hobbs2005}. Compared to pulsars, the small sample of confirmed black hole X-ray binaries suggests slower birth velocities, collectively having a distribution described by a Gaussian with mean $107 \pm 16~\mathrm{km/s}$ and standard deviation $56 \pm 14~\mathrm{km/s}$ \citep{Atri2019}. In contrast, population models find that black hole X-ray binaries require comparable velocity kicks, but greater linear momentum kicks, compared to their neutron star counterparts to achieve the observed displacements out of the Galactic plane \citep{Repetto2012, Repetto2017}.

In the natal kick model, the birth velocity discussed above is equivalent to the post-supernova systemic velocity $\mathbf{v}_{\mathrm{sys}}$. But in general, $\mathbf{v}_{\mathrm{sys}}$ is \textit{not} the same as the natal kick velocity $\mathbf{v}_{\mathrm{k}}$, which combined with mass loss during the supernova event determines $\mathbf{v}_{\mathrm{sys}}$. For the special case of no mass loss ($f = 0$), Equation \eqref{eqn:vsys1} shows that $v_{\mathrm{k}}$ and $v_{\mathrm{sys}}$ are equivalent. For the opposite scenario of no natal kick ($v_{\mathrm{k}} = 0$) and only mass loss, the systemic velocity is \citep[e.g.,][]{Nelemans1999},
\begin{equation}
v_{\mathrm{sys}} = f \frac{M_{2}^{\prime}}{M^{\prime}} v_{\mathrm{orb}}, \label{eqn:vsysMassLoss}
\end{equation}
and the binary remains bound if $f < 0.5$ \citep{Blaauw1961}. Therefore, Equation \eqref{eqn:vsysMassLoss} implies a firm upper limit of $v_{\mathrm{sys}} < 0.5 v_{\mathrm{orb}}$ for the contribution of supernova mass loss alone to the systemic velocity. Exploring a wide range of component masses more appropriate to V4641 Sgr (see \S\ref{sec:bse}), we find that $v_{\mathrm{sys}} \lesssim 0.15 v_{\mathrm{orb}}$ is a more realistic maximum contribution to $v_{\mathrm{sys}}$ due to mass loss alone. This means that mass loss from a wide binary with $v_{\mathrm{orb}} \sim 100~\mathrm{km/s}$ leads to a negligible $v_{\mathrm{sys}} \lesssim 15~\mathrm{km/s}$ \citep[e.g.,][]{Nelemans1999}, while mass loss from a close binary with $v_{\mathrm{orb}} \sim 1000~\mathrm{km/s}$ can produce a substantial $v_{\mathrm{sys}} \lesssim 150~\mathrm{km/s}$. Thus, if the pre-supernova relative orbital speed $v_{\mathrm{orb}}$ was not too large, such that the binary was not ultra-compact and/or the primary was not ultra-massive, then $v_{\mathrm{k}} \simeq v_{\mathrm{sys}}$ is a reasonable approximation. Notably, this approximation improves for a more massive companion star because, for a given amount of mass lost from the primary star during the supernova, there is less fractional mass lost $f$ from the binary.

Ultimately, we want the natal kick velocity distribution $P( v_{\mathrm{k}} )$, as needed by the natal kick model. The $v_{\mathrm{k}} \simeq v_{\mathrm{sys}}$ approximation justified above implies the equivalence of $P( v_{\mathrm{k}} )$ to the systemic velocity distribution $P( v_{\mathrm{sys}} )$ at the time of black hole birth. But $P( v_{\mathrm{sys}} )$ is equivalent to the natal peculiar velocity distribution $P( v_{\mathrm{pec}} )$ if we assume local Galactic rotation for the pre-supernova system. Therefore, we take $P( v_{\mathrm{k}} ) \simeq P( v_{\mathrm{pec}} )$, which we observationally constrained in Figure \ref{fig:Pvpec}, as the natal kick velocity distribution specific to V4641 Sgr.

Finally, we reiterate that substantial impulsive mass loss from a system with an extremely large $v_{\mathrm{orb}}$ can weaken our critical approximation $v_{\mathrm{k}} \simeq v_{\mathrm{sys}}$. However, a typical natal kick has magnitude $v_{\mathrm{k}} \sim 100~\mathrm{km/s}$, which applied to an aligned system with $v_{\mathrm{orb}} \sim 1000~\mathrm{km/s}$ will generally not produce the large spin-orbit misalignments of interest (see \S\ref{sec:constraints}). Therefore, this $v_{\mathrm{k}} \simeq v_{\mathrm{sys}}$ approximation is in-line with the goals of this paper and allows for a system-specific estimate of $P( v_{\mathrm{k}} )$. What's more, this custom kick distribution is an improvement over the alternative approaches of adopting a natal kick distribution that is either heavily model-dependent (i.e., based on core-collapse physics), generic (i.e., based on proper motions of neutron star or black hole populations), or arbitrary (e.g., a uniform distribution).

%===========================================================================
%===========================================================================
% APPLYING THE MISALIGNMENT MODEL
\section{Applying the Misalignment Model}
\label{sec:constraints}
To summarize \S \ref{sec:model}, the supernova kick model takes $v_{\mathrm{orb}}$ as the conditional input parameter and uses constraints on the velocity magnitudes $\left\{ \right.$$v_{\pm}$, $v_{\mathrm{bound}}$, $v_{\mathrm{eff}}$, $v_{\mathrm{sys}}$$\left. \right\}$ cast in terms of the input parameters $\left\{ \right.$$M_{1}$, $M_{1}^{\prime}$, $M_{2}^{\prime}$, $e^{\prime}$, $a^{\prime}$$\left. \right\}$. These constraints restrict the integration region $R$ in $\left( v_{\mathrm{k}}, \phi \right)$-space when calculating the conditional density $P\left( \theta_{0} | v_{\mathrm{orb}} \right)$ for the initial spin-orbit misalignment angle $\theta_{0}$. One must also specify the natal kick velocity distribution $P( v_{\mathrm{k}} )$ for V4641 Sgr (see Figure \ref{fig:Pvpec} and \S\ref{sec:vk}).

% FIGURE 7:
%\begin{comment}
\begin{figure*}[!t]
  \begin{center}
    \includegraphics[width=0.495\textwidth]{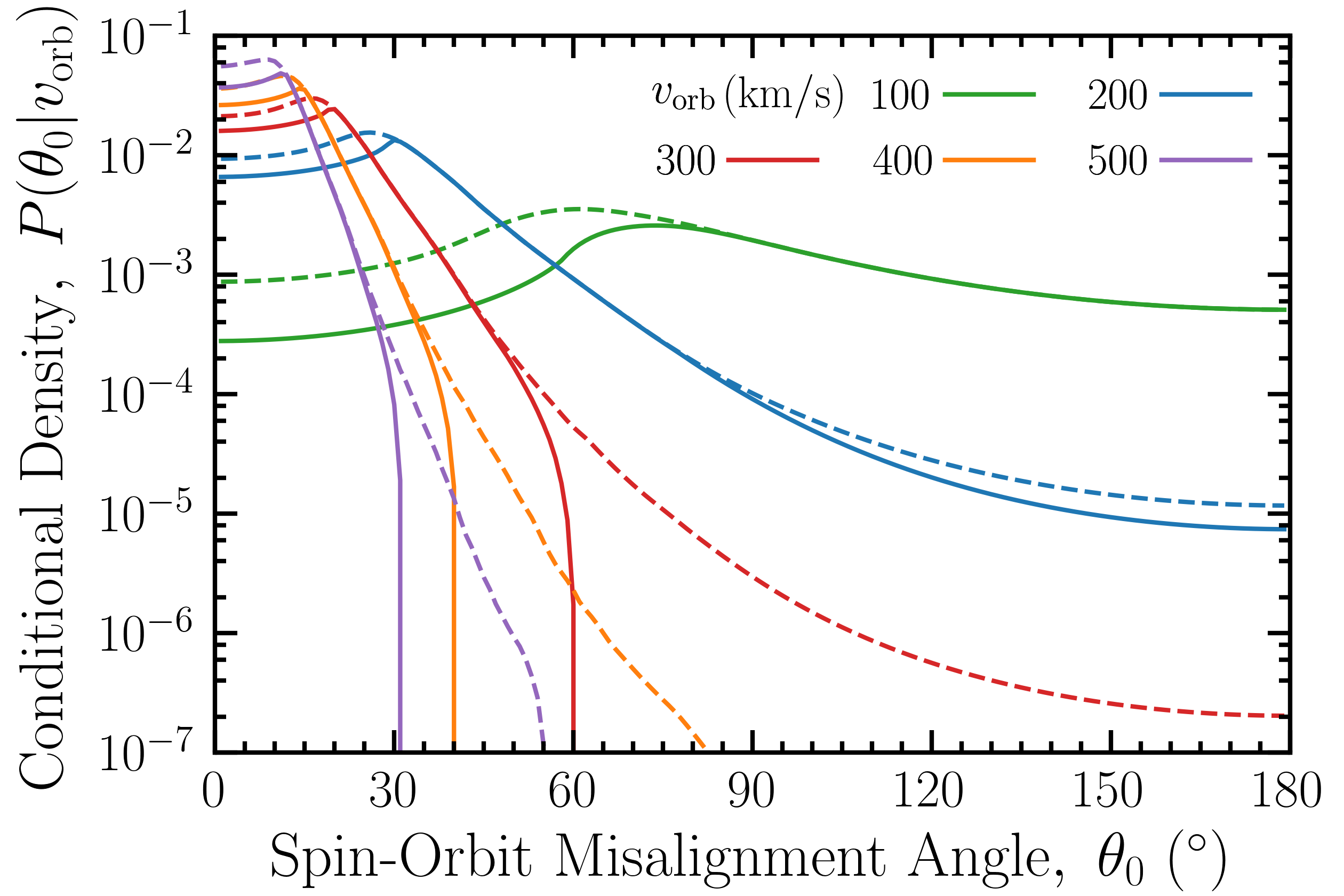}
    \hfill
    \includegraphics[width=0.495\textwidth]{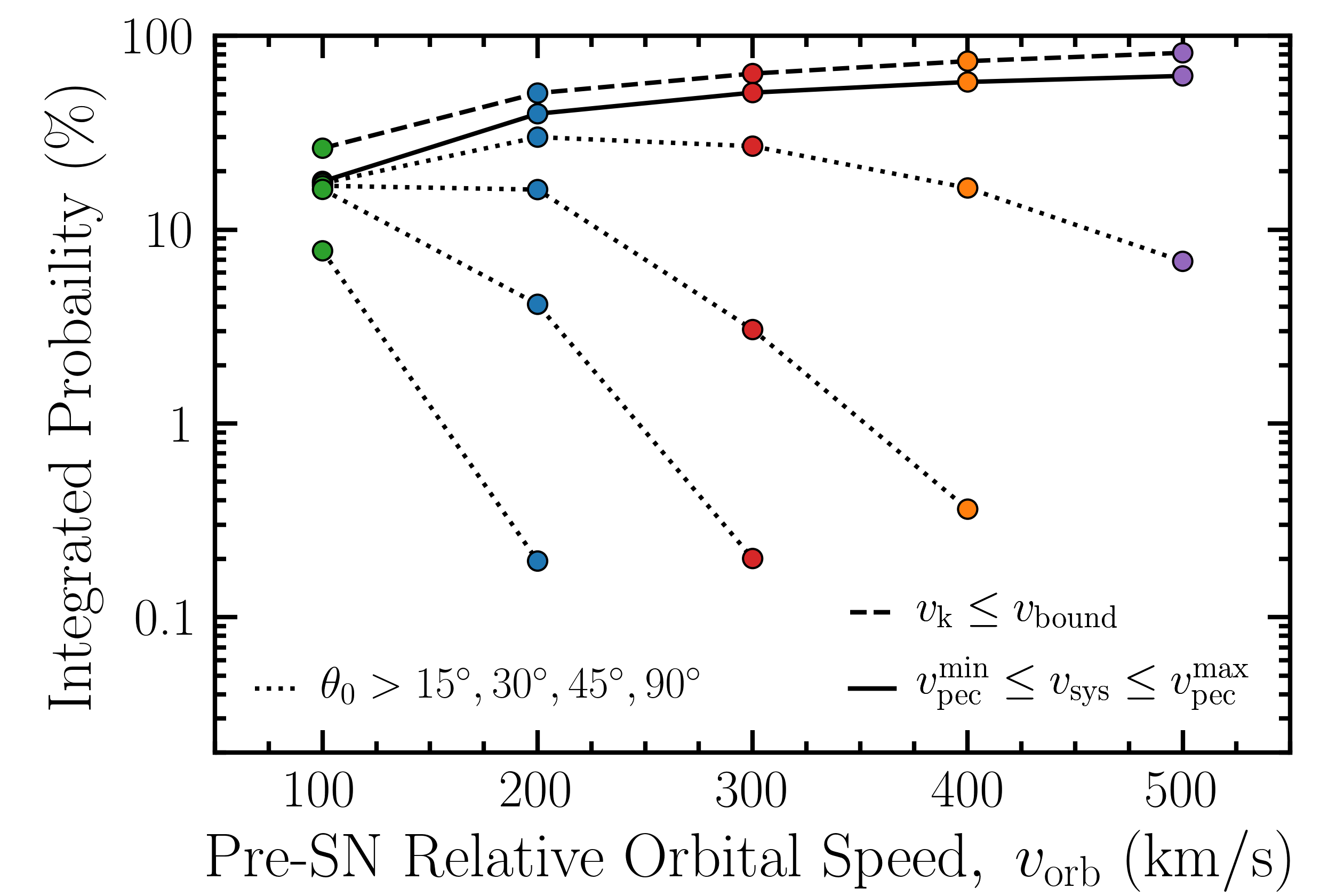}
  \caption{\textit{Left panel}: Conditional density functions $P( \theta_{0} | v_{\mathrm{orb}})$ of the spin-orbit misalignment angle $\theta_{0}$ at the time of black hole birth, and for different choices of the pre-supernova relative orbital speed between the stars $v_{\mathrm{orb}} = [100,~200,~300,~400,~500]~{\mathrm{km/s}}$ (\textit{line colors}). All curves adopt the reference model parameters: $M_{1} = 10~M_{\odot}$, $M_{1}^{\prime} = 6.4~M_{\odot}$, $M_{2}^{\prime} = 2.9~M_{\odot}$. The \textit{line style} indicates the constraints enforced on ($v_{\mathrm{k}}, \phi$)-space when calculating $P( \theta_{0} | v_{\mathrm{orb}})$: \textit{dashed lines} enforce $\min[v_{\pm}] \le v_{\mathrm{k}} \le \max[v_{\pm}]$ and $v_{\mathrm{k}} \le v_{\mathrm{bound}}$, while \textit{solid lines} further restrict $v_{\mathrm{pec}}^{\mathrm{min}} \le v_{\mathrm{sys}} \le v_{\mathrm{pec}}^{\mathrm{max}}$. Increasing $v_{\mathrm{orb}}$ can lead to a turnover in $P( \theta_{0} | v_{\mathrm{orb}})$, meaning there exists a maximum spin-orbit misalignment $\theta_{0}^{\mathrm{max}}$ that the kick model can produce (e.g., $\theta_{0}^{\mathrm{max}} = 60^{\circ}$ for $v_{\mathrm{orb}} = 300~\mathrm{km/s}$). \\
\textit{Right panel}: Each \textit{dot} shows the integrated probability $\int_{\theta_{0}^{\mathrm{min}}}^{180^{\circ}} P( \theta_{0} | v_{\mathrm{orb}} ) d\theta_{0}$ of its respective curve from the \textit{left panel}, displayed as a function of $v_{\mathrm{orb}}$ (\textit{dot color}). The \textit{line style} connecting a set of dots follows the same convention as the \textit{left panel}, signifying the $(v_{\mathrm{k}}, \phi)$-space constraints used in calculating $P( \theta_{0} | v_{\mathrm{orb}})$. Setting the lower integration bound to be $\theta_{0}^{\mathrm{min}} = 0$ and calculating $P( \theta_{0} | v_{\mathrm{orb}})$ over all physically allowable $(v_{\mathrm{k}}, \phi)$-space necessarily yields 100\% integrated probability (\textit{not shown}). Requiring the system to remain bound following the supernova kick naturally reduces the integrated probability (\textit{dashed line}). Enforcing the additional $v_{\mathrm{sys}}$ constraints lowers the integrated probabilities further (\textit{solid line}). From \textit{top} to \textit{bottom}, the \textit{dotted lines} show the probability of producing a system that satisfies all of these constraints and is misaligned by at least $\theta_{0}^{\mathrm{min}} = [15^{\circ},~30^{\circ},~45^{\circ},~90^{\circ}]$.}
  \vspace{-2mm}
  \label{fig:Pi_vorb}
  \end{center}
\end{figure*}
%\end{comment}

To calculate $P( \theta_{0} | v_{\mathrm{orb}} )$ for each $\theta_{0}$, we specify values for the input parameters and impose the constraints,\footnote{We use the \texttt{SciPy} function \texttt{nquad} for numerical integrations, using 181 bins in $\phi \in [-90^{\circ}, 90^{\circ}]$ and 179 bins in $\theta_{0} \in [1^{\circ}, 179^{\circ}]$. In \S\ref{sec:bse}, we replace \texttt{nquad} with an approach that discretizes the integration region $R$ in ($v_{\mathrm{k}}, \phi$)-space onto a grid, calculates the integrand in each grid zone, and then sums the results together, which is much faster and is accurate enough for our purposes.}
\begin{align}
\min\left[ v_{+}, v_{-} \right] \le v_{\mathrm{k}}& \le \max\left[ v_{+}, v_{-} \right] \label{eqn:vkphi_vpm} \\
v_{\mathrm{eff}} \le v_{\mathrm{k}}& \le v_{\mathrm{bound}} \label{eqn:vkphi_veff_vbound} \\
v_{\mathrm{pec}}^{\mathrm{min}} \le v_{\mathrm{sys}}& \le v_{\mathrm{pec}}^{\mathrm{max}}. \label{eqn:vkphi_vsys}
\end{align}
For the moment, we do not use the constraint $v_{\mathrm{k}} \ge v_{\mathrm{eff}}$, which would require specifying $e^{\prime}$ and $a^{\prime}$. Here, we reintroduce the peculiar velocity $\mathbf{v}_{\mathrm{pec}}$, which is the residual velocity after subtracting off the local Galactic rotation from the systemic velocity $\mathbf{v}_{\mathrm{sys}}$. Assuming the pre-supernova system was in local Galactic rotation, the peculiar and systemic velocities are equivalent. For V4641 Sgr, we constrained the minimum and maximum peculiar velocity magnitude immediately post-supernova to be $v_{\mathrm{pec}}^{\mathrm{min}} = 78~\mathrm{km/s}$ and $v_{\mathrm{pec}}^{\mathrm{max}} = 202~\mathrm{km/s}$ (see \S \ref{sec:vpec}).

As a demonstration, we apply the supernova kick model to a reference binary system with input parameters: $M_{1} = 10~M_{\odot}$, $M_{1}^{\prime} = 6.4~M_{\odot}$, $M_{2}^{\prime} = 2.9~M_{\odot}$. We chose $M_{1}^{\prime}$ and $M_{2}^{\prime}$ to match the component masses of V4641 Sgr today, but will show in \S \ref{sec:bse} that many combinations of post-supernova system parameters can potentially evolve to a state consistent with V4641 Sgr today. The \textit{left panel} of Figure \ref{fig:Pi_vorb} shows $P(\theta_{0} | v_{\mathrm{orb}})$ for different choices of $v_{\mathrm{orb}} = [100,~200,~300,~400,~500]~\mathrm{km/s}$ (\textit{line colors}) and subject to different constraints (\textit{line styles}), while the \textit{right panel} shows various integrated probabilities corresponding to these $P(\theta_{0} | v_{\mathrm{orb}})$ distributions.

Imposing only the constraint $\min[ v_{\pm} ] \le v_{\mathrm{k}} \le \max[ v_{\pm} ]$ amounts to integrating over all physically permissible $( v_{\mathrm{k}}, \phi )$-space. We confirmed that, for a $P(\theta_{0} | v_{\mathrm{orb}})$ distribution subject to only this constraint, integrating over all spin-orbit misalignment angles evaluates to unity. This means that $P(\theta_{0} | v_{\mathrm{orb}})$ is properly normalized, but it is not a physically meaningful distribution due to its inclusion of unbound systems, where the concept of a spin-orbit misalignment does not make sense.

Therefore, we apply the additional constraint that, following the instantaneous supernova kick and mass loss episode, the binary system remains bound, $v_{\mathrm{k}} \le v_{\mathrm{bound}}$. The \textit{dashed lines} in Figure \ref{fig:Pi_vorb} (\textit{left panel}) show the resulting spin-orbit misalignment angle distributions, which are physically meaningful. Integrating each of these distributions over all misalignment angles then gives the probability that the system remains bound (\textit{right panel}; \textit{dashed line} connecting the \textit{dots}). Binary survival probabilities diminish with decreasing $v_{\mathrm{orb}}$ because these more loosely bound pre-supernova systems are easier to disrupt for a given kick. For our isotropic kick model and reference set of input parameters, the binary remains intact following the supernova 82\% of the time for a close binary with $v_{\mathrm{orb}} = 500~\mathrm{km/s}$, but only 26\% of the time for a wide binary with $v_{\mathrm{orb}} = 100~\mathrm{km/s}$.

% FIGURE 8:
%\begin{comment}
\begin{figure*}[!t]
  \begin{center}
  \includegraphics[width=0.495\textwidth]{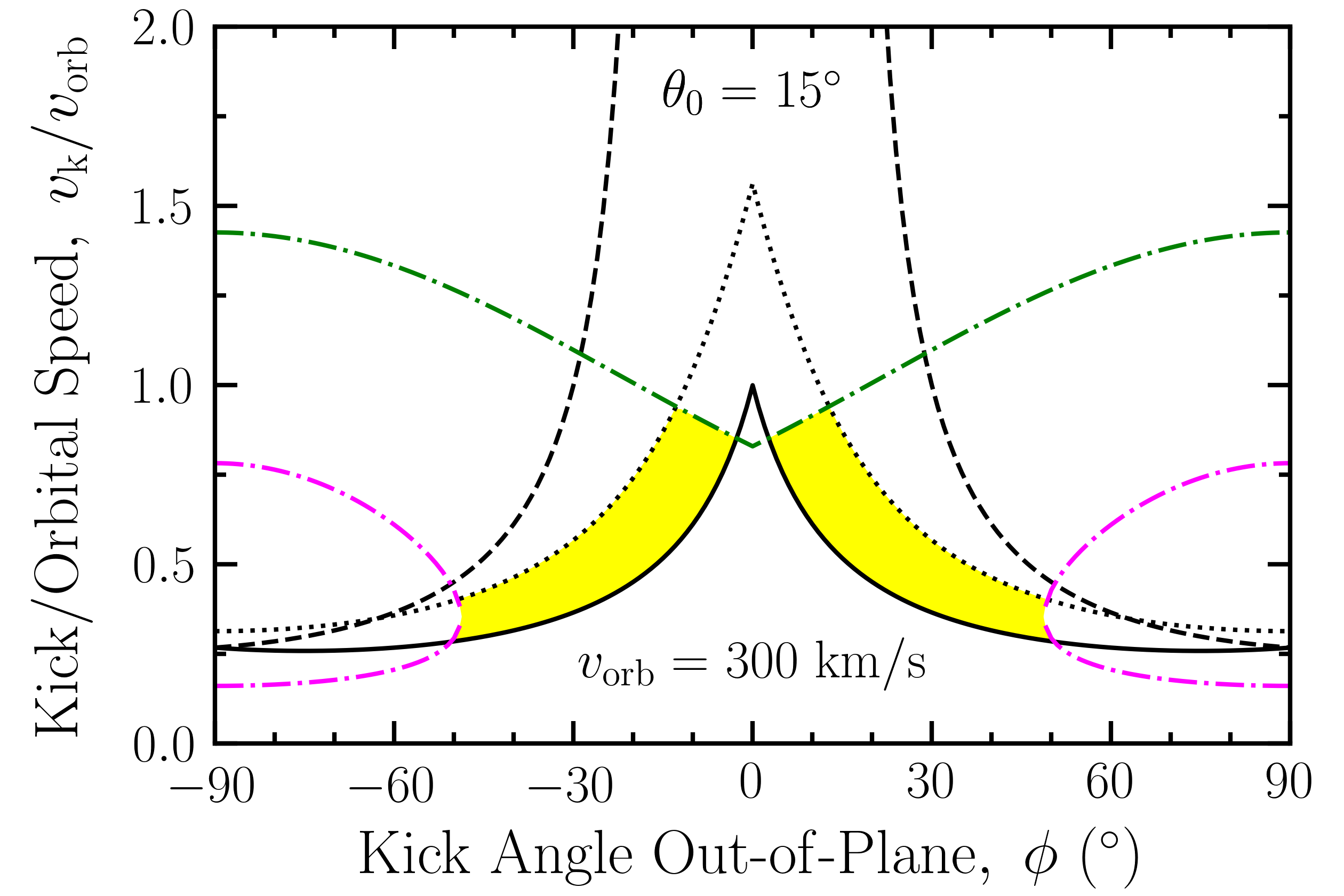}
  \hfill
  \includegraphics[width=0.495\textwidth]{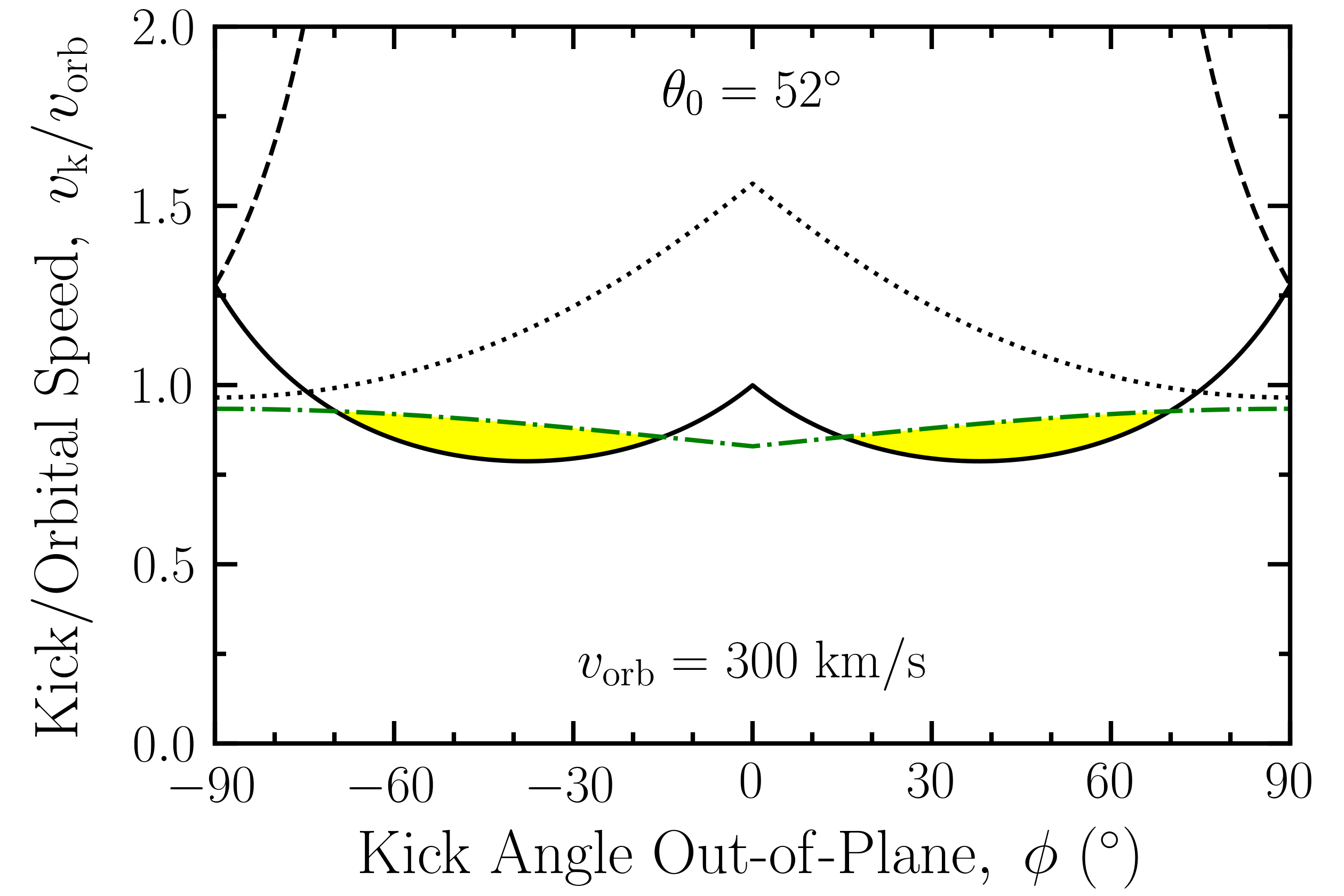}
  \caption{The \textit{yellow area} shows the integration region $R$ in $( v_{\mathrm{k}}, \phi )$-space that can produce an initial spin-orbit misalignment of $\theta_{0} = 15^{\circ}$ (\textit{left panel}) and $\theta_{0} = 52^{\circ}$ (\textit{right panel}) for a pre-supernova relative orbital speed $v_{\mathrm{orb}} = 300~\mathrm{km/s}$ and the reference model parameters: $M_{1} = 10~M_{\odot}$, $M_{1}^{\prime} = 6.4~M_{\odot}$, $M_{2}^{\prime} = 2.9~M_{\odot}$. In other words, these integration regions correspond to $\theta_{0} = 15^{\circ}$ and $\theta_{0} = 52^{\circ}$ for the \textit{red solid line} in Figure \ref{fig:Pi_vorb}. The supernova kick velocity magnitude $v_{\mathrm{k}}$ is scaled to $v_{\mathrm{orb}}$, and $\phi$ is the kick angle out of the binary plane (see Figure \ref{fig:kickdiagram}). The different curves show $v_{+}$ (\textit{solid black line}), $v_{-}$ (\textit{dashed black line}), $v_{\mathrm{bound}}$ (\textit{dotted black line}), and the constraints from $v_{\mathrm{pec}}^{\mathrm{min}} = 78~\mathrm{km/s}$ (\textit{dash-dotted magenta line}) and $v_{\mathrm{pec}}^{\mathrm{max}} = 202~\mathrm{km/s}$ (\textit{dash-dotted green line}). This upper limit on the peculiar velocity holds the most constraining power for ruling out the production of large spin-orbit misalignment angles.}
  \label{fig:vkphi}
  \end{center}
\end{figure*}
%\end{comment}

The \textit{solid lines} in Figure \ref{fig:Pi_vorb} (\textit{left panel}) show the $P(\theta_{0} | v_{\mathrm{orb}})$ distributions that also incorporate the constraints on the systemic velocity, $v_{\mathrm{pec}}^{\mathrm{min}} \le v_{\mathrm{sys}} \le v_{\mathrm{pec}}^{\mathrm{max}}$. Unsurprisingly, restricting the $v_{\mathrm{sys}}$ range of the post-supernova system further reduces the formation probability (\textit{right panel}; \textit{solid line} connecting the \textit{dots}). The lower limit $v_{\mathrm{sys}} \ge v_{\mathrm{pec}}^{\mathrm{min}}$ is responsible for reducing $P( \theta_{0} | v_{\mathrm{orb}} )$ at small misalignments, producing an uptick feature. The upper limit $v_{\mathrm{sys}} \le v_{\mathrm{pec}}^{\mathrm{max}}$ is responsible for reducing $P(\theta_{0} | v_{\mathrm{orb}})$ at large misalignments, which can become severe enough to truncate the distribution at $\theta_{0}^{\mathrm{max}}$. With increasing $v_{\mathrm{orb}}$, we see that the $P(\theta_{0} | v_{\mathrm{orb}})$ turnover progresses to lower $\theta_{0}^{\mathrm{max}}$ values, beyond which the natal kick model cannot produce (for that set of input parameters). We emphasize that the constraint $v_{\mathrm{sys}} \le v_{\mathrm{pec}}^{\mathrm{max}}$ is what allows us to rule out the natal kick model as the origin of large spin-orbit misalignments.\footnote{\citet{Martin2010} incorrectly state in \S 4.1 that out-of-plane kicks preferentially unbind systems with higher $v_{\mathrm{orb}}$, ``so cannot give rise to a large post-explosion misalignment angle''; thus, failing to attribute this effect to the $v_{\mathrm{sys}}$ upper limit.}

From \textit{top} to \textit{bottom}, the \textit{dotted lines} in Figure \ref{fig:Pi_vorb} (\textit{right panel}) show the probability of producing a spin-orbit misalignment $\theta_{0} > 15^{\circ}$, $30^{\circ}$, $45^{\circ}$, $90^{\circ}$ for a given $v_{\mathrm{orb}}$ and the reference parameter set. A missing \textit{dot} means the natal kick model cannot produce a misalignment that large because the $v_{\mathrm{sys}}$ upper limit truncates the $P( \theta_{0} | v_{\mathrm{orb}} )$ distribution. However, these integrated probabilities do not necessarily follow a monotonic trend with $v_{\mathrm{orb}}$. This makes it difficult to determine the value of $v_{\mathrm{orb}}$ that maximizes the probability of producing a spin-orbit misalignment angle greater than some value. Similarly, the $P( \theta_{0} | v_{\mathrm{orb}} )$ distributions in the \textit{left panel} intersect each other, which precludes a simple determination of the $v_{\mathrm{orb}}$ value that maximizes the probability of creating a given $\theta_{0}$. However, the truncation of $P( \theta_{0} | v_{\mathrm{orb}} )$ does follow a predictable trend of decreasing $\theta_{0}^{\mathrm{max}}$ with increasing $v_{\mathrm{orb}}$, which we will take advantage of in \S\ref{sec:bse}.

Using our reference binary system and taking $v_{\mathrm{orb}} = 300~\mathrm{km/s}$ as an example, Figure \ref{fig:vkphi} shows how the constraints discussed above conspire to pare down the integration region $R$ (\textit{yellow area}) in $( v_{\mathrm{k}}, \phi )$-space that can produce a misalignment angle of $\theta_{0} = 15^{\circ}$ (\textit{left panel}) and $\theta_{0} = 52^{\circ}$ (\textit{right panel}). For this specific set of input parameters, the constraint from the $v_{\mathrm{sys}}$ upper limit (\textit{green dash-dotted line}) lies below both the $\min[v_{\pm}]$ curve (\textit{black solid line}) and the $v_{\mathrm{bound}}$ curve (\textit{black dotted line}) when $\theta_{0} > 60^{\circ}$, which is also evident from the turnover of the \textit{red solid line} in Figure \ref{fig:Pi_vorb}. In this case, there is only a 0.04\% chance that a natal kick will produce a spin-orbit misalignment consistent with V4641 Sgr (i.e., $\theta_{0} > 52^{\circ}$). Generally, the probability of producing $\theta_{0}$ greater than some value increases as $v_{\mathrm{orb}}$ decreases.

% FIGURE 9:
%\begin{comment}
\begin{figure}[!t]
  \begin{center}
    \includegraphics[width=0.495\textwidth]{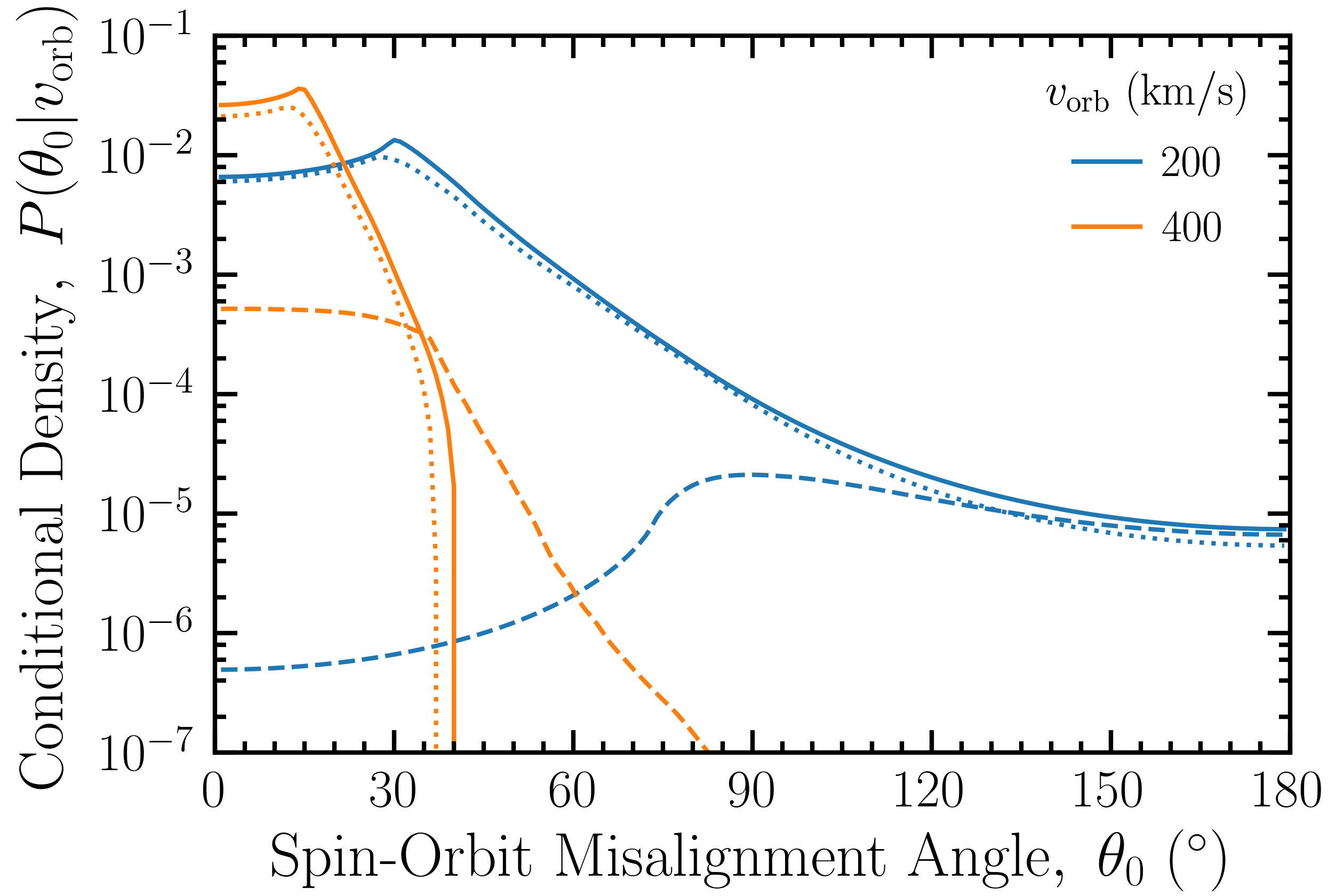}
  \caption{Effect of varying individual model input parameters on $P( \theta_{0} | v_{\mathrm{orb}})$, the conditional density function of the spin-orbit misalignment angle $\theta_{0}$ at black hole birth. All curves enforce the constraints: $\min[v_{\pm}] \le v_{\mathrm{k}} \le \max[v_{\pm}]$, $v_{\mathrm{k}} \le v_{\mathrm{bound}}$, and $v_{\mathrm{pec}}^{\mathrm{min}} \le v_{\mathrm{sys}} \le v_{\mathrm{pec}}^{\mathrm{max}}$. We show results for two different pre-supernova relative orbital speeds: $v_{\mathrm{orb}} = 200~{\mathrm{km/s}}$ (\textit{blue lines}) and $400~{\mathrm{km/s}}$ (\textit{orange lines}). \textit{Solid lines} adopt the reference model parameters: $M_{1} = 10~M_{\odot}$, $M_{1}^{\prime} = 6.4~M_{\odot}$, $M_{2}^{\prime} = 2.9~M_{\odot}$. The \textit{line style} shows the effect of changing only a single parameter in the reference model. Changing $M_{1}$ to $15~M_{\odot}$ (\textit{dotted lines}) is negligible, while changing $M_{2}^{\prime}$ to $20~M_{\odot}$ (\textit{dashed lines}) is dramatic.}
  \vspace{-4mm}
  \label{fig:Pi_M1_M2}
  \end{center}
\end{figure}
%\end{comment}

Figure \ref{fig:Pi_M1_M2} varies one parameter at a time in our reference binary system (\textit{solid lines}) to assess its effect on the resulting $P( \theta_{0} | v_{\mathrm{orb}} )$ distribution. Replacing $M_{1} = 10~M_{\odot}$ with $M_{1} = 15~M_{\odot}$ for the black hole progenitor (\textit{dotted lines}) shifts the $P( \theta_{0} | v_{\mathrm{orb}} )$ distributions downward because more mass loss leads to more unbound systems, but the effect is relatively weak and we stick with $M_{1} = 10~M_{\odot}$ going forward. To support this choice, we suppose the black hole progenitor was the naked helium core of a massive main-sequence star whose hydrogen envelope was lost during binary evolution. Helium core masses in the 8--15 $M_{\odot}$ range are expected to produce a 3--15 $M_{\odot}$ black hole \citep{Fryer1999, Heger2003}.

Making the post-supernova system much more massive by replacing $M_{2}^{\prime} = 2.9~M_{\odot}$ with $M_{2}^{\prime} = 20~M_{\odot}$ (\textit{dashed lines}) can have a strong effect on the spin-orbit misalignment angle distribution. For the $v_{\mathrm{orb}} = 400~\mathrm{km/s}$ case, large $\theta_{0}$ values that were unattainable with $M_{2}^{\prime} = 2.9~M_{\odot}$ (\textit{orange solid line}) become possible with $M_{2}^{\prime} = 20~M_{\odot}$ (\textit{orange dashed line}). For the $v_{\mathrm{orb}} = 200~\mathrm{km/s}$ case, the lower limit $v_{\mathrm{sys}} \ge v_{\mathrm{pec}}^{\mathrm{min}}$ enters in to drastically reduce the probability of small $\theta_{0}$ values (\textit{blue dashed line}).

To reiterate, spin-orbit misalignment probability distributions must be calculated on a case-by-case basis. We do this in \S\ref{sec:bse} for a comprehensive set of possible V4641 Sgr progenitors, and determine whether to accept or reject the natal kick model in each case.

%===========================================================================
%===========================================================================
% TESTING THE MISALIGNMENT MODEL
\section{Testing the Misalignment Model}
\label{sec:bse}
The natal kick model (see \S\ref{sec:model}) connects the immediate pre-/post-supernova epochs and yields the \textit{initial} spin-orbit misalignment distribution (see \S \ref{sec:constraints}). The system then evolves over time to become V4641 Sgr as we observe it \textit{today}. The task at hand, then, is to connect V4641 Sgr today to its immediate post-supernova state, which is the relevant epoch for the natal kick model.

% FIGURE 10:
%\begin{comment}
\begin{figure*}[!t]
  \begin{center}
  \includegraphics[width=\textwidth]{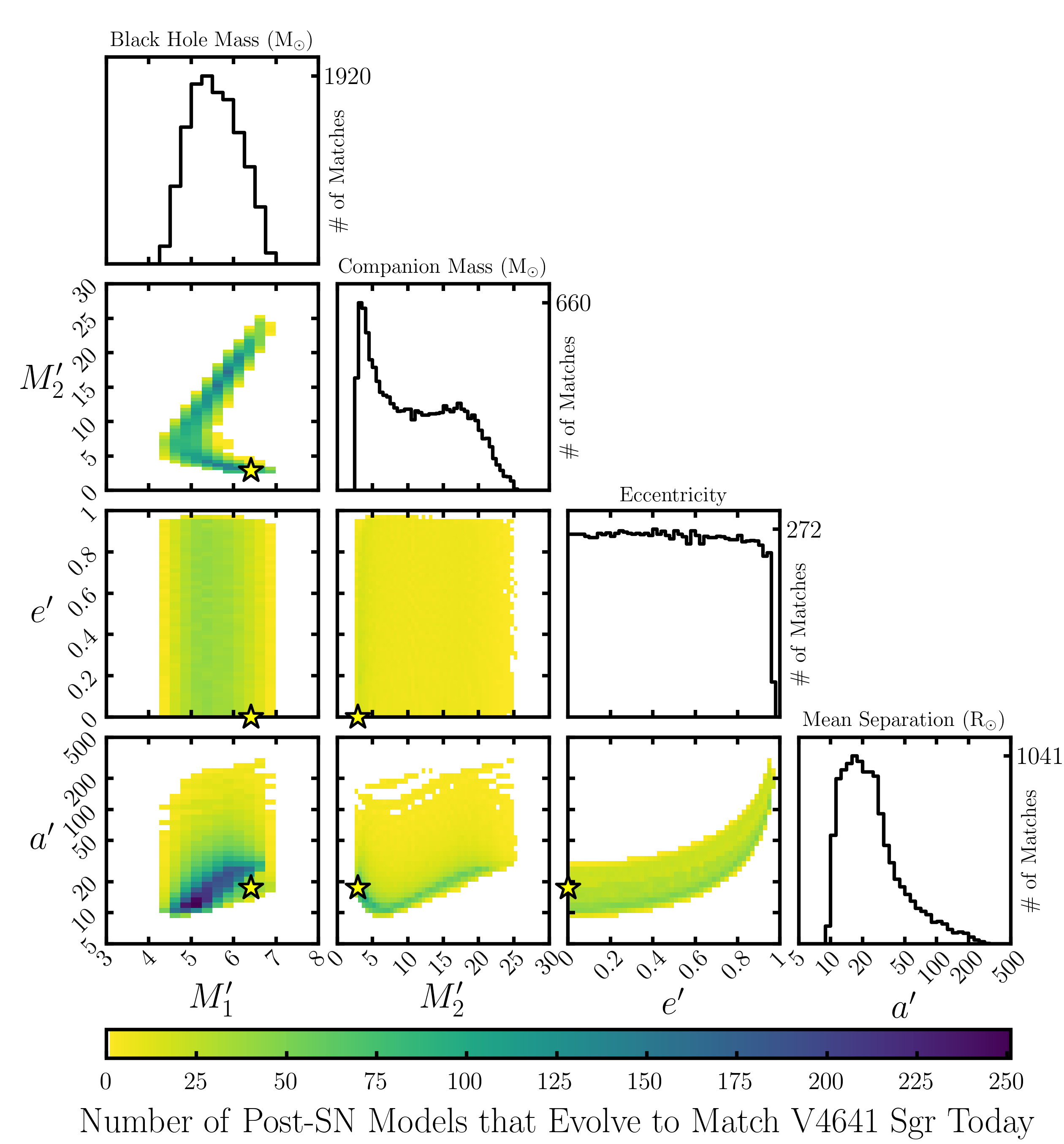}
  \caption{The two-dimensional histograms show the number of BSE models on the input parameter grids that evolve to match V4641 Sgr today according to several criteria (see text), but not imposing any restrictions on the spin-orbit misalignment angle. Starting from the immediate post-supernova state with parameters $\left\{\right.$$M_{1}^{\prime}$, $M_{2}^{\prime}$, $e^{\prime}$, $a^{\prime}$$\left.\right\}$, each BSE model evolves forward in time. In total, 12,585 BSE models evolved from the post-supernova epoch to match V4641 Sgr today (\textit{yellow star}). The one-dimensional histograms show the number of BSE matches within each input parameter bin, plotted on a linear vertical scale from zero to the marked peak of the distribution. \textbf{Disclaimer:} These histograms are \textit{not} probability densities. This is because we have no prior information on the BSE input parameter set $\left\{\right.$$M_{1}^{\prime}$, $M_{2}^{\prime}$, $e^{\prime}$, $a^{\prime}$$\left.\right\}$; therefore, we cannot make any probabilistic statements about the post-supernova properties of V4641 Sgr. However, we are sampling the complete parameter space because the adopted bounds comfortably contain the input parameters of every matching BSE model.}
  \label{fig:bse}
  \end{center}
\end{figure*}
%\end{comment}

% TABLE 4:
\setlength{\tabcolsep}{2pt}
\begin{table}[!t]
\centering
\begin{tabular}{c c c c c c c c}
\toprule
\texttt{mass1} & \texttt{mass2} & \texttt{tphysf} & \texttt{tb} & \texttt{kstar1} & \texttt{kstar2} & \texttt{z} & \texttt{ecc} \\
$M_{1}^{\prime}$ & $M_{2}^{\prime}$ & 13.8e3 & $P^{\prime}$ & $14$ & $1$ & $0.02$ & $e^{\prime}$ \\
\midrule
\texttt{neta} & \texttt{bwind} & \texttt{hewind} & \texttt{alpha1} & \texttt{lambda} & & & \\
0.5 & 0.0 & 1.0 & 1.0 & 0.5 & & & \\
\midrule
\texttt{ceflag} & \texttt{tflag} & \texttt{ifflag} & \texttt{wdflag} & \texttt{bhflag} & \texttt{nsflag} & \texttt{mxns} & \texttt{idum} \\
0 & 1 & 0 & 0 & 0 & 1 & 3 & $-$ \\
\midrule
\texttt{pts1} & \texttt{pts2} & \texttt{pts3} & & & & & \\
0.05 & 0.01 & 0.02 & & & & & \\
\midrule
\texttt{sigma} & \texttt{beta} & \texttt{xi} & \texttt{acc2} & \texttt{epsnov} & \texttt{eddfac} & \texttt{gamma} & \\
$-$ & 0.125 & 1.0 & 1.5 & 0.001 & 1.0 & -1.0 & \\
\midrule
\bottomrule
\end{tabular}
\caption{Inputs to the BSE code (see source code \texttt{bse.f} for parameter descriptions). Each BSE run adopts a different set of post-supernova system parameters $\left\{\right.$$M_{1}^{\prime}$, $M_{2}^{\prime}$, $e^{\prime}$, $P^{\prime}$$\left.\right\}$, while all other inputs remain fixed across all BSE runs. Several BSE inputs are irrelevant to the post-supernova system evolution we consider (e.g., common envelope parameters).}
\vspace{-2mm}
\label{tab:bse}
\end{table}

To relate these two epochs, we evolve different post-supernova states using the Binary-Star Evolution (BSE) code \citep{Hurley2002, Hurley2000, Tout1997},\footnote{BSE code website: \url{http://astronomy.swin.edu.au/\~jhurley/}} then we determine what parameter combinations can reproduce V4641 Sgr as observed today. The BSE code includes prescriptions to handle various important processes in binary evolution (e.g., tidal effects, stellar winds, gravitational radiation, magnetic braking, Roche-lobe overflow). Acknowledging the inherent uncertainty in stellar evolution models, our intention here is to use the BSE code as a tool to place informed constraints on the post-supernova system. Later on in \S\ref{sec:CEE}, we will examine the prescriptions in binary population synthesis models, and their implications, more critically.

We start by initializing individual binary systems consisting of the newborn black hole and the companion star at the post-supernova epoch. We take the companion to be on the zero age main sequence (ZAMS), an assumption that may not be valid if the companion was comparable in mass to the ZAMS progenitor of the black hole. The BSE input parameter grid is linearly-spaced in $M_{1}^{\prime} \in [3, 8]~M_{\odot}$ (20 bins), $M_{2}^{\prime} \in [1, 30]~M_{\odot}$ (58 bins), $e^{\prime} \in [0, 1]$ (50 bins), and logarithmically-spaced in $a^{\prime} \in [1, 1000]~R_{\odot}$ (60 bins).\footnote{These parameter ranges came from experimenting on a coarse grid and finding the ``matches'' that evolve to become consistent with V4641 Sgr today as described in this section. The component masses and eccentricity of these matches span a narrow range, while the orbital separation spans three decades, which justifies using linear and logarithmic bin spacings, respectively.} Table \ref{tab:bse} lists the values used for all other BSE inputs, which are mostly set to their defaults. The BSE code then evolves each set of initial conditions forward in time until either 13.8 Gyr elapse or the system reaches an end state (e.g., the companion star dies, the binary unbinds, the stars merge).

Treating each BSE model as a potential progenitor to V4641 Sgr, we begin the process of eliminating the post-supernova configurations that are inconsistent with the V4641 Sgr observables. At some point in the evolution of a given BSE model, we require that the post-supernova system parameters evolve to match those of V4641 Sgr today, to within their uncertainties (see Table \ref{tab:V4641}): $M_{1}^{\prime} \rightarrow M_{\bullet} = 6.4 \pm 0.6~M_{\odot}$, $M_{2}^{\prime} \rightarrow M_{\star} = 2.9 \pm 0.4~M_{\odot}$, $e^{\prime} \rightarrow e = 0.0^{+0.1}$,  $a^{\prime} \rightarrow a = 17.5 \pm 1.0~R_{\odot}$. To avoid putting too much stock in the BSE models, we match to the orbital separation $a$ instead of the tightly constrained binary orbital period $P = 2 \pi a^{3/2} / \sqrt{GM} = 2.817 \pm 0.002~\mathrm{days}$. In addition, the companion star must be filling its Roche lobe and crossing the Hertzsprung Gap to be considered consistent with V4641 Sgr today. We further require that at least 8.9 Myr elapsed since the moment of the supernova event (i.e., the starting time for the BSE models), which is the minimum age of the black hole in V4641 Sgr based on its most recent Galactic plane crossing (see \S\ref{sec:vpec}).

Of the initial 3,480,000 BSE models, Figure \ref{fig:bse} shows the parameter space occupied by the 12,585 post-supernova systems that survive this first cut as candidate V4641 Sgr progenitors. Interestingly, a very wide range in the companion star mass ($M_{2}^{\prime} \simeq 2.5$--$25~M_{\odot}$) can evolve to match V4641 Sgr today, but $M_{2}^{\prime}$ is not strongly correlated with the other BSE inputs. Successful matches have post-supernova mean separations $a^{\prime} \sim 10$--$100~R_{\odot}$ and span all eccentricities $e^{\prime}$, with wider mean separations associated with higher eccentricities. Figure \ref{fig:bse} further shows that the input parameter space for the BSE models is complete because no successful matches bump up against the $\left\{\right.$$M_{1}^{\prime}$, $M_{2}^{\prime}$, $e^{\prime}$, $a^{\prime}$$\left.\right\}$ extrema.

% FIGURE 11:
%\begin{comment}
\begin{figure}[!t]
  \begin{center}
  \includegraphics[width=0.495\textwidth]{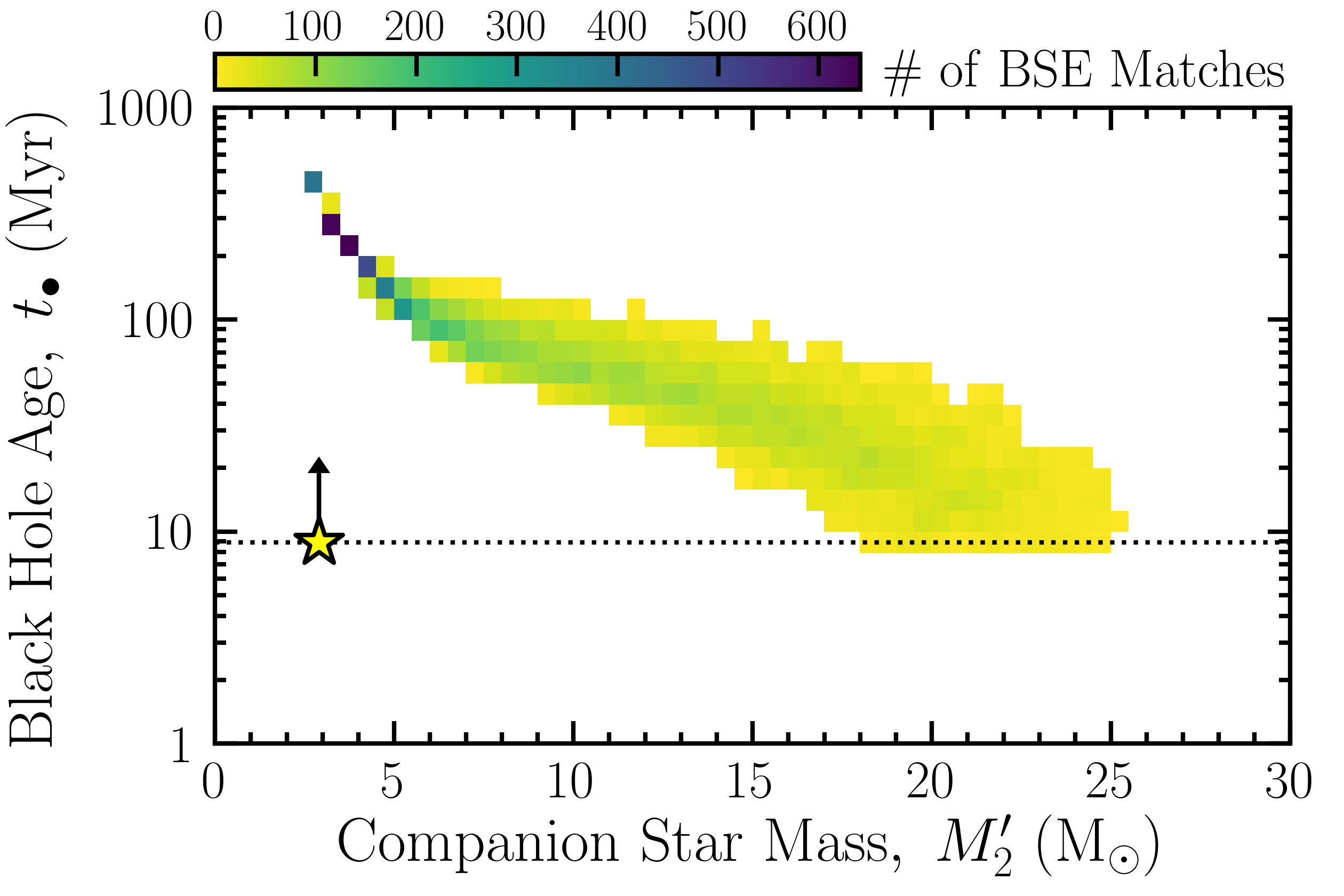}
  \caption{Black hole age $t_{\bullet}$, taken as the time elapsed from the immediate post-supernova initialization of a BSE model to when its system parameters match those of V4641 Sgr today, as a function of the companion star mass $M_{2}^{\prime}$. We also require $t_{\bullet} \ge 8.9~\mathrm{Myr}$ (\textit{dotted line}; see \S\ref{sec:vpec}).}
  \label{fig:M2tBH}
  \end{center}
\end{figure}
%\end{comment}

% FIGURE 12:
%\begin{comment}
\begin{figure*}[!t]
  \begin{center}
  \includegraphics[width=0.495\textwidth]{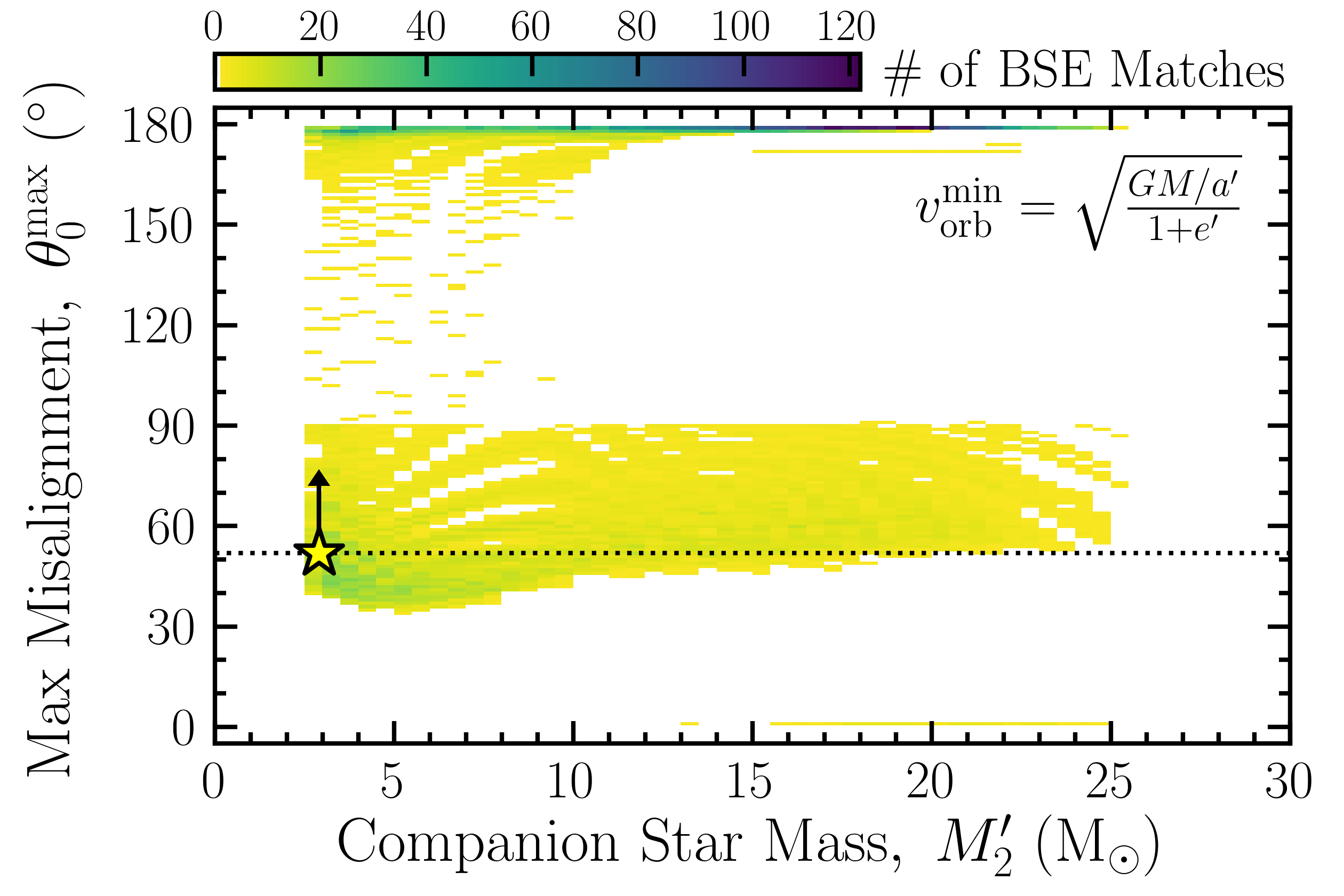}
  \hfill
  \includegraphics[width=0.495\textwidth]{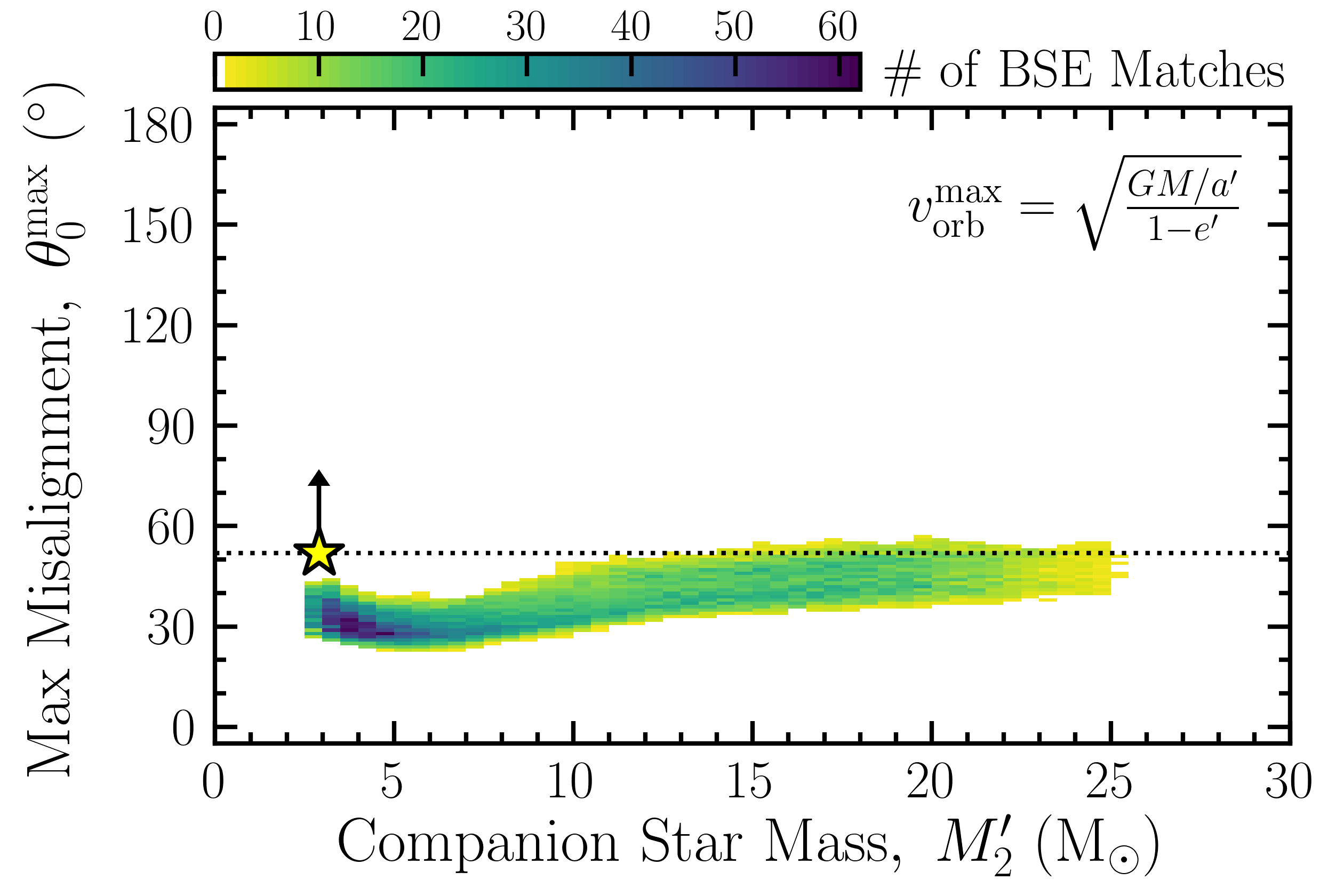}
  \caption{Maximum initial spin-orbit misalignment $\theta_{0}^{\mathrm{max}}$ that can be produced by the natal kick model, determined by using the post-supernova parameters from each BSE model that evolved to match V4641 Sgr today, as a function of companion star mass $M_{2}^{\prime}$. The turnover (or lack thereof) in the $P( \theta_{0} | v_{\mathrm{orb}} )$ distribution determines $\theta_{0}^{\mathrm{max}}$, and we show results from choosing either the minimum or maximum pre-supernova orbital speed: $v_{\mathrm{orb}}^{\mathrm{min}}$ (\textit{left panel}) and $v_{\mathrm{orb}}^{\mathrm{max}}$ (\textit{right panel}). For $v_{\mathrm{orb}}^{\mathrm{min}}$, most BSE matches have a non-zero probability of producing a $\theta_{0}$ that exceeds the $52^{\circ}$ minimum in V4641 Sgr today (\textit{yellow star}; \textit{dotted line}). While for $v_{\mathrm{orb}}^{\mathrm{max}}$, essentially none of the matching BSE models can produce such an extreme spin-orbit misalignment.}
  \label{fig:i0max}
  \end{center}
\end{figure*}
%\end{comment}

Each BSE model began just after the supernova event that gave birth to the black hole. For all BSE matches, Figure \ref{fig:M2tBH} shows the black hole age at the time when the system evolved to match V4641 Sgr today, as a function of the companion star mass $M_{2}^{\prime}$. BSE models with an initially low-mass companion star evolve for several hundred Myr before becoming consistent with V4641 Sgr today. Binaries with a high mass companion evolve more quickly to match V4641 Sgr, although the BSE code has these massive stars ($M_{2}^{\prime} \gtrsim 10~M_{\odot}$) surviving much longer than the lifetime expected for their mass in an isolated evolution. Taken at face value, the BSE models imply that the age of V4641 Sgr is inversely proportional to the initial mass of the companion star.

Crucially, for a BSE model to remain a viable progenitor to V4641 Sgr, the natal kick model must be able to produce a large enough initial spin-orbit misalignment $\theta_{0}$, using the BSE parameters $\left\{\right.$$M_{1}^{\prime}$, $M_{2}^{\prime}$, $e^{\prime}$, $a^{\prime}$$\left.\right\}$ as inputs. At this juncture, we stress that for any of these ``matching'' BSE models, we can calculate the probability of producing some $\theta_{0}$ from a natal kick, but not the probability that the BSE model represents the post-supernova configuration of V4641 Sgr.\footnote{In other words, the two-dimensional histograms we present should not be interpreted as joint probability densities.} Instead, our approach going forward is to apply the kick model to all possible progenitor systems informed by the BSE models, regardless of their (unknown) relative likelihoods, and ask if each one can be conservatively ruled out.

In \S\ref{sec:constraints}, we saw that choosing the pre-supernova relative orbital speed $v_{\mathrm{orb}}$ is the crux of determining the maximum spin-orbit misalignment $\theta_{0}^{\mathrm{max}}$ resulting from a natal kick. The allowable $v_{\mathrm{orb}}$ range is,
\begin{equation}
v_{\mathrm{orb}}^{\mathrm{min}} = \sqrt{\frac{G M / a^{\prime}}{1 + e^{\prime}}} \le v_{\mathrm{orb}} \le \sqrt{\frac{G M / a^{\prime}}{1 - e^{\prime}}} = v_{\mathrm{orb}}^{\mathrm{max}}, \label{eqn:vorbRange}
\end{equation}
which follows from the requirement that the total energy of the post-supernova binary be greater than the effective potential energy (see Appendix \ref{sec:appA}).

We specify $v_{\mathrm{orb}}$ and apply the natal kick model to each set of matching BSE input parameters $\left\{ \right.$$M_{1}^{\prime}$, $M_{2}^{\prime}$, $e^{\prime}$, $a^{\prime}$$\left. \right\}$, enforcing all constraints on the integration region when calculating the conditional density $P( \theta_{0} | v_{\mathrm{orb}} )$, including $v_{\mathrm{k}} \ge v_{\mathrm{eff}}$ (see Equations \ref{eqn:vkphi_vpm}--\ref{eqn:vkphi_vsys}). From the resulting $P( \theta_{0} | v_{\mathrm{orb}})$ distribution, we determine $\theta_{0}^{\mathrm{max}}$ --- the maximum spin-orbit misalignment that the natal kick model can produce (see \S\ref{sec:constraints}) --- as the largest $\theta_{0}$ that satisfies $\int_{\theta_{0}}^{180^{\circ}} P( \theta_{0} | v_{\mathrm{orb}} ) d\theta_{0} > 0$. For the choices $v_{\mathrm{orb}}^{\mathrm{min}}$ (\textit{left panel}) and $v_{\mathrm{orb}}^{\mathrm{max}}$ (\textit{right panel}), Figure \ref{fig:i0max} shows the histogram of results from applying this exercise to all BSE matches, as a function of the companion star mass $M_{2}^{\prime}$. Most of the $P( \theta_{0} | v_{\mathrm{orb}}^{\mathrm{min}})$ distributions extend to $\theta_{0} = 180^{\circ}$ without turning over, meaning there is a non-zero probability, however small, of producing $\theta_{0} = 180^{\circ}$. Conversely, all of the $P( \theta_{0} | v_{\mathrm{orb}}^{\mathrm{max}})$ distributions turn over to give tight constraints on $\theta_{0}^{\mathrm{max}}$.
 
The spin-orbit misalignment constraint, which we enforce on each matching BSE model, is to require a non-zero probability that $\theta_{0}$ exceed some minimum value,
\begin{equation}
\int_{\theta_{0}^{\mathrm{min}}}^{180^{\circ}} P\left( \theta_{0} \left| v_{\mathrm{orb}} \right) \right. d\theta_{0} > 0. \label{eqn:PiTest}
\end{equation}
As a first step, we set $\theta_{0}^{\mathrm{min}}$ to the minimum spin-orbit misalignment of V4641 Sgr today, $\theta_{\mathrm{pro}}^{\mathrm{min}} = 52^{\circ}$ (see Table \ref{tab:misalign}). Later on in \S\ref{sec:align}, we will replace this choice with an estimate for the larger spin-orbit misalignment at the time of black hole birth. Of the 12,585 matching BSE models, we are still left with 9,882 possible progenitor models for $v_{\mathrm{orb}}^{\mathrm{min}}$, but only 277 for $v_{\mathrm{orb}}^{\mathrm{max}}$. These numbers follow from counting how many BSE models lie above the \textit{dotted line} in Figure \ref{fig:i0max}. Taking $\theta_{0}^{\mathrm{min}} = \theta_{\mathrm{ret}}^{\mathrm{min}} = 88^{\circ}$ instead, the surviving BSE models reduce dramatically to 4,956 for $v_{\mathrm{orb}}^{\mathrm{min}}$, while no BSE models survive for $v_{\mathrm{orb}}^{\mathrm{max}}$.

If we can eliminate all of the BSE matches, by increasing $\theta_{0}^{\mathrm{min}}$ and/or $v_{\mathrm{orb}}$, then we can rule out the natal kick model as the origin of the V4641 Sgr spin-orbit misalignment. To make headway, then, we must replace our conservative choices $\theta_{0}^{\mathrm{min}} = \theta_{\mathrm{pro}}^{\mathrm{min}}$ and $v_{\mathrm{orb}} = v_{\mathrm{orb}}^{\mathrm{min}}$ with more realistic constraints. First (\S\ref{sec:align}), we will use the accretion history of each BSE model to estimate the initial spin-orbit misalignment $\theta_{0}$, which was larger than that of V4641 Sgr today. Second (\S\ref{sec:CEE}), we will estimate $v_{\mathrm{orb}}$ from expectations of common envelope evolution.

%----------------------------------------------------------------------------------------------------
% EVOLUTION TOWARD SPIN-ORBIT ALIGNMENT
\subsection{Evolution Toward Spin-Orbit Alignment}
\label{sec:align}
\begin{figure}[!t]
  \begin{center}
  \includegraphics[width=0.495\textwidth]{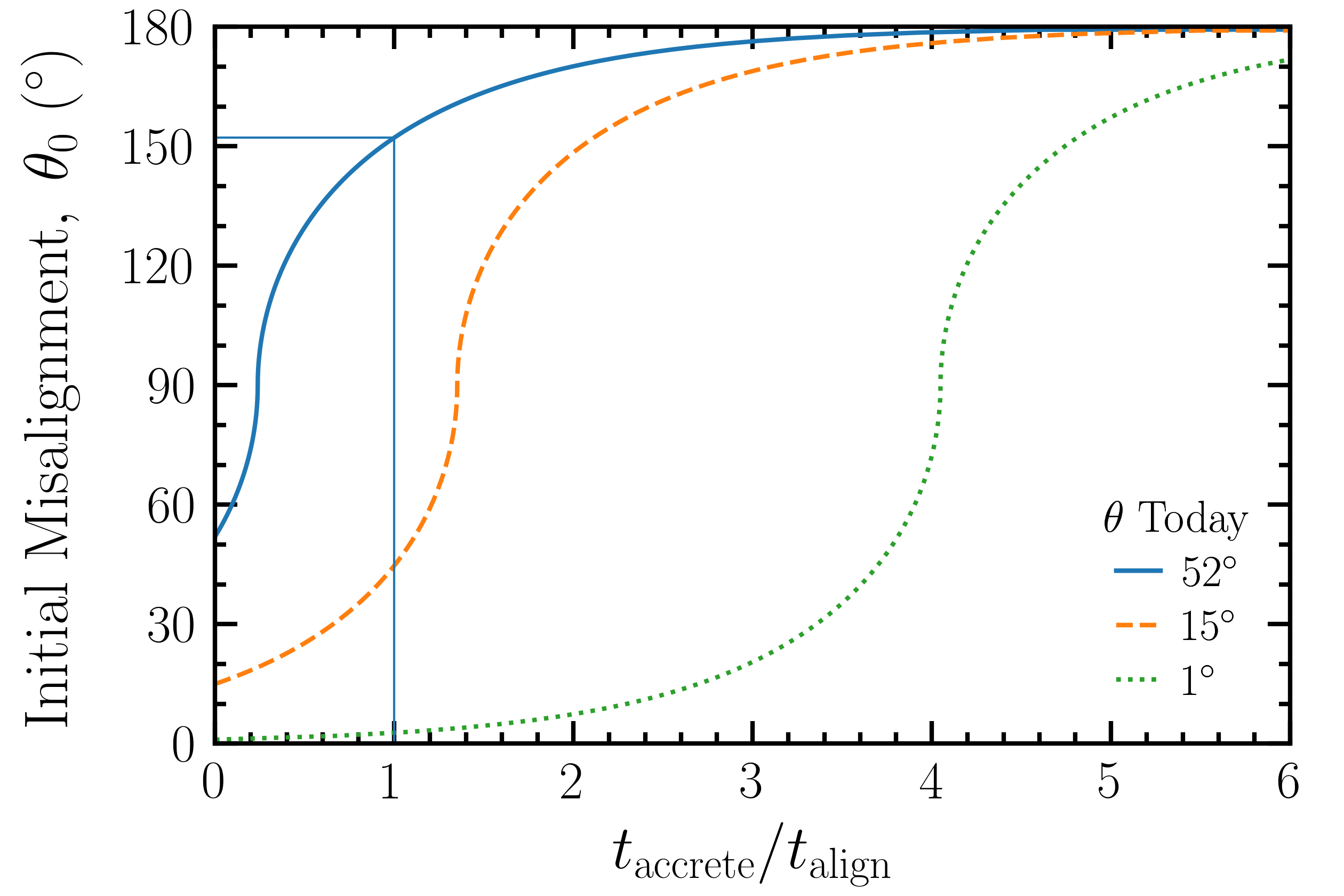}
  \caption{Plot showing the initial spin-orbit misalignment angle $\theta_{0}$, given the spin-orbit misalignment today $\theta$ (\textit{line color/style}) and knowledge of how long the system has been steadily accreting $t_{\mathrm{accrete}}$, measured in units of the alignment timescale $t_{\mathrm{align}}$. If the V4641 Sgr black hole accreted mass in a steady state for one alignment timescale, its spin-orbit misalignment today of $\theta > 52^{\circ}$ implies its initial spin-orbit misalignment was $\theta_{0} > 152^{\circ}$.}
  \label{fig:i0}
  \end{center}
\end{figure}

The conventional expectation is that a spin-orbit misalignment in a black hole X-ray binary will evolve toward alignment over time. The alignment mechanism is the Lense-Thirring reaction torque exerted by the misaligned accretion flow on the black hole that acts to gradually align its spin with the total angular momentum of the system, which is dominated by the binary orbital component \citep[e.g.,][]{King2005}. Figure \ref{fig:i0} shows the relationship between the spin-orbit misalignment today, $\theta$, and in the past, $\theta_{0}$, if the black hole steadily accretes mass at a rate $\dot{M}$ for a time $t_{\mathrm{accrete}}$ relative to the alignment timescale \citep{Martin2007, Martin2008a},
\begin{align}
t_{\mathrm{align}} &= \frac{\left( 1 + \beta \right)^{- \beta / (1 + \beta)}}{\sqrt{2} \cos\left\{ \pi / \left[ 4 \left( 1 + \beta \right) \right] \right\}} \frac{\Gamma\left\{ 1 / \left[ 2 \left( 1+ \beta \right) \right] \right\}}{\Gamma\left\{ \left( 1 + 2 \beta \right) / \left[ 2 \left( 1+ \beta \right) \right] \right\}} \nonumber \\ 
&\times \frac{3 \nu_{1}}{\sqrt{2} \dot{M}} \sqrt{\frac{a_{\ast} c M_{1}}{\nu_{2} G}}, \label{eqn:talign}
\end{align}
where $\beta = 3/4$ for the standard $\alpha$-disk model \citep{ShakuraSunyaev1973}. The azimuthal and vertical viscosities $\nu_{1}$ and $\nu_{2}$, with corresponding effective viscosity parameters $\alpha_{1}$ and $\alpha_{2}$, are given by \citep{WijersPringle1999},
\begin{align}
\nu_{1,2} &= 6.40 \times 10^{15}~\alpha_{1,2}^{4/5} \left( \frac{M_{1}}{M_{\odot}} \right)^{-1/4} \left( \frac{\dot{M}}{10^{-8}~M_{\odot}/\mathrm{yr}} \right)^{3/10} \nonumber \\ 
&\times \left( \frac{R}{10^{11}~\mathrm{cm}} \right)^{7/10}~\mathrm{cm^{2} / s}, \label{eqn:nu1nu2}
\end{align}
and we evaluate $R$ at the warp radius $R_{\mathrm{warp}}$, where the aligning torque is strongest \citep{Martin2008a},
\begin{align}
R_{\mathrm{warp}} &= 8.29 \times 10^{7}~\alpha_{2}^{-16/35} a_{\ast}^{4/7} \left( \frac{M_{1}}{M_{\odot}} \right)^{9/7} \nonumber \\
&\times \left( \frac{\dot{M}}{10^{-8}~M_{\odot} / \mathrm{yr}} \right)^{-6/35}~\mathrm{cm}. \label{eqn:Rwarp}
\end{align}
Given the duration of an accretion episode relative to the alignment timescale $t_{\mathrm{accrete}} / t_{\mathrm{align}}$, as well as the spin-orbit misalignment $\theta$ at the end of the episode, we can deduce the spin-orbit misalignment $\theta_{0}$ at the beginning of the episode by following \S4 of \citet{Martin2008a}.

% FIGURE 13:
%\begin{comment}
%\end{comment}

% FIGURE 14:
%\begin{comment}
\begin{figure}[!t]
  \begin{center}
  \includegraphics[width=0.495\textwidth]{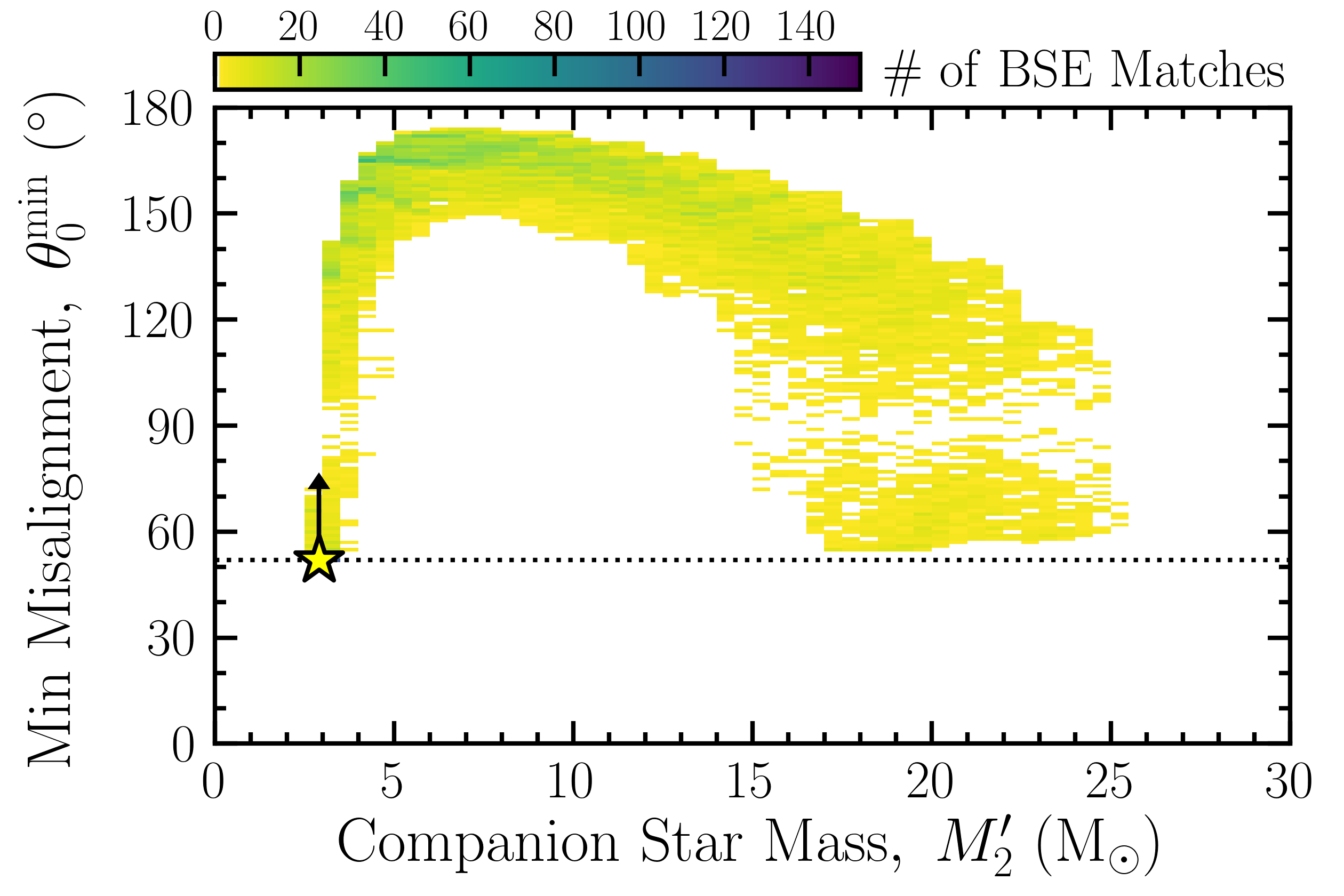}
  \caption{Minimum initial spin-orbit misalignment $\theta_{0}^{\mathrm{min}}$, as a function of initial companion star mass $M_{2}^{\prime}$. For each matching BSE model, we determine $\theta_{0}^{\mathrm{min}}$ by using its accretion history to conservatively estimate $t_{\mathrm{accrete}} / t_{\mathrm{align}}$, combined with the condition that the V4641 Sgr spin-orbit misalignment today be its minimum value $\theta^{\mathrm{min}} = 52^{\circ}$. Most BSE matches imply that $\theta$ was much larger in the past.}
  \label{fig:i0MinM2}
  \end{center}
\end{figure}
%\end{comment}

Up until now, we ruled out a BSE model if a natal kick could not produce a spin-orbit misalignment as extreme as $\theta > 52^{\circ}$ in V4641 Sgr today. This was a conservative approach because the misalignment $\theta_{0}$ at black hole birth is the relevant point of comparison, \textit{not} its smaller value $\theta$ today. But now, we will estimate $\theta_{0}$ for each matching BSE model using the equations above. Formally, the $t_{\mathrm{align}}$ expression is only appropriate for steady state, thin disk accretion in the approximation of a small misalignment \citep[e.g.,][]{ScheuerFeiler1996}, but we apply it to large misalignments hoping to roughly estimate the temporal evolution in $\theta$, and ultimately $\theta_{0}$. 

% FIGURE 15:
%\begin{comment}
\begin{figure*}[!t]
  \begin{center}
  \includegraphics[width=0.495\textwidth]{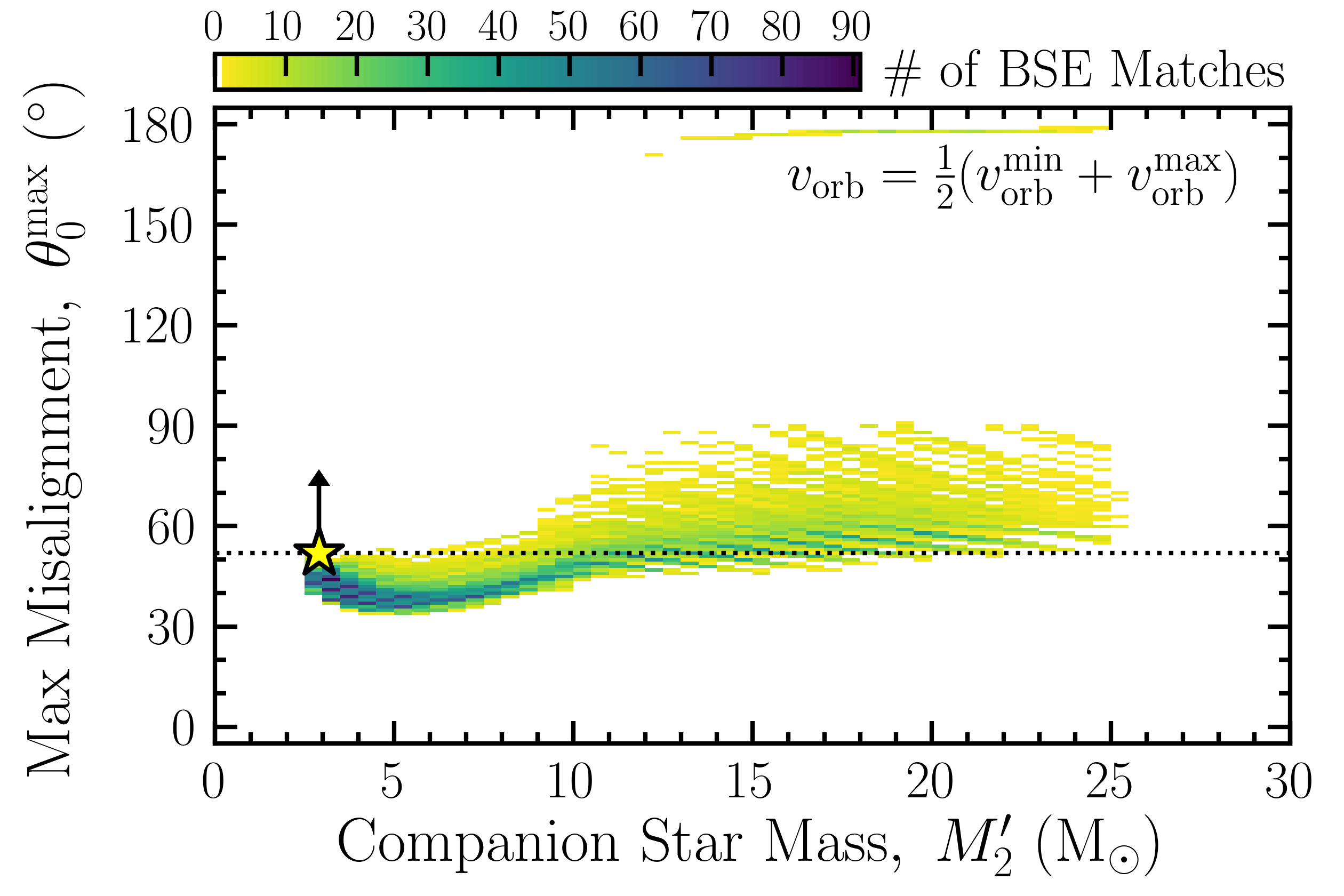}
  \hfill
  \includegraphics[width=0.495\textwidth]{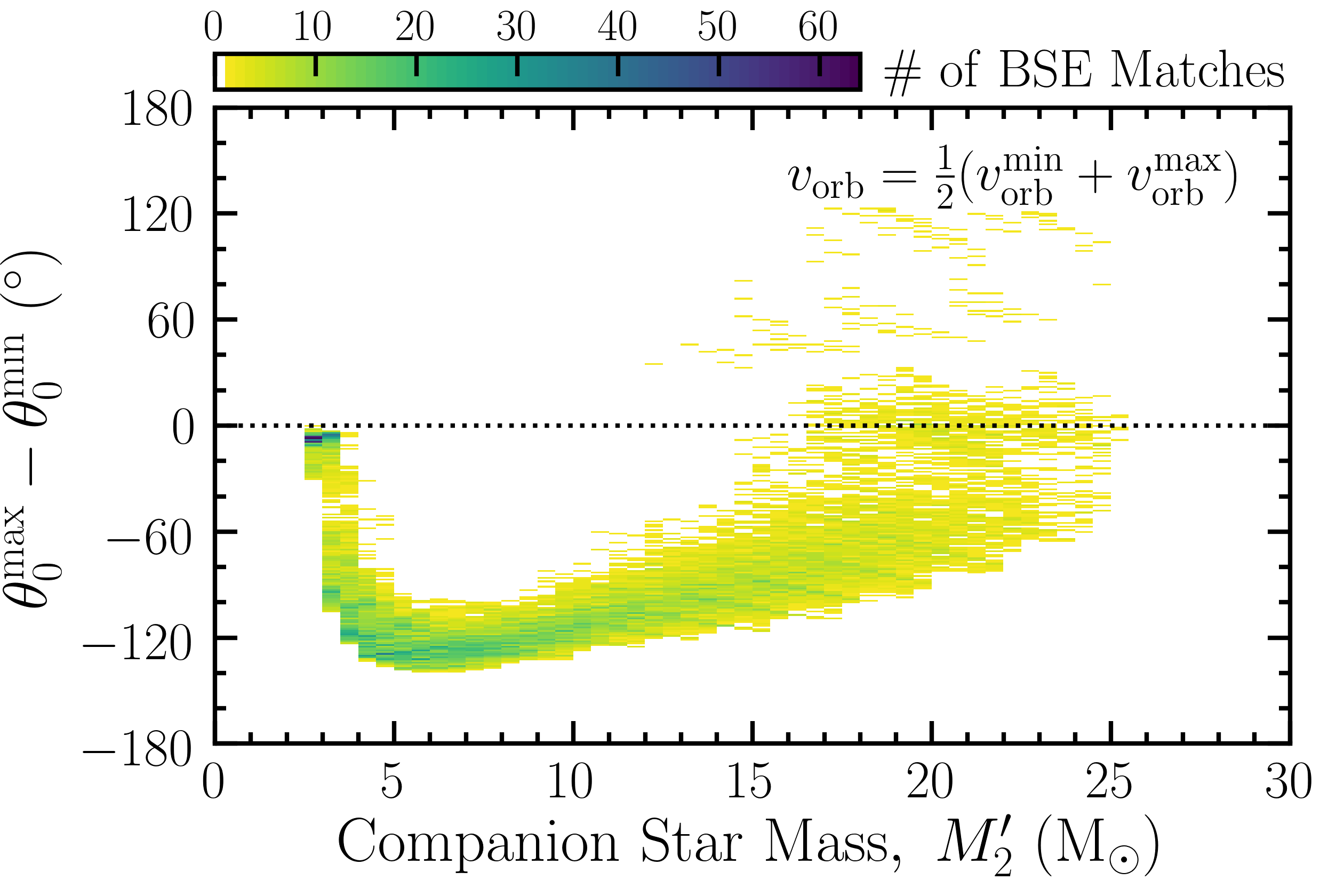}
  \caption{\textit{Left panel}: Same as Figure \ref{fig:i0max}, but using the intermediate pre-supernova relative orbital speed $v_{\mathrm{orb}} = ( v_{\mathrm{orb}}^{\mathrm{min}} + v_{\mathrm{orb}}^{\mathrm{max}} ) / 2$. \textit{Right panel}: Difference between the maximum initial spin-orbit misalignment $\theta_{0}^{\mathrm{max}}$ attainable with a natal kick and the minimum initial spin-orbit misalignment $\theta_{0}^{\mathrm{min}}$ based on the accretion history of each BSE model, as a function of the companion star initial mass $M_{2}^{\prime}$. The natal kick model is ruled out as the origin of the initial spin-orbit misalignment in V4641 Sgr for the BSE models below the \textit{dotted line}. For this intermediate choice of $v_{\mathrm{orb}}$, there are 500 BSE models above the \textit{dotted line} that can produce the requisite misalignment, but they all have a very high companion star mass compared to its value today of $M_{\star} = 2.9 \pm 0.4~M_{\odot}$.}
  \label{fig:vorbHalf_i0MaxDiff}
  \end{center}
\end{figure*}
%\end{comment}

Using the accretion history of a given BSE model, we determine the epochs when the Eddington-scaled mass accretion rate onto the black hole is $0.01 \le \dot{M} / \dot{M}_{\mathrm{Edd}} \le 0.3$, with radiative efficiency $\eta = 0.1$. This is the appropriate $\dot{M}$ range for a geometrically thin, optically thick disk \citep[e.g.,][]{Esin1997, LaorNetzer1989}, as assumed when calculating $t_{\mathrm{align}}$. Knowing the temporal evolution of $\dot{M}$ and the black hole mass $M_{1}$ gives the duration of each accretion episode $t_{\mathrm{accrete}}$. To avoid overestimating the initial spin-orbit misalignment $\theta_{0}$, we intentionally minimize $t_{\mathrm{accrete}} / t_{\mathrm{align}}$ by choosing a maximal black hole spin $a_{\ast} = 1$ and large isotropic viscosities $\alpha_{1} = \alpha_{2} = 0.2$ when calculating $t_{\mathrm{align}}$ for an accretion episode. Starting from the moment when a BSE model matches the system parameters of V4641 Sgr, we conservatively initialize $\theta$ to $\theta_{\mathrm{pro}}^{\mathrm{min}} = 52^{\circ}$ and work backwards in time, calculating the progressive evolution of $\theta$ from each accretion episode until arriving at $\theta_{0}^{\mathrm{min}}$, the minimum spin-orbit misalignment at black hole birth.

% FIGURE 16:
%\begin{comment}
\begin{figure*}[!t]
  \begin{center}
  \includegraphics[width=0.495\textwidth]{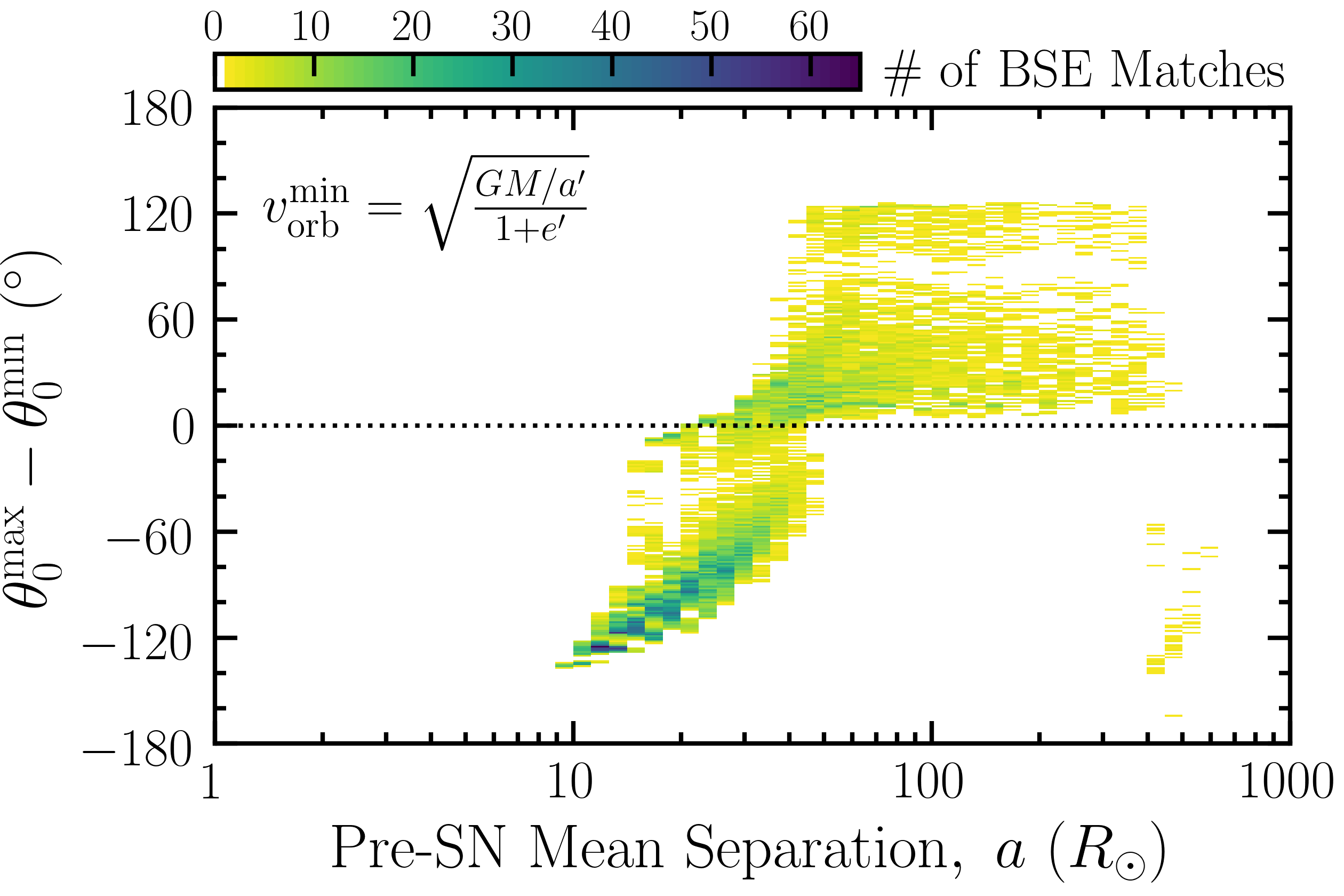}
  \hfill
  \includegraphics[width=0.495\textwidth]{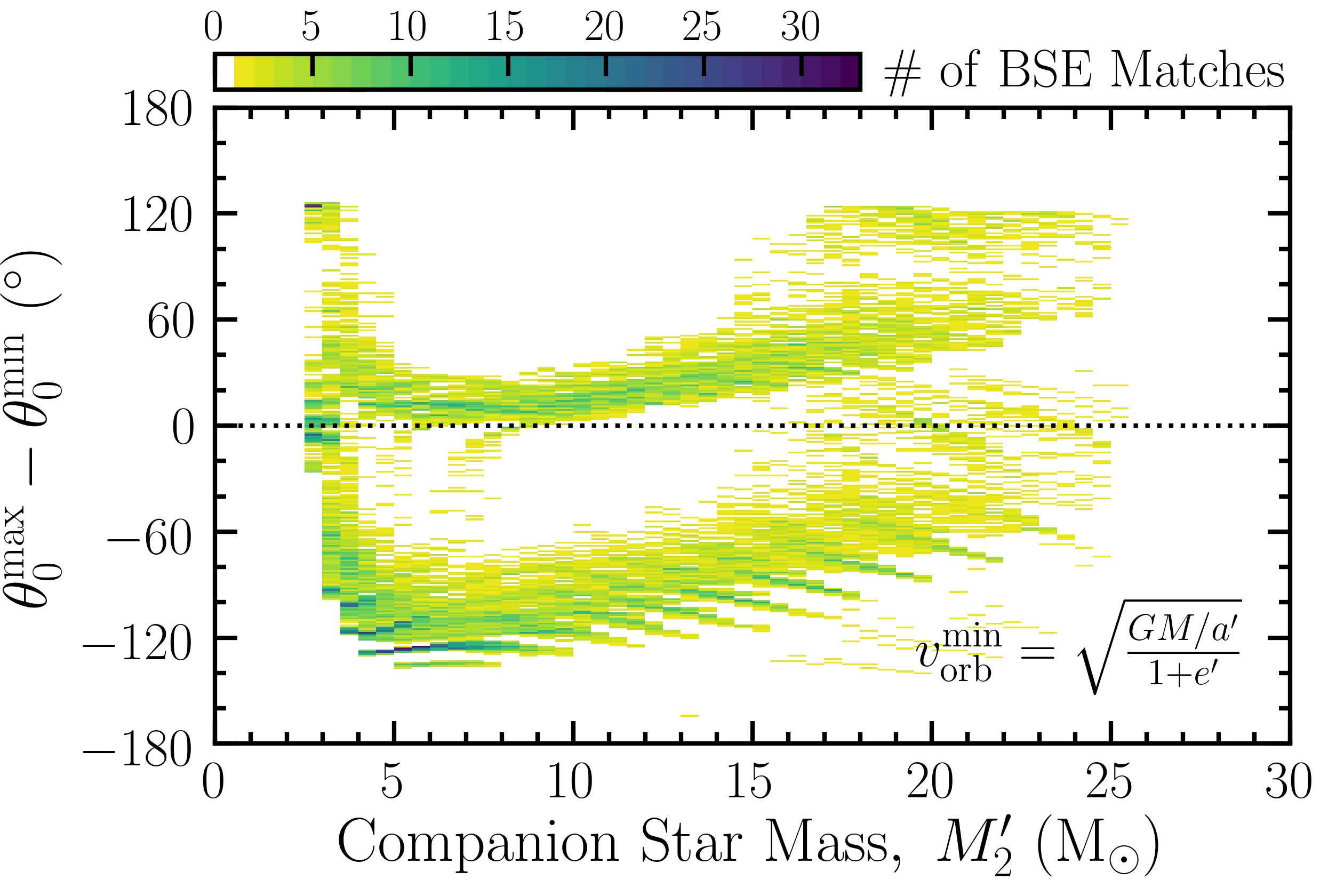}
  \caption{Difference in the initial spin-orbit misalignment extrema $\theta_{0}^{\mathrm{max}} - \theta_{0}^{\mathrm{min}}$, where we set the pre-supernova relative orbital velocity $v_{\mathrm{orb}}$ to its minimum value when applying the natal kick model to a set of BSE input parameters. This results in the largest possible $\theta_{0}^{\mathrm{max}}$ value that a natal kick can produce (see \S\ref{sec:constraints}), meaning that BSE matches lying below the \textit{dotted line} can be firmly ruled out as possible V4641 Sgr progenitors. Only wide pre-supernova mean separations of $a \simeq 20$--$400~R_{\odot}$ can be made consistent with a natal kick as the origin of the initial spin-orbit misalignment in V4641 Sgr (\textit{left panel}), while a broad range in companion star initial masses $M_{2}^{\prime}$ can work (\textit{right panel}). These statements do not hold true as $v_{\mathrm{orb}}$ increases.}
  \label{fig:vorbMin_i0Diff}
  \end{center}
\end{figure*}
%\end{comment}

Figure \ref{fig:i0MinM2} shows the $\theta_{0}^{\mathrm{min}}$ estimates for all matching BSE models, as a function of initial companion star mass $M_{2}^{\prime}$. The majority of BSE matches --- 10,944 out of 12,585 --- have accretion histories that, when used to estimate alignment history, imply a retrograde spin-orbit misalignment in the past (i.e., $\theta_{0} > 90^{\circ}$). Prograde initial misalignments are possible for the bimodal $M_{2}^{\prime}$ ranges 2.5--4.0~$M_{\odot}$ and 15--25~$M_{\odot}$. Notably, these $\theta_{0}^{\mathrm{min}}$ constraints come from the post-supernova accretion history and are independent of the natal kick model. We also emphasize our conservative approach, which strives to drive $\theta_{0}^{\mathrm{min}}$ to be as small as possible; thus, likely underestimating the initial spin-orbit misalignment.

As before, we apply the natal kick model to each matching BSE model to determine the maximum initial spin-orbit misalignment $\theta_{0}^{\mathrm{max}}$ that can be produced. The \textit{left panel} of Figure \ref{fig:vorbHalf_i0MaxDiff} shows the result, where we chose a pre-supernova orbital speed $v_{\mathrm{orb}}$ intermediate between the allowed extrema from Equation \ref{eqn:vorbRange}. Of the 12,585 matching BSE models, a natal kick can produce a spin-orbit misalignment $> 52^{\circ}$ in 5,323 cases, but this comparison is to the misalignment of V4641 Sgr today, \textit{not} its initial misalignment immediately post-supernova.

Instead, the \textit{right panel} of Figure \ref{fig:vorbHalf_i0MaxDiff} compares $\theta_{0}^{\mathrm{max}}$ for each matching BSE model to its corresponding estimate for the minimum initial spin-orbit misalignment $\theta_{0}^{\mathrm{min}}$ from Figure \ref{fig:i0MinM2}. BSE models below the \textit{dotted line} where $\theta_{0}^{\mathrm{max}} < \theta_{0}^{\mathrm{min}}$ can be ruled out as the origin of the V4641 Sgr spin-orbit misalignment because even the maximum misalignment that can be produced by a natal kick is not as large as the minimum misalignment implied by the accretion history of the system. For this intermediate $v_{\mathrm{orb}}$, no BSE models are  consistent with the estimated initial spin-orbit misalignment for V4641 Sgr, unless we are willing to seriously consider the prospect that the companion star was once $\gtrsim 5$ times its current mass. Increasing $v_{\mathrm{orb}}$ always reduces the number of successful BSE models due to our approach of using the turnover in $P( \theta_{0} | v_{\mathrm{orb}} )$ to find $\theta_{0}^{\mathrm{max}}$, as discussed in \S\ref{sec:constraints}.

The natal kick model can produce larger spin-orbit misalignments by decreasing $v_{\mathrm{orb}}$, which is equivalent to widening the pre-supernova binary orbit. The choice $v_{\mathrm{orb}} = v_{\mathrm{orb}}^{\mathrm{min}}$ leads to the largest possible $\theta_{0}^{\mathrm{max}}$ (see \S\ref{sec:constraints}), and consequently the largest difference ($\theta_{0}^{\mathrm{max}} - \theta_{0}^{\mathrm{min}}$) because $\theta_{0}^{\mathrm{min}}$ is independent of $v_{\mathrm{orb}}$. In other words, $v_{\mathrm{orb}} = v_{\mathrm{orb}}^{\mathrm{min}}$ gives the natal kick model the best chance of producing an extreme spin-orbit misalignment. For this optimistic case, Figure \ref{fig:vorbMin_i0Diff} shows where the BSE matches congregate in the parameter space of ($\theta_{0}^{\mathrm{max}} - \theta_{0}^{\mathrm{min}}$) \textit{vs.} the pre-supernova mean separation $a$ (\textit{left panel}) or \textit{vs.} the companion star initial mass $M_{2}^{\prime}$ (\textit{right panel}). By virtue of lying above the \textit{dotted line}, fairly wide pre-supernova separations of $a \simeq 20$--$400~R_{\odot}$ can be made consistent with the initial spin-orbit misalignment, but close separations $a \lesssim 20$ cannot. A broad range in $M_{2}^{\prime}$ can also work and although not shown, these successful BSE matches also require post-supernova eccentricities $e^{\prime} \simeq 0.5$--$1$ and mean separations $a^{\prime} \simeq 20$--$200~R_{\odot}$.

We found that a natal kick can explain the origin of the extreme spin-orbit misalignment in V4641 Sgr, if the companion star was once incredibly massive and/or the pre-supernova binary orbit was wide. We disfavor the prospect of intense evolution in the companion star mass (see \S\ref{sec:origin}), and now explore the possibility of a wide separation resulting from common envelope evolution.

%----------------------------------------------------------------------------------------------------
% COMMON ENVELOPE EVOLUTION
\subsection{Common Envelope Evolution}
\label{sec:CEE}
The natal kick model connects the immediate pre-/post-supernova states, without explicitly assuming a binary formation pathway prior to the supernova event. Here, we demonstrate that V4641 Sgr appears to be inconsistent with the current understanding of common envelope evolution for massive stars, which is thought to be the dominant formation channel for close binaries containing a black hole \citep[e.g.,][]{Paczynski1976}.

Common envelope evolution commences when a red supergiant star of mass $M_{1}^{0}$ overfills its Roche lobe radius $R_{\mathrm{L}1}^{0}$ and expands in size beyond the initial mean separation $a_{0}$ to engulf the companion star of lesser mass $M_{2}^{0}$. Unstable mass transfer ensues, and some fraction $\alpha_{\mathrm{ce}}$ of the liberated orbital energy $\Delta E_{\mathrm{orb}}$ from the in-spiral of the supergiant core and companion star gets deposited into the common envelope \citep{LivioSoker1988}, whose binding energy $E_{\mathrm{bind}}$ is parametrized by an envelope structure factor $\lambda$ \citep{deKool1990}. The common envelope is ejected quickly on a timescale of the initial orbital period, leaving behind the helium core of mass $M_{1}$ and the companion star of mass $M_{2} \simeq M_{2}^{0}$ in a close binary with final mean separation $a_{\mathrm{f}}$ given by,
\begin{equation}
\frac{a_{\mathrm{f}}}{a_{0}} = \frac{M_{1} M_{2}}{M_{1}^{0} \left( M_{1}^{0} - M_{1} \right)} \left[ \frac{M_{2}}{\left( M_{1}^{0} - M_{1} \right)} + \frac{2}{\lambda \alpha_{\mathrm{ce}} r_{\mathrm{L}1}^{0}} \right]^{-1}, \label{eqn:af_a0} 
\end{equation}
which follows from the standard energy budget formalism, $E_{\mathrm{bind}} = \alpha_{\mathrm{ce}} \Delta E_{\mathrm{orb}}$ \citep{IbenTutukov1984, Webbink1984}. Here, the mass of the ejected envelope is $(M_{1}^{0} - M_{1})$, while $r_{\mathrm{L}1}^{0} = R_{\mathrm{L}1}^{0} / a_{0}$ relates the orbital separation and the Roche lobe radius of the supergiant star at the onset of common envelope evolution. The Roche lobe radius for some separation $a$ is \citep{Eggleton1983},
\begin{equation}
\frac{R_{\mathrm{L}}}{a} \simeq \frac{0.49 q^{2/3}}{0.6 q^{2/3} + \ln\left( 1 + q^{1/3}\right)}, \label{eqn:Roche}
\end{equation}
where $q$ is the mass ratio of the star whose $R_{\mathrm{L}}$ is desired, relative to its companion (e.g., $q = M_{1}^{0} / M_{2}^{0}$ for $R_{\mathrm{L}1}^{0} / a_{0}$).

To kick off the common envelope phase, the initial separation $a_{0}$ must be close enough for the more massive star (i.e., the envelope donor) to fill its Roche lobe. In this situation, conservative mass transfer dictates that the orbit proceeds to shrink because the donor-to-accretor mass ratio $q = M_{1}^{0} / M_{2}^{0} > 1$ \citep[e.g.,][]{FKR2002}. Therefore, by setting $R_{\mathrm{L}}$ in Equation \eqref{eqn:Roche} to the radius of the donor star $R_{1}^{0}$, we obtain the maximum initial mean separation $a_{0}^{\mathrm{max}}$ that will result in a common envelope phase. Combining $a_{0}^{\mathrm{max}}$ with Equation \eqref{eqn:af_a0} then gives the maximum final mean separation $a_{\mathrm{f}}^{\mathrm{max}}$ at the end of the common envelope phase. From this moment up until the supernova event, we assume that the mean separation and masses do not change, and that tidal interactions circularize the detached close binary. Under these assumptions, $a_{\mathrm{f}}^{\mathrm{max}}$ effectively places a lower limit on the pre-supernova relative orbital speed, $v_{\mathrm{orb}}^{\mathrm{min}} = \sqrt{G M / a_{\mathrm{f}}^{\mathrm{max}}}$, for use in the natal kick model.

Calculating $a_{\mathrm{f}}^{\mathrm{max}}$ this way requires specifying the parameters $\left\{ a_{0},~\lambda,~\alpha_{\mathrm{ce}} \right\}$ and the masses $\left\{ M_{1}^{0},~M_{1},~M_{2} \right\}$. As a demonstration, we choose representative masses for the red supergiant star $M_{1}^{0} = 30~M_{\odot}$ and its helium core $M_{1} = 12~M_{\odot}$. We start by making it as easy as possible for the natal kick model to produce the large spin-orbit misalignment in V4641 Sgr, which means making conservative choices that push $v_{\mathrm{orb}}^{\mathrm{min}}$ to its lowest possible value (see \S\ref{sec:constraints}). With this in mind, we calculate $a_{0}^{\mathrm{max}}$ as described above, using the maximum radius of a red supergiant star $R_{1}^{0} = 1500~R_{\odot}$ \citep[e.g.,][]{Levesque2017}. This extreme situation corresponds to the widest possible initial separation, such that the largest possible red supergiant star would just barely fill its Roche lobe and trigger a common envelope phase. For this scenario, the widest possible final orbital separation $a_{\mathrm{f}}^{\mathrm{max}}$ then follows from the energy budget formalism (Equation \ref{eqn:af_a0}).

% FIGURE 17:
%\begin{comment}
\begin{figure}[!t]
  \begin{center}
  \includegraphics[width=0.495\textwidth]{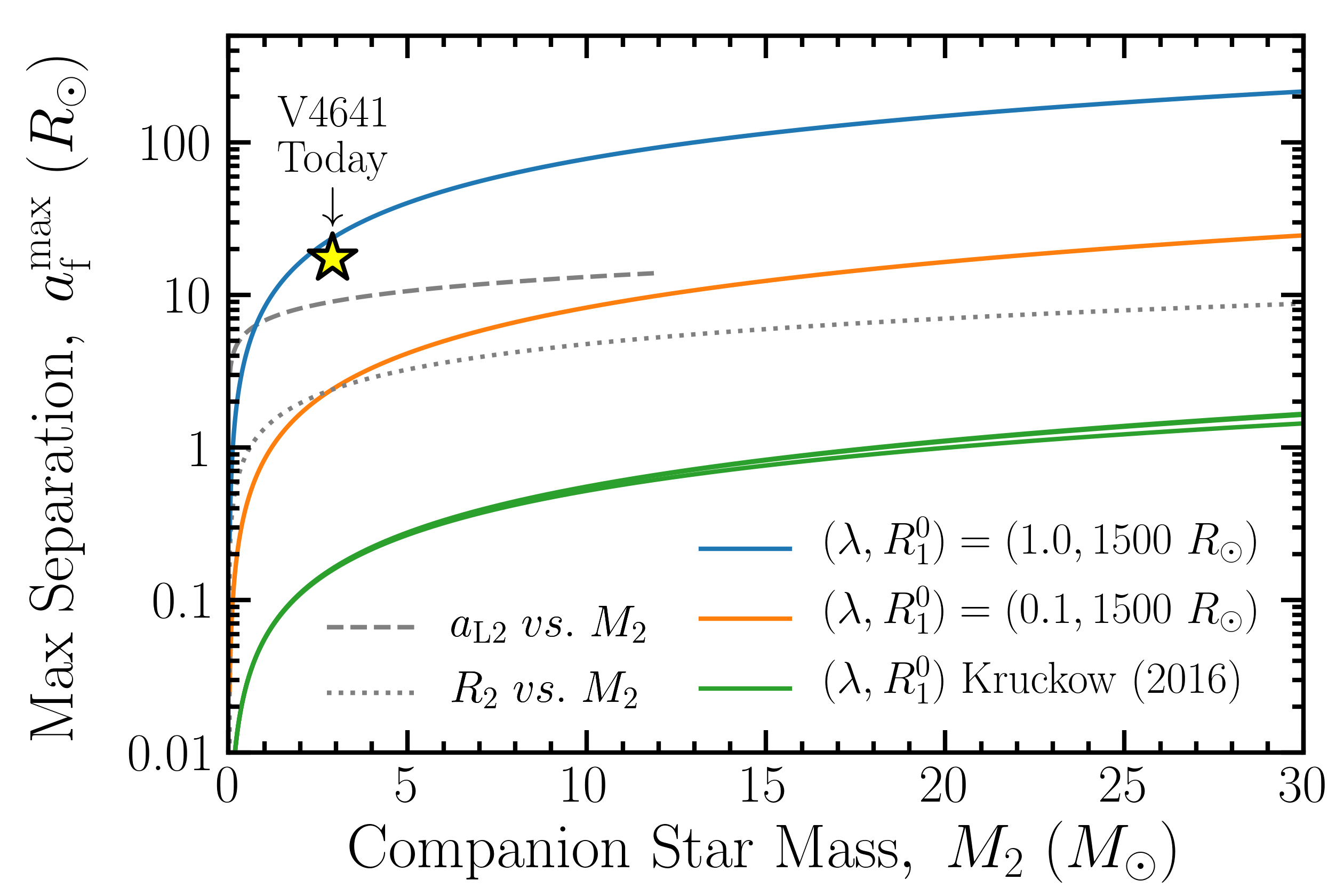}
  \caption{Common envelope energy budget predictions for the maximum final mean separation $a_{\mathrm{f}}^{\mathrm{max}}$, as a function of companion star mass $M_{2}$. All curves adopt $\alpha_{\mathrm{ce}} = 0.5$, and \textit{line colors} map to different $(\lambda, R_{1}^{0})$-pairs. Reducing $\lambda$ from 1.0 (\textit{blue line}) to 0.1 (\textit{orange line}) also reduces $a_{\mathrm{f}}^{\mathrm{max}}$ by an order of magnitude, for a fixed initial donor radius $R_{1}^{0} = 1500~R_{\odot}$. For the $\lambda = 0.1$ case, $a_{\mathrm{f}}^{\mathrm{max}}$ can be smaller than the separation where the companion fills its Roche lobe to initiate mass transfer and further orbital shrinkage (\textit{dashed line}). We caution that predictions based on the energy formalism are questionable, as $(\lambda, R_{1}^{0})$-pairs motivated by stellar structure models (\textit{green lines}) imply final separations much smaller than the radius of the companion star (\textit{dotted line}).}
  \label{fig:vorbM2}
  \end{center}
\end{figure}
%\end{comment}

% FIGURE 18:
%\begin{comment}
\begin{figure*}[!t]
  \begin{center}
  \includegraphics[width=0.495\textwidth]{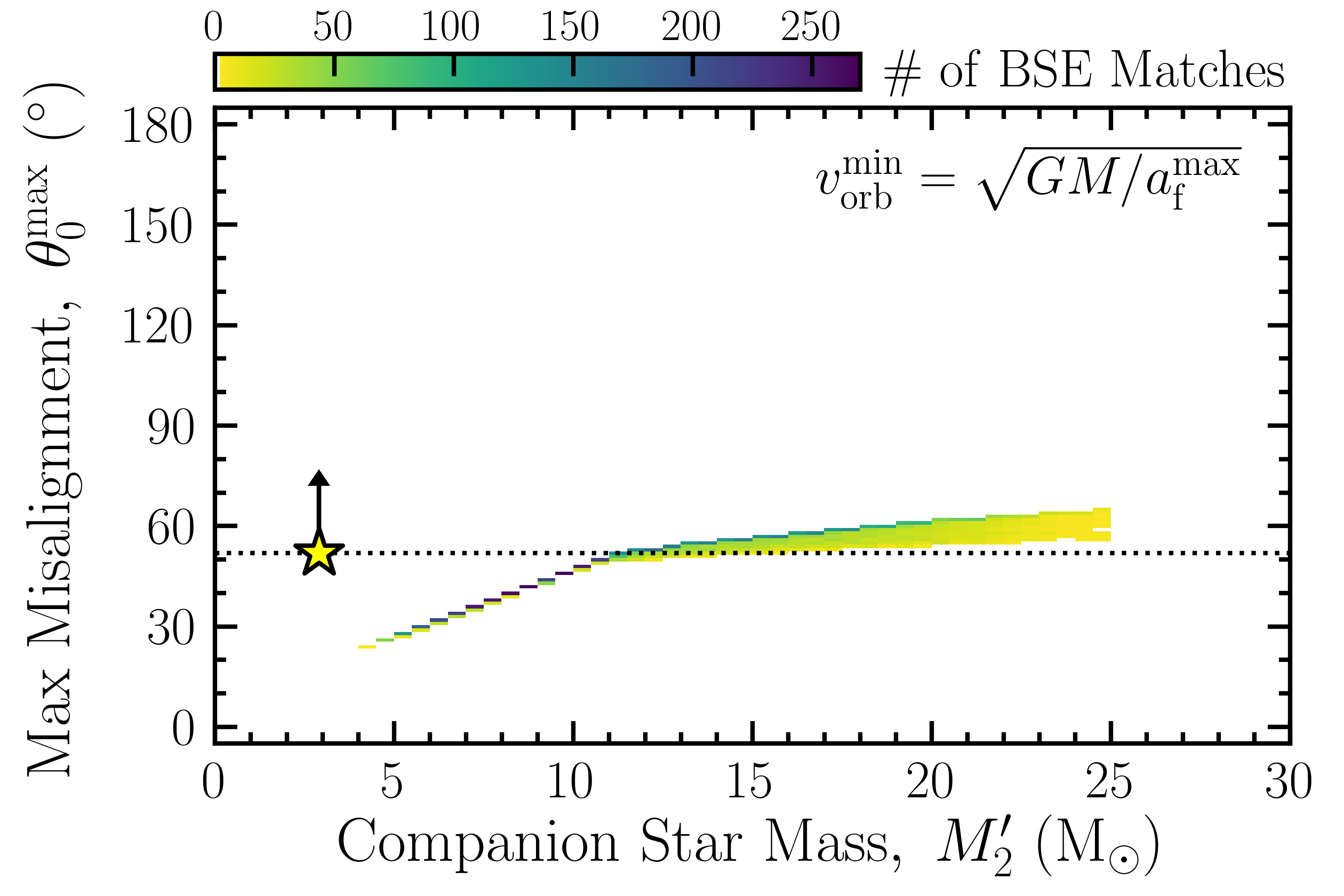}
  \hfill
  \includegraphics[width=0.495\textwidth]{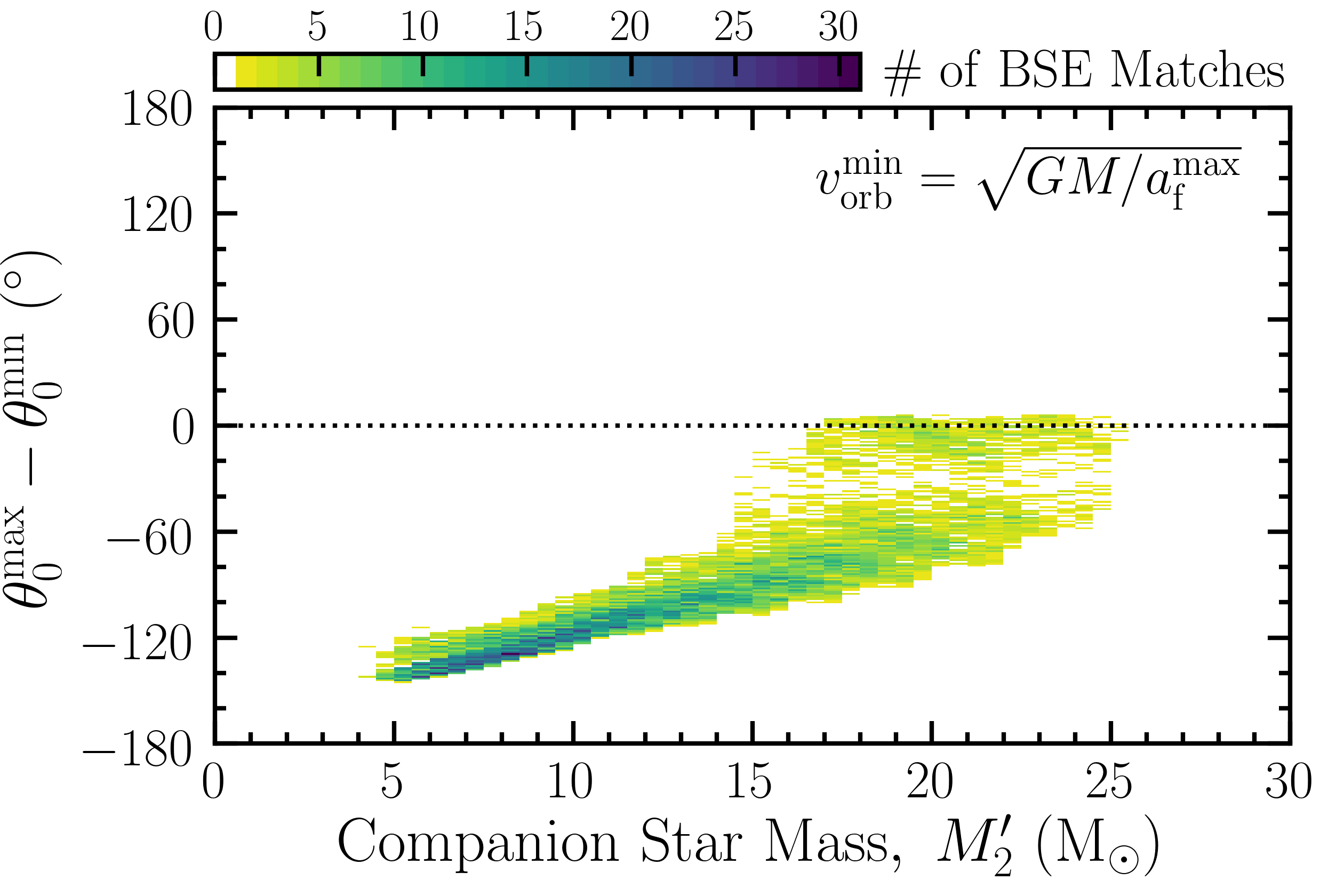}
  \caption{Same as Figure \ref{fig:vorbHalf_i0MaxDiff}, but using the minimum pre-supernova relative orbital speed $v_{\mathrm{orb}}^{\mathrm{min}} = \sqrt{G M / a_{\mathrm{f}}^{\mathrm{max}}}$, as predicted by the common envelope energy budget formalism with $\lambda \alpha_{\mathrm{ce}} = 0.1$ and $R_{1}^{0} = 1500~R_{\odot}$. Of the 12,585 matching BSE models, we plot the 8,655 where $v_{\mathrm{orb}}^{\mathrm{min}}$ did not violate the hard limits of Equation \eqref{eqn:vorbRange}. \textit{Left panel}: Applying the natal kick model leaves 5,076 BSE models above the \textit{dotted line} that can technically produce a spin-orbit misalignment $> 52^{\circ}$, but these require that the V4641 Sgr companion star was $\gtrsim 4$ times more massive than it is today (\textit{yellow star}). No BSE models can be made consistent with $\theta_{0} > 64^{\circ}$. \textit{Right panel}: There are 161 BSE models that can, in the most optimistic of circumstances, achieve a maximum spin-orbit misalignment $\theta_{0}^{\mathrm{max}}$ from a natal kick that just barely exceeds the minimum value $\theta_{0}^{\mathrm{min}}$ implied by their accretion histories. These all require an incredibly high initial companion star mass compared to today, which is seemingly unrealistic.}
  \label{fig:vorbCEE_i0MaxDiff}
  \end{center}
\end{figure*}
%\end{comment}

Figure \ref{fig:vorbM2} shows the predicted $a_{\mathrm{f}}^{\mathrm{max}}$ as a function of companion star mass $M_{2}$, for the choices $\lambda \alpha_{\mathrm{ce}} = 0.5$ (\textit{blue line}) and $\lambda \alpha_{\mathrm{ce}} = 0.05$ (\textit{orange line}), where we stress that our extreme choice $R_{1}^{0} = 1500~R_{\odot}$ makes these final separations especially wide. Population synthesis studies routinely choose $\lambda \alpha_{\mathrm{ce}} \simeq 0.5$ \citep[e.g.,][]{Fragos2010}, even for massive star common envelope evolution, which gives a final mean separation of 10's to 100's of solar radii. Reducing $\lambda \alpha_{\mathrm{ce}}$ by an order of magnitude is problematic because the implied final separation can become so close that the companion star is Roche lobe-filling (\textit{dashed gray line}), such that conservative mass transfer continues to shrink the binary. The bottom line is that the maximum final separation $a_{\mathrm{f}}^{\mathrm{max}}$, and by extension $v_{\mathrm{orb}}^{\mathrm{min}}$, is sensitive to the product $\lambda \alpha_{\mathrm{ce}}$.

As an application to V4641 Sgr, for each matching BSE model we set $\lambda \alpha_{\mathrm{ce}} = 0.1$, calculate $v_{\mathrm{orb}}^{\mathrm{min}} = \sqrt{G M / a_{\mathrm{f}}^{\mathrm{max}}}$ as described above,\footnote{We discard a BSE model if $v_{\mathrm{orb}}^{\mathrm{min}} \lessgtr \sqrt{\frac{G M / a^{\prime}}{1 \pm e^{\prime}}}$ (Equation \ref{eqn:vorbRange}).} and determine the maximum spin-orbit misalignment angle $\theta_{0}^{\mathrm{max}}$ that the natal kick model can produce (see Figure \ref{fig:i0max}). Collecting the results together in Figure \ref{fig:vorbCEE_i0MaxDiff}, the \textit{left panel} shows that the only BSE models remaining as possible V4641 Sgr progenitors (i.e., $\theta_{0}^{\mathrm{max}} > 52^{\circ}$) require the companion star to have lost an incredible amount of mass over time. That is, the companion mass today is $M_{\star} = 2.9 \pm 0.4~M_{\odot}$, while the potential progenitors have companion masses $M_{2} \simeq 11$--$25~M_{\odot}$ at the time of black hole birth. The \textit{right panel} shows that effectively all BSE models are ruled out if we use their accretion histories to account for the tendency toward spin-orbit alignment over time. We point out that no BSE models are consistent with the V4641 Sgr spin-orbit misalignment today if we push this common envelope efficiency product $\lambda \alpha_{\mathrm{ce}}$ down to 0.05 (Figure \ref{fig:vorbM2}, \textit{orange line}).

Figure \ref{fig:vorbCEE_i0MaxDiff} shows the failure of a common envelope phase followed by a natal kick to explain the spin-orbit misalignment in V4641 Sgr. To demonstrate the robustness of this result, derived from $\lambda \alpha_{\mathrm{ce}} = 0.1$ and $R_{1}^{0} = 1500~R_{\odot}$, we now address theoretical estimates of $\alpha_{\mathrm{ce}}$ and $\lambda$ for massive stars to show that our choices were highly conservative given the current understanding.

To be relevant to the black hole progenitor star in V4641 Sgr, we need $\alpha_{\mathrm{ce}}$ and $\lambda$ values associated with red supergiants. Hydrodynamical simulations of massive stars undergoing common envelope evolution find $\alpha_{\mathrm{ce}} < 0.5$ \citep[e.g.,][]{TaamRicker2010}. Stellar evolution models of $\simeq20$--$40~M_{\odot}$ stars find that $\lambda$ decreases with the expanding stellar radius, approximately as a power-law with an exponent between $-2/3$ and $-1$ \citep[e.g.,][]{Kruckow2016, Wang2016}. Rather than fixing $R_{1}^{0} = 1500~R_{\odot}$ at the onset of the common envelope evolutionary phase as done above to very conservatively estimate $a_{\mathrm{f}}^{\mathrm{max}}$, we consult Figure 1 of \citet{Kruckow2016} for three representative $(\lambda, R_{1}^{0})$-pairs: $(1, 10~R_{\odot})$; $(0.1, 100~R_{\odot})$; $(0.01, 1000~R_{\odot})$. Using these $(\lambda, R_{1}^{0})$-pairs determined from stellar structure models and $\alpha_{\rm ce} = 0.5$, Figure \ref{fig:vorbM2} (\textit{green lines}) shows that the energy budget formalism predicts the maximum final separation to be nearly an order of magnitude smaller than the radius of the companion star (\textit{gray dotted line}).\footnote{We estimate the companion star radius from the mass-radius empirical relation $R_{\star} = 1.33 \left( M_{\star} / M_{\odot} \right)^{0.555} R_{\odot}$ for main-sequence stars with mass $M_{\star} > 1.66~{M_{\odot}}$ \citep{DemircanKahraman1991}.}

Taken at face value, inserting constraints on $\alpha_{\mathrm{ce}}$ and $(\lambda, R_{1}^{0})$ for massive stars into the energy budget formalism results in coalescence, barring a mechanism to halt the in-spiral. To skirt this issue, applications of common envelope evolution for massive stars either choose efficiencies much larger than theoretically predicted \citep[e.g., $\lambda \alpha_{\mathrm{ce}} \simeq 0.5$;][]{Fragos2010}, or artificially stop the in-spiral at the separation where the helium core of the donor star fills its Roche lobe \citep[e.g.,][]{Kruckow2016}. Many open questions remain (see \S \ref{sec:origin}), but the simplistic prescriptions adopted by binary evolution models are in serious tension with results from hydrodynamical simulations and stellar structure calculations.

%===========================================================================
%===========================================================================
% DISCUSSION
\section{Discussion}
\label{sec:disc}
Supernova kicks imparted to natal black holes and neutron stars garner support from X-ray binaries and pulsars with high Galactic latitudes and large peculiar velocities. However, we found in \S\ref{sec:bse} that the natal kick model struggles to explain the extreme spin-orbit misalignment $\theta > 52^{\circ}$ in the microquasar V4641 Sgr, which warrants scrutinizing the model assumptions. One assumption is that the pre-supernova binary system achieved spin-orbit alignment, presumably through tidal synchronization following a common envelope phase (\S\ref{sec:envelope}). Another assumption is that the supernova kick does not reorient the angular momentum of the remnant from that of its progenitor; that is, the remnant receives an on-center, linear momentum kick (\S\ref{sec:collapse}). Both of these assumptions hold up to our scrutiny below, leading us to examine the assumption that the jet axis traces the black hole spin axis, which we also deem to be reasonable for V4641 Sgr (\S\ref{sec:tracers}). Having established support for the main assumptions of the natal kick model, we consider alternative origins of the V4641 Sgr spin-orbit misalignment (\S\ref{sec:origin}).

%----------------------------------------------------------------------------------------------------
% COMMON ENVELOPE EVOLUTION AND TIDAL SPIN-UP
\subsection{Common Envelope Evolution and Tidal Spin-Up}
\label{sec:envelope}
The dominant formation channel for close binaries is thought to be common envelope evolution \citep{Paczynski1976, TaamSandquist2000, Ivanova2013}. The story goes that the surface of an evolved star expands beyond its Roche lobe and engulfs its companion in their now ``common envelope.'' The efficient transfer of orbital angular momentum and energy to this common envelope causes the binary to in-spiral, shrinking the separation from $a_{0} \sim 100$--$1000~R_{\odot}$ down to $a_{\mathrm{f}} \sim 1$--$10~R_{\odot}$ \citep[e.g.,][]{RickerTaam2012}. The in-spiral halts when the common envelope is completely ejected, which occurs on a timescale comparable to the initial orbital period \citep[e.g.,][]{Terman1995, TaamRicker2010}, and leaves behind a detached binary consisting of the companion star and the stripped helium core of the primary star. The spins of the core and companion are not thought to change much during this short-lived in-spiral \citep[e.g.,][]{Ivanova2002}. Subsequent tidal synchronization circularizes the binary and extracts orbital angular momentum to spin-up the core and companion on a timescale much shorter than the helium core burning phase \citep[e.g.,][]{vandenHeuvelYoon2007}. Therefore, if tidal spin-up dominates the rotation of the helium core, then the assumption made by the natal kick model of a spin-orbit aligned pre-supernova system is justified.

Asteroseismology of red giant stars reveals the rotation rate of the radiative core to be $\sim10$ times faster than the surface \citep[e.g.,][]{Beck2012, Aerts2019}, but this is $\sim100$ times slower than previously predicted by angular momentum transport mechanisms \citep[e.g.,][]{Marques2013, Cantiello2014}. Encouragingly, recent implementations of the Tayler-Spruit dynamo mechanism for angular momentum transport find red giant core rotation rates in rough agreement with observations \citep{Fuller2019, Spruit2002}. Extrapolating this implementation to red supergiant stars, which are thought to be the progenitors to X-ray binaries, stellar evolution models predict slow-rotating cores \citep{FullerMa2019}, which if correct would validate the notion that tidal spin-up brings the core into spin-orbit alignment. Unfortunately, asteroseismic measurements do not yet exist for the rotation rates of the convective cores in red supergiant stars. Therefore, we cannot rule out the possibility of a fast-rotating core that is misaligned to the binary orbit and whose angular momentum dominates over the aligned component received from tidal spin-up. In this scenario, the core remains misaligned up until the supernova event, which would invalidate the natal kick model assumption of initial spin-orbit alignment.

The aforementioned stellar evolution models predict that black holes born from the core collapse of isolated red supergiant stars are essentially non-spinning, although a post-common envelope close binary can produce moderate spins through tidal spin-up of the helium star \citep{FullerMa2019}. This theoretical expectation of low natal black hole spins is inconsistent with measurements of near-maximal spins in X-ray binaries \citep{Reynolds2014, McClintock2014}, which one expects to be natal unless the black hole subsequently accretes a sizable fraction of its natal mass during the X-ray binary phase \citep{Bardeen1970, KingKolb1999}. The extreme growth needed to achieve high spins would require donor star masses that were much larger in the past, along with highly efficient mass transfer, while also implying spin-orbit alignment \citep[e.g.,][]{FragosMcClintock2015}.\footnote{Accreting millisecond pulsars can reach spin parameters up to $a_{\ast} \simeq 0.4$, limited by gravitational radiation due to the quadrupole of the accreting neutron star \citep[e.g.,][]{Bildsten1998, Chakrabarty2003}. Given that the lever arm of the neutron star surface is similar to the innermost stable circular orbit of a stellar mass black hole, perhaps black hole spin-up by accretion is conceivable.} Therefore, if stellar mass black holes are born from red supergiants with slow-rotating cores, then high spins and high spin-orbit misalignments appear to be mutually exclusive, contrary to observations (see Table \ref{tab:misalign}). This uncomfortable predicament can be avoided if red supergiants turn out to have fast-rotating cores \citep[e.g.,][]{Meynet2015}.

%----------------------------------------------------------------------------------------------------
% NATAL KICKS FROM CORE-COLLAPSE SUPERNOVAE
\subsection{Natal Kicks from Core-Collapse Supernovae}
\label{sec:collapse}
To explain both the short spin periods ($\sim0.01$--$1~\mathrm{s}$) and fast space velocities ($\sim100$--$1000~\mathrm{km/s}$) of radio pulsars, \citet{SpruitPhinney1998} proposed delivering an off-center momentum impulse, or a series of them, to the proto-neutron star. This scenario is supported by the geometric constraints on the double pulsar PSR J0737--3039 \citep{Burgay2003}, whose binary orbit is seen almost perfectly edge-on at $i_{\mathrm{orb}} = 88\fdg7^{+0.5}_{-0.8}$ \citep{Kramer2006}. During orbital conjunction, the magnetosphere of the younger pulsar B eclipses the older/recycled pulsar A. Using a geometric eclipse model \citep{LyutikovThompson2005}, and considering the relativistic precession of pulsar B's spin axis about the total angular momentum of the system, \citet{Breton2008} deduced a spin-orbit misalignment for pulsar B of $\theta_{\mathrm{B}} = 130^{\circ} \pm 1^{\circ}$ ($3\sigma$). The spin of pulsar A is misaligned to the orbital angular momentum by $\theta_{\mathrm{A}} <14^{\circ}$ \citep[$3\sigma$;][]{Ferdman2008}. \citet{Farr2011} attributes the small misalignment of pulsar A to a natal kick to pulsar B, which changed the orbital angular momentum of the pre-kick system (assumed to be aligned; see \S\ref{sec:envelope}). To explain pulsar B's severe \textit{retrograde} spin-orbit misalignment, \citet{Farr2011} concludes that pulsar B must have tumbled to its misaligned configuration after receiving a natal kick directed off-center by $\sim1$--$10~\mathrm{km}$. Conceivably, an asymmetric core-collapse spanning thousands of kilometers in radial extent could miss the bullseye by a few kilometers.

Although plausible for radio pulsars, a similar off-center kick is an implausible explanation for either the spin magnitudes or spin-orbit misalignments of black holes in X-ray binaries. This follows from working out the change in the dimensionless spin parameter $\Delta a_{\ast} = \Delta J c / (G M^{2})$ associated with the change in angular momentum $\Delta J$ that results from an off-center momentum impulse \citep[see][]{SpruitPhinney1998},
\begin{equation}
\Delta a_{\ast} \simeq 0.01 \left( \frac{v}{200~\mathrm{km/s}} \right) \left( \frac{\sin\alpha}{0.5} \right) \left( \frac{f_{\Omega} R_{\mathrm{NS}}}{3 \cdot 10~\mathrm{km}} \right) \left( \frac{M_{\odot}}{M} \right),
\end{equation}
taking typical values of the proto-remnant mass $M$ and radius $R = f_{\Omega} R_{\mathrm{NS}}$ at the time of impulse, and the kick velocity magnitude $v$, directed at an angle $\alpha$ from the center of mass position vector with impact parameter $R \sin\alpha$. Because $\Delta a_{\ast}$ is negligibly small, we conclude that an off-center velocity kick comparable in magnitude to the observed peculiar velocity of V4641 Sgr influenced neither its natal black hole spin magnitude nor direction.

%----------------------------------------------------------------------------------------------------
% JET AXIS AS A TRACER OF THE BLACK HOLE SPIN AXIS
\subsection{Jet Axis as a Tracer of the Black Hole Spin Axis}
\label{sec:tracers}
From a theoretical perspective, the jet is expected to either propagate along or precess about the black hole spin axis. A favored model for what begets jets appeals to a poloidal magnetic field that penetrates the ergosphere, producing a Poynting flux that accelerates material into jets along the black hole spin axis \citep{BlandfordZnajek1977}. An alternative model supposes the inner disk regions support a strong poloidal field, which magneto-centrifugally launches jets that propagate parallel to the inner disk rotational axis \citep{BlandfordPayne1982}. This inner disk may or may not be aligned to the black hole spin axis. A misaligned, geometrically thin ($H/R < \alpha$) inner disk is expected to align quickly on a sub-viscous timescale \citep{BardeenPetterson1975, PapaloizouPringle1983}. However, a misaligned, geometrically thick ($H/R > \alpha$) inner flow is expected to remain misaligned \citep[e.g.,][]{IvanovIllarionov1997, Lubow2002} and precess about the black hole spin axis \citep[e.g.,][]{Fragile2007}, presumably causing the associated jet to precess as well. In another scenario, the jets get redirected by surrounding material, such as a disk wind \citep[e.g.,][]{Begelman2006c}, which decouples the jet orientation from the black hole spin axis.

From an observational perspective, the dynamics and morphologies of microquasar jets offer clues about their utility as black hole spin tracers. In XTE J1550--564, the well-studied X-ray and radio jets from multiple outbursts do not show any evidence of precession (i.e., the spatial extent transverse to the jet axis is $\lesssim1^{\circ}$), which makes a compelling case for jet propagation along the black hole spin axis \citep{SteinerMcClintock2012}. Similarly, the position angle symmetry of the two-sided jet ejection from H1743--322 is consistent with jet-spin alignment expectations \citep{Corbel2005, Steiner2012a}. Also in line with jet-spin alignment, the 1994 outburst of GRO J1655--40 showcased a two-sided radio jet composed of multiple ejections at a $\sim$constant position angle \citep{HjellmingRupen1995}, with deviations from a straight jet axis of only $2^{\circ}$, perhaps indicating mild jet precession with a 3-day period. However, the 2015 outburst of V404 Cyg displayed spatially-resolved, relativistic jets with rapid (hours--days) and large (between $-31^{\circ}$ and $6^{\circ}$) position angle variations, interpreted as Lense-Thirring precession about the black hole spin axis \citep{MillerJones2019} --- see also \citet{Tetarenko2017, Tetarenko2019a}. In SS 433, interactions with outflowing disk material far from the (alleged) black hole redirect the jet, invalidating its use as a spin axis proxy and causing it to precesses on a $20^{\circ}$ half-angle cone with a 162-day period \citep{Begelman2006c}.

The issue at hand is whether the V4641 Sgr jet axis inclination constraint of $i_{\mathrm{jet}} < 16^{\circ}$ \citep[see \S\ref{sec:ijet};][]{Orosz2001} --- derived from the apparent super-luminal motion of a single, spatially extended/elongated, radio emitting ejection --- reliably traces the black hole spin axis. Testing whether the position angle of the jet axis $\phi_{\mathrm{jet}}$ varies between ejections would be helpful \citep[e.g.,][]{MillerJones2019}, but extended jet emission has not been observed since the 1999 outburst where $\phi_{\mathrm{jet}} = 162^{\circ}$ \citep{Hjellming2000}. Lacking spatial information, we turn to the spectral domain for clues. The X-ray spectrum of V4641 Sgr shows a broad emission feature centered on $\sim 6.5~\mathrm{keV}$, which is unusual in its large equivalent width (up to $2~\mathrm{keV}$), rapid variability ($\lesssim$days), and persistence over four orders of magnitude in X-ray luminosity spanning outburst and quiescence \citep[e.g.,][]{Gallo2014}. One possible interpretation of this feature is a blend of Doppler shifted/boosted iron lines from unresolved X-ray jets that precess in-and-out of our line-of-sight, which is consistent with the low jet inclination angle constraint \citep{Gallo2014}.

Alternatively, \citet{Miller2002} interpreted this spectral feature as the broad iron line signature from X-ray disk reflection and found $i_{\mathrm{disk}} = 43^{\circ} \pm 15^{\circ}$, assuming a fixed Galactic column density. Curiously, this $i_{\mathrm{disk}}$ value lies intermediate between $i_{\mathrm{jet}} < 16^{\circ}$ and $i_{\mathrm{orb}} = 72\fdg3 \pm 4\fdg1$. The implied large disk-jet misalignment is inconsistent with the expectation of alignment out to $\sim1000~{R_{\mathrm{g}}}$ in V4641 Sgr \citep{Martin2008b}, which is well-beyond the iron line emitting regions in the disk reflection interpretation. Further challenges to a disk reflection interpretation come from uncertainty in whether the broad feature originates from an inner disk and the variable obscuration known to be associated with V4641 Sgr \citep[e.g.,][]{MaitraBailyn2006}.

Simply put, the observational evidence above suggests that the jets in V4641 Sgr propagate along or precesses about an axis, which is theoretically expected to be the black hole spin axis, inclined slightly to our line-of sight. Therefore, we reasonably conclude that the spin-orbit misalignment is at least several tens of degrees, but we can't be sure. Perhaps the jet precesses on a cone with a very wide opening angle or distant material deflects the jet's trajectory.

Finally, we address the recent claim made for both geometrically thick and thin disks that jets propagate along the rotational axis of a tilted flow --- \textit{not} the black hole spin axis --- based on general relativistic, magnetohydrodynamic simulations of tilted, precessing accretion disks \citep{Liska2018a, Liska2019}. In the thick disk simulations, the jet propagates along the disk rotational axis and they precess together about the black hole spin axis. However, the majority of the analysis focused on disk regions that had not achieved a steady state.\footnote{Inflow equilibrium is established interior to the disk radius associated with the viscous timescale, $R_{\mathrm{visc}} \sim [\alpha (H / R)^{2} (t / t_{\mathrm{g}}) ]^{2/3} R_{\mathrm{g}}$, where $\alpha$ is the effective viscosity parameter, $H$ is the disk scale-height, $t_{\mathrm{g}} = R_{\mathrm{g}} / c$ is the gravitational time, and $R_{\mathrm{g}} = G M / c^{2}$ is the gravitational radius \citep[e.g.,][]{FKR2002}. Inflow equilibrium is only plausibly established out to $\lesssim 50~R_{\mathrm{g}}$ for the thick disk simulations of \citet{Liska2018a}, using $H/R \sim 0.3$, $t = 1.2 \times 10^{5}~t_{\mathrm{g}}$, and $\alpha \sim 0.03$ estimated from the reported $\beta \equiv p_{\mathrm{g}} / p_{\mathrm{mag}} \simeq 18$--$35$ at $R \sim 100~R_{\mathrm{g}}$ \citep[see Table 2 of][]{Salvesen2016a}.} In the thin disk simulations, the innermost disk regions align to the black hole spin, but the gas with comparatively ``negligible angular momentum'' composing the corona remains misaligned. A jet launches parallel to the black hole spin vector, but gets reoriented after propagating only a few gravitational radii. The claim is that interactions with coronal torques reorient the jet, but supportive calculations of these torques are not provided. We note that several systems in Table \ref{tab:misalign} are inconsistent with jet propagation along the outer disk rotational axis, supposing the outer disk and binary orbital planes coincide.

%----------------------------------------------------------------------------------------------------
% ORIGIN OF THE V4641 SGR SPIN-ORBIT MISALIGNMENT
\subsection{Origin of the V4641 Sgr Spin-Orbit Misalignment}
\label{sec:origin}
The natal kick model struggles to explain the impressive $\theta > 52^{\circ}$ spin-orbit misalignment in V4641 Sgr, if the system experienced common envelope evolution that behaves according to the standard energy budget formalism (see \S\ref{sec:CEE}). To avoid this conclusion, either the the envelope structure parameter $\lambda$ and/or the efficiency parameter $\alpha_{\rm ce}$ for unbinding the envelope would have to be a least an order of magnitude greater than their current theoretical predictions. Such large departures from current estimates might be possible by considering the sensitivity of $\lambda$ to the density gradient at the core-envelope boundary \citep[e.g.,][]{TaurisDewi2001, Ivanova2013, Fragos2019} and/or a nuclear energy contribution to unbinding the envelope \citep[e.g.,][]{Podsiadlowski2010}.

Technically, we found that in the most optimistic circumstances a common envelope evolutionary channel could be made compatible with V4641 Sgr if the now-$3~M_{\odot}$ companion star was once very massive (see Figure \ref{fig:vorbCEE_i0MaxDiff}). The companion star today has a B9III classification and is consistent in most respects with generic comparison stars (e.g., the same mass, effective temperature, surface gravity, position on the HR diagram). The notable differences are that the V4641 Sgr companion has $\gtrsim6$ times solar abundance of N and Na \citep{Sadakane2006} and a large de-projected rotational velocity of $v_{\mathrm{rot}} = 106.5 \pm 3.0~\mathrm{km/s}$ \citep{MacDonald2014}, while the comparison stars 14 Cyg (B9III) and $\nu$ Cap (B9IV) have solar abundances in N and Na and projected rotational velocities of $v_{\mathrm{rot}} \sin( i_{\star} ) = 31~\mathrm{km/s}$ \citep{Adelman1999} and $24~\mathrm{km/s}$ \citep{Royer2007}, respectively. Accepting the alternative scenario that the companion mass was once $\sim20~M_{\odot}$ requires an explanation for why it is now masquerading as a relatively typical B9III star.

Having reasonably ruled out a common envelope evolutionary pathway, we consider the possibility of close binary formation through a stable mass transfer channel \citep[e.g.,][]{Langer2019}, whereby the primary (i.e., donor) star expands to fill its Roche lobe around the time that helium-core burning commences. Generally speaking, Roche lobe-overflow mass transfer for a donor-to-accretor mass ratio of $q < q_{\mathrm{crit}}$ is stable, while $q > q_{\mathrm{crit}}$ leads to orbital shrinkage and dynamically unstable mass transfer (e.g., common envelope evolution).

Stable mass transfer calculations for giant donors with convective envelopes find $q_{\mathrm{crit}} \simeq 1.5$--2.2, and even larger $q_{\mathrm{crit}} \simeq 4$ for radiative envelopes \citep{PavlovskiiIvanova2015, Misra2020}. Supposing the progenitor to the black hole in V4641 Sgr was a red supergiant with mass $M_{1} \simeq 20$--$40~M_{\odot}$ and a convection zone that penetrated from the surface deep into the interior, an appropriate choice for the critical mass ratio is $q_{\mathrm{crit}} \simeq 2$. In this scenario, stable mass transfer is only conceivable if the companion star was incredibly massive ($M_{2} > 10~M_{\odot}$) compared to its mass today ($M_{\star} \simeq 3~M_{\odot}$). However, if mass transfer initiated while the envelope was still radiative, then the companion mass constraint relaxes to $M_{2} > 5~M_{\odot}$. This should be considered a firm lower limit on the required companion mass, which would continue to accrete the donor envelope until reaching critical rotation, at which point the mass transfer transitions to being non-conservative \citep[e.g.,][]{deMink2009}.

Interestingly, the surface composition of the companion star shows no evidence of pollution from $\alpha$-process elements \citep{Sadakane2006}. This merely indicates that there is no obvious evidence that V4641 Sgr evolved from a close binary that experienced a supernova event, unlike the case for GRO J1655--40 \citep{Israelian1999}. This lack of contamination is consistent with alternative formation channels, but does not constitute definitive evidence against a binary evolution pathway. For instance, one could attribute the lack of companion star pollution to a ``dark'' explosion that only ejected the N- and Na-rich outer layers of the helium star \citep{Sadakane2006}, but this requires that the supernova retained most of its mass, implying a weak kick and consequently a small spin-orbit misalignment. Other possibilities are that, in its subsequent post-supernova evolution, the companion star lost its ejecta-rich surface layers or this polluted material sunk and is not being efficiently dredged up to the surface.

The appeal of formation channels that do not appeal to binary evolution is their ability to accommodate arbitrary initial spin-orbit misalignments. This is because the angular momenta of the orbit and black hole are decoupled upon binary formation. Unfortunately, we can only speculate on alternative origin stories for the close binary V4641 Sgr and its spin-orbit misalignment, such as dynamical capture during a Galactic plane crossing or triple star evolution \citep[e.g.,][]{EggletonVerbunt1986, Naoz2016}. However, these scenarios are not easily testable with existing observations and suffer from low expected formation rates.

%===========================================================================
%===========================================================================
% SUMMARY AND CONCLUSIONS
\section{Summary and Conclusions}
\label{sec:sumconc}
If the jet from the microblazar V4641 Sgr propagates along the spin axis of the black hole, then its angular momentum is misaligned to that of the binary orbit by $\theta > 52^{\circ}$ today (\S\ref{sec:V4641}), but the initial misalignment $\theta_{0}$ was likely even larger in the past (\S\ref{sec:align}). As a possible explanation for the origin of this extreme spin-orbit misalignment, we applied a natal kick model (\S\ref{sec:model}; \S\ref{sec:constraints}) to potential progenitor systems determined from a grid of binary evolution models (\S\ref{sec:bse}), subject to several constraints on V4641 Sgr today (\S\ref{sec:V4641}) and in the past (\S\ref{sec:vpec}). We found that a standard common envelope phase followed by a natal kick struggles to explain the origin of the spin-orbit misalignment in V4641 Sgr (\S\ref{sec:CEE}). This conclusion gains strength from the conservative theme of our analysis that consistently stacked the deck in favor of a natal kick as the misalignment production mechanism:
\begin{itemize}
\setlength\itemsep{0mm}
\item Our approach was to rule out every conceivable V4641 Sgr progenitor configuration individually.
\item We considered a progenitor to be possible if a natal kick gave the requisite spin-orbit misalignment with non-zero probability, no matter how small.
\item We underestimated this requisite spin-orbit misalignment by adopting the lower limit observed today ($\theta > 52^{\circ}$) and under-predicting the larger initial spin-orbit misalignment $\theta_{0}$.
\item We maximized the pre-supernova separation by choosing a common envelope efficiency product $\lambda \alpha_{\mathrm{ce}} = 0.1$ and a maximal radius for the red supergiant progenitor $R_{1}^{0} = 1500~R_{\odot}$. These parameter choices result in post-common envelope separations an order of magnitude larger than those based on current theoretical predictions.
\item We adopted a natal kick velocity distribution based on observables specific to the V4641 Sgr system and basic Galactic dynamics. This approach has the major advantage of removing the uncertainty associated with choosing a kick distribution and allowed us to constrain the natal systemic velocity extrema.
\end{itemize}
Even with these leniencies, Figure \ref{fig:vorbCEE_i0MaxDiff} shows that a natal kick applied to a close binary that formed through a common envelope channel effectively fails to explain the origin of the V4641 Sgr spin-orbit misalignment. Models where the companion star was incredibly massive ($\simeq20~M_{\odot}$) at the time of the supernova can just barely be made consistent with the very minimum required misalignment. To be taken seriously, such a scenario must explain the subsequent evolution of the companion into the ordinary $3~M_{\odot}$ B9III star observed today, which we deem to be exceedingly unlikely. Stable mass transfer pathways to close binary formation also require large initial companion masses (\S\ref{sec:origin}). The natal kick model assumptions appear to be sound (\S\ref{sec:envelope}; \S\ref{sec:collapse}), leaving us to conclude that large spin-orbit misalignments in black hole X-ray binaries could be evidence against a common envelope evolutionary pathway, as it is currently understood. An alternative interpretation of our results would be to refute the conventional wisdom that the jet axis traces the black hole spin axis (\S\ref{sec:tracers}).

Table \ref{tab:misalign} shows that $\theta > 10^{\circ}$ for most of the X-ray binaries with spin-orbit misalignment constraints, while $\theta$ is formally unconstrained for the systems that are consistent with spin-orbit alignment. This starkly contradicts the expectation of widespread spin-orbit alignment resulting from common envelope evolution \citep{Fragos2010} and/or substantial growth from accretion during an X-ray binary phase \citep{FragosMcClintock2015}. One also expects the angular momentum of the inner disk to align to the black hole spin \citep[e.g.,][]{BardeenPetterson1975}. Invoking these two expectations, the disk continuum fitting community routinely adopts the binary orbital inclination in place of the unknown inner disk inclination when measuring black hole spin \citep[e.g.,][]{McClintock2014}. Our results challenge the validity of this spin-orbit alignment expectation, which likely contributes to tensions between different black hole spin measurement techniques. Embracing misaligned accretion flows might also improve our understanding of X-ray binary variability, such as quasi-periodic oscillations \citep[e.g.,][]{Ingram2009}, precessing jets \citep[e.g.,][]{MillerJones2019}, and state transitions \citep{NixonSalvesen2014}.

Although most X-ray binaries are not black hole merger progenitors, their spin-orbit misalignment is more directly accessible. Therefore, X-ray binaries can be used to test the natal kick model, at least as it applies to lower mass black holes. We advocate opening a critical eye to applications of the natal kick model and its variants, especially binary population synthesis studies. Arguably, population models are of limited utility because, by design, they cannot produce the substantial population of black hole X-ray binaries inferred to have large spin-orbit misalignments based on jet geometries and kinematics \citep[see Table \ref{tab:misalign}; e.g.,][]{Fragos2010}.

Binary population synthesis models that incorporate variants of the natal kick model are commonplace for studying black hole merger statistics in the gravitational wave era \citep[e.g.,][]{Banerjee2019}. The observational diagnostic of black hole mergers that is accessible to population studies is the effective spin parameter $\chi_{\mathrm{eff}}$, which is the vector sum of each black hole's dimensionless spin, mass-weighted and projected onto the binary orbital angular momentum vector. However, using $\chi_{\mathrm{eff}}$ to test the natal kick model is difficult because the pre-merger spin orientations of each black hole are inaccessible from $\chi_{\mathrm{eff}}$ and its distribution is debated \citep[e.g.,][]{LIGO2019, Zackay2019, Venumadhav2019}. We encourage exercising healthy skepticism of population studies that rely on observationally unconfirmed prescriptions for common envelope evolution and natal kicks.

We showed that observational constraints on V4641 Sgr permitted the first refutation of the natal kick model applied to a black hole, to our knowledge. In our opinion, several important questions need to be addressed:
\begin{itemize}
\setlength\itemsep{0mm}
\item Are the predictions from binary population synthesis models of massive star evolution correct (e.g., extended lifetimes, extreme mass loss)?
\item What is the link between natal black hole spin and core rotation in red supergiant stars?
\item What is the orbital separation at the end of the common envelope phase for massive stars?
\item As formation channels for close binaries, what is the relative importance of common envelope evolution \textit{vs.} dynamical capture/encounters?
\end{itemize}
Addressing these questions will improve the systematic uncertainties of binary population synthesis models, and consequently the reliability of black hole merger population studies, which are in vogue at the moment.

%===========================================================================
%===========================================================================
% ACKNOWLEDGEMENTS
\section*{Acknowledgements}
%\acknowledgements
We appreciate the reviewer's thoughtful comments on our work. We thank Omer Blaes, Eric Coughlin, Chris Fryer, Natasha Ivanova, Emily Levesque, Jessica Lu, Rebecca Martin, Jordan Mirocha, Roman Rafikov, and Paul Ricker for productive discussions, along with Jo Bovy, Sylvain Chaty, Tassos Fragos, Jim Fuller, Chase Kimball, Matthias Kruckow, James Miller-Jones, and Jack Steiner for helpful e-mail exchanges.

GS acknowledges support through a National Science Foundation Astronomy \& Astrophysics Postdoctoral Fellowship under award AST-1602169. SP acknowledges support through a Summer Undergraduate Research Fellowship, coordinated by the College of Creative Studies at the University of California, Santa Barbara and funded by Mr. and Mrs. Rodriguez. This work received support from the U.S. Department of Energy through the Los Alamos National Laboratory (LANL), operated by Triad National Security, LLC, for the National Nuclear Security Administration of the U.S. Department of Energy (Contract No. 89233218CNA000001). LANL assigned this article the document release number LA-UR-19-31444.

\software{\texttt{Astropy} \citep{Astropy2013, Astropy2018}, \texttt{BSE} \citep{Hurley2002}, \texttt{galpy} \citep{Bovy2012}, \texttt{Matplotlib} \citep{matplotlib}, \texttt{NumPy} \citep{numpy}, \texttt{SciPy} \citep{scipy}.}

%===========================================================================
%===========================================================================
% BIBLIOGRAPHY
\bibliographystyle{aasjournal}
\bibliography{/Users/salvesen/ms/bib/salvesen}

%===========================================================================
%===========================================================================
% APPENDIX
\appendix
\label{sec:app}
Outlined in \S\ref{sec:model}, the natal kick model offers a mechanism to produce a misalignment between the pre- and post-supernova binary orbital angular momenta. This misalignment is equivalent to the spin-orbit misalignment angle $\theta_{0}$, under two assumptions: (1) the pre-supernova system is spin-orbit aligned and (2) the kick imparts no angular momentum to the black hole. Below, we present the natal kick model comprehensively (Appendix \ref{sec:appA}) and derive the conditional density $P( \theta_{0} | v_{\mathrm{orb}})$ for the spin-orbit misalignment angle $\theta_{0}$ subject to several constraints on the pre-supernova relative orbital speed $v_{\mathrm{orb}}$ (Appendix \ref{sec:appB}). Ultimately, we use $P( \theta_{0} | v_{\mathrm{orb}})$ to test the viability of the natal kick model as a spin-orbit misalignment production mechanism (\S\ref{sec:constraints}; \S\ref{sec:bse}).

We direct the reader to the references listed at the beginning of \S\ref{sec:model} for the original derivations. Thoroughness is our main motivation for this Appendix, but we also correct several minor mistakes peppered throughout the literature and introduce a velocity constraint from requiring the total energy of the post-supernova binary system to exceed its effective potential energy (Equation \ref{eqn:veffAppendix}).
\bigskip

%----------------------------------------------------------------------------------------------------
% NATAL KICK MODEL
\section{Natal Kick Model}
\label{sec:appA}
Figure \ref{fig:kickdiagram} shows the system geometry and dynamics at the moment of the supernova. We adopt a rectangular $( x, y, z )$ coordinate system with unit vectors $( \ihat, \jhat, \khat )$ and an origin at the center of mass of the pre-supernova binary system. The angles $\phi$ and $\omega$ describe the polar and azimuthal directions, respectively, of the kick velocity $\mathbf{v}_{\mathrm{k}}$. The pre-supernova binary system has mass $M = M_{1} + M_{2}$ and a relative orbital speed between the two stars $v_{\mathrm{orb}}$. The post-supernova binary system has mass $M^{\prime} = M_{1}^{\prime} + M_{2}^{\prime}$, eccentricity $e^{\prime}$, and mean orbital separation $a^{\prime}$. All references to the pre- and post-supernova states correspond to conditions immediately before and after the supernova event, taken to be instantaneous, where the prime symbol $(^{\prime})$ denotes the post-supernova epoch.

Just prior to the supernova, star 1 (the black hole progenitor) has mass $M_{1}$ located at $\mathbf{r}_{1} = r_{1} (+\jhat)$ with velocity $\mathbf{v}_{1} = v_{1} (-\ihat)$, while star 2 (the companion) has mass $M_{2}$ located at $\mathbf{r}_{2} = r_{2} (-\jhat)$ with velocity $\mathbf{v}_{2} = v_{2} (+\ihat)$. Working in the pre-supernova center of mass frame, the equations for the center of mass position ($\mathbf{r}_{\mathrm{CM}} \equiv 0$) and velocity ($\mathbf{v}_{\mathrm{CM}} \equiv 0$) of the pre-supernova binary system are given respectively by,
\begin{align}
0 &= M_{1} \mathbf{r}_{1} + M_{2} \mathbf{r}_{2} \label{eqn:comx} \\
0 &= M_{1} \mathbf{v}_{1} + M_{2} \mathbf{v}_{2}. \label{eqn:comv}
\end{align}
The instantaneous separation between the two stars is $r$ and the position vector from star 2 to star 1 is,
\begin{equation}
\mathbf{r} = \mathbf{r}_{1} - \mathbf{r}_{2} = r (+\jhat), \label{eqn:rVec}
\end{equation}
which when combined with Equation \eqref{eqn:comx} gives,
\begin{align}
\mathbf{r}_{1} &= \frac{M_{2}}{M} \mathbf{r} \label{eqn:r1pre} \\
\mathbf{r}_{2} &= \frac{-M_{1}}{M} \mathbf{r}. \label{eqn:r2pre}
\end{align}
The orbital velocity of star 1 relative to star 2 just prior to the supernova is,
\begin{equation} 
\mathbf{v}_{\mathrm{orb}} = \mathbf{v}_{1} - \mathbf{v}_{2} = v_{\mathrm{orb}} (-\ihat), \label{eqn:vorbVec}
\end{equation}
which when combined with Equation \eqref{eqn:comv} gives,
\begin{align}
\mathbf{v}_{1} &= \frac{M_{2}}{M} \mathbf{v}_{\mathrm{orb}} \label{eqn:v1pre} \\
\mathbf{v}_{2} &= \frac{-M_{1}}{M} \mathbf{v}_{\mathrm{orb}}. \label{eqn:v2pre}
\end{align}

The total energy of the pre-supernova binary is,
\begin{align}
E &= - \frac{G M_{1} M_{2}}{r} + \frac{1}{2} M_{1} | \mathbf{v}_{1} |^{2} + \frac{1}{2} M_{2} | \mathbf{v}_{2} |^{2} \nonumber \\
&= - \frac{G M_{1} M_{2}}{r} + \frac{1}{2} \mu \left| \mathbf{v}_{\mathrm{orb}} \right|^{2}, \label{eqn:Epre}
\end{align}
where $\mu = M_{1} M_{2} / M$ is the reduced mass and we used Equations \eqref{eqn:v1pre}--\eqref{eqn:v2pre} to relate $\mathbf{v}_{\mathrm{orb}}$ to $\mathbf{v}_{1}$ and $\mathbf{v}_{2}$. The total energy of the pre-supernova binary can also be expressed in terms of the mean separation $a$ between the stars as,
\begin{equation}
E = - \frac{1}{2} \frac{G M_{1} M_{2}}{a}. \label{eqn:EpreCirc}
\end{equation}
We assume each star follows a circular orbit around the pre-supernova center of mass, presumably the end state of a common envelope evolutionary channel, which implies that $a = r$. Making this substitution and equating \eqref{eqn:Epre} to \eqref{eqn:EpreCirc} gives the instantaneous separation between the stars,
\begin{equation}
r = \frac{G M}{v_{\mathrm{orb}}^{2}}, \label{eqn:rMag}
\end{equation}
where $v_{\mathrm{orb}}$ is the magnitude of the relative orbital velocity between the two stars in the pre-supernova binary system.

The supernova then ejects a mass $\Delta M = M_{1} - M_{1}^{\prime}$ from star 1 and imparts a velocity kick,
\begin{equation}
\mathbf{v}_{\mathrm{k}} = v_{\mathrm{k}} \left[ \cos\left( \phi \right) \cos\left( \omega \right) \ihat + \cos\left( \phi \right) \sin\left( \omega \right) \jhat + \sin\left( \phi \right) \khat \right], \label{eqn:vkVec}
\end{equation}
to the black hole remnant of mass $M_{1}^{\prime}$, while the mass of star 2 is assumed unchanged, $M_{2}^{\prime} = M_{2}$. Continuing to work in the pre-supernova center of mass frame, the post-supernova velocities of the black hole and star 2 are,
\begin{align}
\mathbf{v}_{1}^{\prime} &= \mathbf{v}_{1} + \mathbf{v}_{\mathrm{k}} \label{eqn:v1post} \\
\mathbf{v}_{2}^{\prime} &= \mathbf{v}_{2}. \label{eqn:v2post}
\end{align}
The supernova kick combined with mass loss gives the immediate post-supernova binary system as a whole a systemic velocity $\mathbf{v}_{\mathrm{sys}}$, or bulk motion, relative to the pre-supernova center of mass frame. In other words, $\mathbf{v}_{\mathrm{sys}}$ is the velocity of the center of mass of the post-supernova binary system,
\begin{equation}
M^{\prime} \mathbf{v}_{\mathrm{sys}} = M_{1}^{\prime} \mathbf{v}_{1}^{\prime} + M_{2}^{\prime} \mathbf{v}_{2}^{\prime}. \label{eqn:vpostCoM}
\end{equation}
Inserting Equations \eqref{eqn:v1post} and \eqref{eqn:v2post} for $\mathbf{v}_{1}^{\prime}$ and $\mathbf{v}_{2}^{\prime}$ into Equation \eqref{eqn:vpostCoM}, using $\mathbf{v}_{1} = (- M_{2} / M) v_{\mathrm{orb}} \ihat$ and $\mathbf{v}_{2} = (M_{1} / M) v_{\mathrm{orb}} \ihat$ from Equations \eqref{eqn:vorbVec}--\eqref{eqn:v2pre}, using the relation $M_{1} - M_{1}^{\prime} = M - M^{\prime}$ implied by $M_{2}^{\prime} = M_{2}$, replacing $\mathbf{v}_{\mathrm{k}}$ with Equation \eqref{eqn:vkVec}, and taking the dot product of $\mathbf{v}_{\mathrm{sys}}$ with itself yields,
\begin{equation}
v_{\mathrm{sys}}^{2} = \frac{M_{1}^{\prime 2}}{M^{\prime 2}} v_{\mathrm{k}}^{2} + 2 f \frac{M_{1}^{\prime} M_{2}^{\prime}}{M^{\prime 2}} v_{\mathrm{k}} v_{\mathrm{orb}} \cos\left( \phi \right) \cos\left( \omega \right) + f^{2} \frac{M_{2}^{\prime 2}}{M^{\prime 2}} v_{\mathrm{orb}}^{2}, \label{eqn:vsys1}
\end{equation}
where $f = 1 - M^{\prime} / M$ is the fractional mass lost in the supernova. 

The assumption of supernova instantaneity means the positions of the binary components remain unchanged between the immediate pre- and post-supernova epochs. Therefore, $\mathbf{r}_{1}^{\prime} = \mathbf{r}_{1}$, $\mathbf{r}_{2}^{\prime} = \mathbf{r}_{2}$, $\mathbf{r}^{\prime} = \mathbf{r}$, $r^{\prime} = r$, and we drop the primes on these instantaneous positions and separations from here onward.

The total energy of the post-supernova binary system is,
\begin{align}
E^{\prime} &= - \frac{G M_{1}^{\prime} M_{2}^{\prime}}{r} + \frac{1}{2} \mu^{\prime} | \mathbf{v}_{\mathrm{orb}}^{\prime} |^{2} \nonumber \\
&= - \frac{G M_{1}^{\prime} M_{2}^{\prime}}{r} + \frac{1}{2} \mu^{\prime} | \mathbf{v}_{\mathrm{orb}} + \mathbf{v}_{\mathrm{k}} |^{2}, \label{eqn:Epost}
\end{align}
where $\mu^{\prime} = M_{1}^{\prime} M_{2}^{\prime} / M^{\prime}$ is the new reduced mass and $\mathbf{v}_{\mathrm{orb}}^{\prime} = \mathbf{v}_{1}^{\prime} - \mathbf{v}_{2}^{\prime}$ is the post-supernova orbital velocity of the newborn black hole relative to star 2. We obtained $\mathbf{v}_{\mathrm{orb}}^{\prime} = \mathbf{v}_{\mathrm{orb}} + \mathbf{v}_{\mathrm{k}}$ from Equations \eqref{eqn:v1post}--\eqref{eqn:v2post} and \eqref{eqn:vorbVec}. Assuming the supernova does not unbind the binary system, the total energy of the post-supernova binary is given by,
\begin{equation}
E^{\prime} = - \frac{1}{2} \frac{G M_{1}^{\prime} M_{2}^{\prime}}{a^{\prime}}, \label{eqn:EpostCirc}
\end{equation}
where, in general, the post-supernova mean separation $a^{\prime} \ne r$ because the new binary orbit can be eccentric.\footnote{Our notation has some subtle differences from \citet{Martin2009}. We distinguish between the instantaneous separation $r$ and the mean orbital separation $a$, such that $r = a = r^{\prime}$ but $a^{\prime} \ne a$, whereas \citet{Martin2009} express the instantaneous separation as $a$, such that $a^{\prime} = a$. Our definitions of $a^{\prime}$ and $v_{\mathrm{orb}}^{\prime}$ are the same as $a_{\mathrm{n}}$ and $v_{\mathrm{n}}$ in \citet{Martin2009}, who instead chose $v_{\mathrm{orb}}^{\prime} = G M^{\prime} / a^{\prime}$ to be the relative orbital speed that the black hole and star 2 would have if their orbit immediately post-supernova was circular.}

The requirement that the post-supernova orbit remain bound places an upper limit on the total energy,
\begin{align}
E^{\prime} \le 0. \label{eqn:EpostUpLim}
\end{align}
The effective potential energy $U_{\mathrm{eff}}^{\prime}$ imposes a lower limit on the total energy of the post-supernova binary system,
\begin{align}
E^{\prime} &\ge U_{\mathrm{eff}}^{\prime} \nonumber \\
&\ge \frac{L^{\prime2}}{2 \mu^{\prime} r^{2}} - \frac{G M_{1}^{\prime} M_{2}^{\prime}}{r}, \label{eqn:EpostLowLim}
\end{align}
where the total angular momentum of the post-supernova binary system about its center of mass is,
\begin{align}
\mathbf{L}^{\prime} &= \mathbf{r} \times \mu^{\prime} \mathbf{v}_{\mathrm{orb}}^{\prime} \nonumber \\
&= \mathbf{r} \times \mu^{\prime} \left( \mathbf{v}_{\mathrm{orb}} + \mathbf{v}_{\mathrm{k}} \right) \nonumber \\
&= \mu^{\prime} r \left( v_{\mathrm{orb}} \khat + \jhat \times \mathbf{v}_{\mathrm{k}} \right). \label{eqn:LpostVec}
\end{align}
Here, we again used $\mathbf{v}_{\mathrm{orb}}^{\prime} = \mathbf{v}_{\mathrm{orb}} + \mathbf{v}_{\mathrm{k}}$, as well as $\mathbf{r} = r (+\jhat)$ and $\mathbf{v}_{\mathrm{orb}} = v_{\mathrm{orb}} (-\ihat)$ from Equations \eqref{eqn:rVec} and \eqref{eqn:vorbVec}. The magnitude of the post-supernova angular momentum can also be expressed in terms of the eccentricity $e^{\prime}$ and mean separation $a^{\prime}$ as,
\begin{equation}
L^{\prime} = \mu^{\prime} \left[ G M^{\prime} a^{\prime} \left( 1 - e^{\prime 2} \right) \right]^{1/2}. \label{eqn:LpostMag}
\end{equation}
Notably, inequality \eqref{eqn:EpostLowLim} restricts the allowable $v_{\mathrm{orb}}$ range to,
\begin{equation}
\frac{G M / a^{\prime}}{1 + e^{\prime}} \le v_{\mathrm{orb}}^{2} \le \frac{G M / a^{\prime}}{1 - e^{\prime}}, \label{eqn:vorbMinMax}
\end{equation}
which follows from substituting for $E^{\prime}$ (Equation \ref{eqn:EpostCirc}), $L^{\prime}$ (Equation \ref{eqn:LpostMag}), $r$ (Equation \ref{eqn:rMag}), and $\mu^{\prime}$.

Manipulating the above expressions for the binary system's orbital angular momentum post-supernova and energy pre-/post-supernova as follows leads to results that we will use in Appendix \ref{sec:appB}. Equating the magnitude $| \mathbf{L^{\prime}} |$ from \eqref{eqn:LpostVec} to $L^{\prime}$ from \eqref{eqn:LpostMag} and squaring this equality gives,
\begin{equation}
\left| v_{\mathrm{orb}} \khat + \jhat \times \mathbf{v}_{\mathrm{k}} \right|^{2} = \frac{G M^{\prime}}{r} \frac{a^{\prime}}{r}\left( 1 - e^{\prime 2} \right). \label{eqn:vorbXvk1}
\end{equation}
Equating \eqref{eqn:Epre} to \eqref{eqn:EpreCirc} for $E$, equating \eqref{eqn:Epost} to \eqref{eqn:EpostCirc} for $E^{\prime}$, and then dividing these equalities by each other gives, 
\begin{equation}
\left| \mathbf{v}_{\mathrm{orb}} + \mathbf{v}_{\mathrm{k}} \right|^{2} = \frac{G M^{\prime}}{r} \left( 2 - \frac{r}{a^{\prime}} \right), \label{eqn:vorb+vk1}
\end{equation}
where we used the substitution $a = r$ from the assumption of binary circularization pre-supernova. Directly inserting Equation \eqref{eqn:vkVec} for the kick velocity $\mathbf{v}_{\mathrm{k}}$ into the left-hand sides of Equations \eqref{eqn:vorbXvk1} and \eqref{eqn:vorb+vk1}, and recalling that $\mathbf{v}_{\mathrm{orb}} = v_{\mathrm{orb}} (-\ihat)$, gives the independent expressions,
\begin{align}
\left| v_{\mathrm{orb}} \khat + \jhat \times \mathbf{v}_{\mathrm{k}} \right|^{2} &= v_{\mathrm{orb}}^{2} - 2 v_{\mathrm{orb}} v_{\mathrm{k}} \cos\left( \phi \right) \cos\left( \omega \right) + v_{\mathrm{k}}^{2} \left[ \sin^{2}\left( \phi \right) + \cos^{2}\left( \phi \right) \cos^{2}\left( \omega \right) \right] \label{eqn:vorbXvk2} \\
\left| \mathbf{v}_{\mathrm{orb}} + \mathbf{v}_{\mathrm{k}} \right|^{2} &= v_{\mathrm{orb}}^{2} - 2 v_{\mathrm{orb}} v_{\mathrm{k}} \cos\left( \phi \right) \cos\left( \omega \right) + v_{\mathrm{k}}^{2}. \label{eqn:vorb+vk2}
\end{align}

The right-hand sides of Equations \eqref{eqn:vorbXvk1}--\eqref{eqn:vorb+vk2} are cast in terms of the supernova kick parameters $\left\{ v_{\mathrm{k}}, \phi, \omega \right\}$ and the binary system parameters $\left\{ v_{\mathrm{orb}}, M, M^{\prime}, e^{\prime}, a^{\prime} \right\}$, recalling that $r = G M / v_{\mathrm{orb}}^{2}$ (Equation \ref{eqn:rMag}).
\bigskip
\newpage
%----------------------------------------------------------------------------------------------------
% CONDITIONAL DENSITY FUNCTION FOR THE SPIN-ORBIT MISALIGNMENT ANGLE 
\section{Conditional Density Function for the Spin-Orbit Misalignment Angle}
\label{sec:appB}
The supernova kick misaligns the post-supernova binary orbital angular momentum $\mathbf{L}^{\prime}$ relative to the pre-supernova binary orbital angular momentum $\mathbf{L} = L (+\khat)$ by the angle $\theta_{0}$, given by,
\begin{equation}
\mathbf{L}^{\prime} \cdot \khat = L^{\prime} \cos( \theta_{0} ). \label{eqn:cosi_def}
\end{equation}
Replacing $\mathbf{L}^{\prime}$ and $L^{\prime} = | \mathbf{L}^{\prime} \cdot \mathbf{L}^{\prime} |^{1/2}$ in Equation \eqref{eqn:cosi_def} with Equation \eqref{eqn:LpostVec}, and then replacing $\mathbf{v}_{\mathrm{k}}$ with Equation \eqref{eqn:vkVec} gives \citep{BrandtPodsiadlowski1995},
\begin{equation}
\cos\left( \theta_{0} \right) = \frac{v_{\mathrm{orb}} - v_{\mathrm{k}} \cos\left( \phi \right) \cos\left( \omega \right)}{\left| v_{\mathrm{k}}^{2} \sin^{2}\left( \phi \right) + \left[ v_{\mathrm{orb}} - v_{\mathrm{k}} \cos\left( \phi \right) \cos\left( \omega \right) \right]^{2} \right|^{1/2}},\label{eqn:cos_i0}
\end{equation}
which can be solved for $\omega$ \citep{Martin2009},
\begin{equation}
\omega = \cos^{-1}\left[ \frac{v_{\mathrm{orb}}}{v_{\mathrm{k}}} \frac{1}{\cos\left( \phi \right)} - \frac{\left| \tan\left( \phi \right) \right|}{\tan\left( \theta_{0} \right)} \right]. \label{eqn:cos_omega}
\end{equation}

The supernova kick is parametrized in terms of the kick velocity magnitude $v_{\mathrm{k}}$ and direction $(\phi, \omega)$, which we take to be independently distributed such that their joint density is $P_{v_{\mathrm{k}}, \phi, \omega}\left( v_{\mathrm{k}}, \phi, \omega \right) = P_{v_{\mathrm{k}}}\left( v_{\mathrm{k}} \right) P_{\phi}\left( \phi \right) P_{\omega}\left( \omega \right)$. The velocity kick magnitude can span $v_{\mathrm{k}} \in [ 0, \infty )$ and we assume its direction angles $\phi \in [ -\pi/2, \pi/2 ]$ and $\omega \in [ 0, 2 \pi )$ are uniformly distributed on a unit sphere, such that $P_{\phi}( \phi ) = \cos( \phi ) / 2$ and $P_{\omega}( \omega ) = 1 / (2 \pi)$.\footnote{\citet{Hurley2002} and others \citep[e.g.,][]{Martin2009, Martin2010} omit the factor of $\frac{1}{2}$ in $P_{\phi}( \phi )$, which leads to the improper normalization $\iint P_{\phi}( \phi) P_{\omega}( \omega ) d\phi d\omega = 2$.}

Changing variables from $\left( v_{\mathrm{k}}, \phi, \omega \right)$ to $\left( v_{\mathrm{k}}, \phi, \theta_{0} \right)$ gives the joint conditional density,
\begin{align}
P_{v_{\mathrm{k}}, \phi, \theta_{0} | v_{\mathrm{orb}}}\left( v_{\mathrm{k}}, \phi, \theta_{0} | v_{\mathrm{orb}} \right) &= P_{v_{\mathrm{k}}, \phi, \omega | v_{\mathrm{orb}}}\left( v_{\mathrm{k}}, \phi, h^{-1}\left( v_{\mathrm{k}}, \phi, \theta_{0} \right) | v_{\mathrm{orb}} \right) \times \left| J\left( v_{\mathrm{k}}, \phi, \theta_{0} | v_{\mathrm{orb}} \right) \right| \nonumber \\
&= P_{v_\mathrm{k}}\left( v_{\mathrm{k}} \right) P_{\phi}\left( \phi \right) P_{\omega | v_{\mathrm{orb}}}\left( h^{-1}\left( v_{\mathrm{k}}, \phi, \theta_{0} \right) | v_{\mathrm{orb}} \right) \times \left| J\left( v_{\mathrm{k}}, \phi, \theta_{0} | v_{\mathrm{orb}} \right) \right|, \label{eqn:Pi_joint}
\end{align}
where Equation \eqref{eqn:cos_omega} provides the inverse transformation function $\omega = h^{-1}\left( v_{\mathrm{k}}, \phi, \theta_{0} \right)$ and the Jacobian of the transformation is,
\begin{align}
J &= \mathrm{det}\left[ \frac{\partial \left( v_{\mathrm{k}}, \phi, \omega \right)}{\partial \left( v_{\mathrm{k}}, \phi, \theta_{0} \right)} \right]
= \mathrm{det}\left[ \begin{array}{@{\hspace{1mm}}c@{\hspace{1mm}}c@{\hspace{1mm}}c@{\hspace{1mm}}c@{\hspace{1mm}}c@{\hspace{1mm}}}
\left. \frac{\partial v_{\mathrm{k}}}{\partial v_{\mathrm{k}}} \right|_{\phi, \theta_{0}} & \left. \frac{\partial v_{\mathrm{k}}}{\partial \phi} \right|_{v_{\mathrm{k}}, \theta_{0}} & \left. \frac{\partial v_{\mathrm{k}}}{\partial \theta_{0}} \right|_{v_{\mathrm{k}}, \phi} \smallskip \\
\left. \frac{\partial \phi}{\partial v_{\mathrm{k}}} \right|_{\phi, \theta_{0}} & \left. \frac{\partial \phi}{\partial \phi} \right|_{v_{\mathrm{k}}, \theta_{0}} & \left. \frac{\partial \phi}{\partial \theta_{0}} \right|_{v_{\mathrm{k}}, \phi} \smallskip \\
\left. \frac{\partial \omega}{\partial v_{\mathrm{k}}} \right|_{\phi, \theta_{0}} & \left. \frac{\partial \omega}{\partial \phi} \right|_{v_{\mathrm{k}}, \theta_{0}} & \left. \frac{\partial \omega}{\partial \theta_{0}} \right|_{v_{\mathrm{k}}, \phi}
\end{array} \right]
= \mathrm{det}\left[ \begin{array}{@{\hspace{1mm}}c@{\hspace{1mm}}c@{\hspace{1mm}}c@{\hspace{1mm}}c@{\hspace{1mm}}c@{\hspace{1mm}}}
1 & 0 & 0 \smallskip \\
0 & 1 & 0 \smallskip \\
0 & 0 & \left. \frac{\partial \omega}{\partial \theta_{0}} \right|_{v_{\mathrm{k}}, \phi}
\end{array} \right]
= \left. \frac{\partial \omega}{\partial \theta_{0}} \right|_{v_{\mathrm{k}}, \phi} \nonumber \\
J &= \frac{- \left| \tan\left( \phi \right) \right|}{\sin\left( h^{-1}\left( v_{\mathrm{k}}, \phi, \theta_{0} \right) \right) \sin^{2}\left( \theta_{0} \right)}. \label{eqn:Ji}
\end{align}
Marginalizing out $v_{\mathrm{k}}$ and $\phi$ in Equation \eqref{eqn:Pi_joint} gives the conditional density for the misalignment angle $\theta_{0}$,
\begin{equation}
P_{\theta_{0} | v_{\mathrm{orb}}}\left( \theta_{0} | v_{\mathrm{orb}} \right) = \frac{1}{2 \pi} \iint\limits_{R}  P_{v_{\mathrm{k}}}\left( v_{\mathrm{k}} \right) \frac{\left| \sin\left( \phi \right) \right|}{\left| \sin\left( h^{-1}\left( v_{\mathrm{k}}, \phi, \theta_{0} \right) \right) \right| \sin^{2}\left( \theta_{0} \right)} dv_{\mathrm{k}} d\phi. \label{eqn:Pi_norm}
\end{equation}
For normalization purposes, we introduced a multiplicative factor of 2 in Equation \eqref{eqn:Pi_norm} such that $\int_{0}^{\pi} P_{\theta_{0} | v_{\mathrm{orb}}}\left( \theta_{0} | v_{\mathrm{orb}} \right) d\theta_{0} = 1$.\footnote{By omitting the $\frac{1}{2}$ factor in $P_{\phi}( \phi )$, \citet{Martin2009, Martin2010} serendipitously obtain the properly normalized expression for $P_{\theta_{0} | v_{\mathrm{orb}}}( \theta_{0} | v_{\mathrm{orb}} )$.} The justification for this is that the change of variables replaced an azimuthal angle $\omega \in [ 0, 2 \pi )$ with a polar angle $\theta_{0} \in [ 0, \pi ]$. In other words, two different $\omega$ values can produce the same $\theta_{0}$ value in Equation \eqref{eqn:cos_i0}.

The integration region $R$ in Equation \eqref{eqn:Pi_norm} defines the allowable $(v_{\mathrm{k}}, \phi)$-space for a given misalignment angle $\theta_{0}$. Mapping the limits $\omega \in [ 0, 2 \pi )$ to an integrable region in $(v_{\mathrm{k}}, \phi)$-space for a given $\theta_{0}$ follows from requiring $\cos( \omega )$ in Equation \eqref{eqn:cos_omega} to be real-valued, which restricts $v_{\mathrm{k}}$ to the range $\min\left[ v_{\pm} \right] \le v_{\mathrm{k}} \le \max\left[ v_{\pm} \right]$, where \citep{Martin2009},
\begin{equation}
v_{\pm} = \frac{v_{\mathrm{orb}}}{\cos\left( \phi \right)} \left[ \frac{\left| \tan\left( \phi \right) \right|}{\tan\left( \theta_{0} \right)} \pm 1 \right]^{-1}. \label{eqn:vpmAppendix}
\end{equation}
Specifying this constraint on $R$ amounts to considering the entire physically permissible region in $(v_{\mathrm{k}}, \phi)$-space for a given $\theta_{0}$ and yields integral unity of the misalignment angle conditional density, $\int_{0}^{\pi} P_{\theta_{0} | v_{\mathrm{orb}}}(\theta_{0} | v_{\mathrm{orb}}) d\theta_{0} = 1$. Next, we enforce more restrictive criteria on $R$ to determine the probability of producing specific misalignment scenarios.

The post-supernova binary systems of interest remain bound ($E^{\prime} < 0$) and have energies exceeding the effective potential ($E^{\prime} > U_{\mathrm{eff}}^{\prime}$). To cast these constraints on $E^{\prime}$ into constraints on $v_{\mathrm{k}}( \phi | \theta_{0} )$ that can be incorporated into the integration region $R$ of Equation \eqref{eqn:Pi_norm}, we appeal to inequalities \eqref{eqn:EpostUpLim} and \eqref{eqn:EpostLowLim}, replacing $E^{\prime}$ and $L^{\prime}$ with Equations \eqref{eqn:Epost} and \eqref{eqn:LpostMag}. The inequalities \eqref{eqn:EpostUpLim} and \eqref{eqn:EpostLowLim} then become, respectively,
\begin{align}
\left| \mathbf{v}_{\mathrm{orb}} + \mathbf{v}_{\mathrm{k}} \right|^{2} &\le 2 \frac{G M^{\prime}}{r} \label{eqn:EpostUpLim2} \\
\left| \mathbf{v}_{\mathrm{orb}} + \mathbf{v}_{\mathrm{k}} \right|^{2} &\ge \frac{G M^{\prime}}{r} \frac{a^{\prime}}{r} \left( 1 - e^{\prime 2} \right). \label{eqn:EpostLowLim2}
\end{align}
After substituting Equation \eqref{eqn:rMag} for $r$ and Equation \eqref{eqn:vorb+vk2} for $| \mathbf{v}_{\mathrm{orb}} + \mathbf{v}_{\mathrm{k}} |^{2}$, inequalities \eqref{eqn:EpostUpLim2} and \eqref{eqn:EpostLowLim2} become,
\begin{align}
0 &\ge v_{\mathrm{k}}^{2} - 2 v_{\mathrm{k}} v_{\mathrm{orb}} \cos\left( \phi \right) \cos\left( \omega \right) + \left( 1 - 2 \frac{M^{\prime}}{M} \right) v_{\mathrm{orb}}^{2} \label{eqn:EpostUpLim3} \\
0 &\le v_{\mathrm{k}}^{2} - 2 v_{\mathrm{k}} v_{\mathrm{orb}} \cos\left( \phi \right) \cos\left( \omega \right) + \left[ 1 - \frac{M^{\prime}}{M} \left( 1 - e^{\prime2} \right) \frac{v_{\mathrm{orb}}^{2}}{GM / a^{\prime}} \right] v_{\mathrm{orb}}^{2}, \label{eqn:EpostLowLim3}
\end{align}
Inserting Equation \eqref{eqn:cos_omega} for $\cos( \omega )$, the inequalities \eqref{eqn:EpostUpLim3} and \eqref{eqn:EpostLowLim3} finally become,
\begin{align}
0 &\ge v_{\mathrm{k}}^{2} + 2 v_{\mathrm{k}} v_{\mathrm{orb}} \frac{\left| \sin\left( \phi \right) \right|}{\tan\left( \theta_{0} \right)} - \left( 1 + 2 \frac{M^{\prime}}{M} \right) v_{\mathrm{orb}}^{2} \label{eqn:EpostUpLim4} \\
0 &\le v_{\mathrm{k}}^{2} + 2 v_{\mathrm{k}} v_{\mathrm{orb}} \frac{\left| \sin\left( \phi \right) \right|}{\tan\left( \theta_{0} \right)} - \left[ 1 + \frac{M^{\prime}}{M} \left( 1 - e^{\prime 2} \right) \frac{v_{\mathrm{orb}}^{2}}{GM / a^{\prime}} \right] v_{\mathrm{orb}}^{2}. \label{eqn:EpostLowLim4}
\end{align}
Both of these inequalities are quadratic in $v_{\mathrm{k}}$ and have real roots --- one negative and one positive. Only the positive root is physically meaningful because $v_{\mathrm{k}} \ge 0$. The positive root from inequality \eqref{eqn:EpostUpLim4} gives the kick velocity required to unbind the binary system \citep{BrandtPodsiadlowski1995, Martin2009},
\begin{equation}
v_{\mathrm{bound}} = v_{\mathrm{orb}} \left[ \sqrt{1 + 2 \frac{M^{\prime}}{M} + \frac{\sin^{2}\left( \phi \right)}{\tan^{2} \left( \theta_{0} \right)}} - \frac{\left| \sin\left( \phi \right) \right|}{\tan\left( \theta_{0} \right)} \right], \label{eqn:vboundAppendix}
\end{equation}
while the positive root from inequality \eqref{eqn:EpostLowLim4} gives the minimum kick velocity required for the energy of the post-supernova binary system to exceed the effective potential,
\begin{equation}
v_{\mathrm{eff}} = v_{\mathrm{orb}} \left[ \sqrt{1 + \frac{M^{\prime}}{M} \left( 1 - e^{\prime 2} \right) \frac{v_{\mathrm{orb}}^{2}}{G M / a^{\prime}} + \frac{\sin^{2}\left( \phi \right)}{\tan^{2} \left( \theta_{0} \right)}} - \frac{\left| \sin\left( \phi \right) \right|}{\tan\left( \theta_{0} \right)} \right]. \label{eqn:veffAppendix}
\end{equation}
Therefore, the permissible energy range $U_{\mathrm{eff}}^{\prime} \le E^{\prime} \le 0$ of the immediate post-supernova binary system translates to the restricted range for the kick velocity magnitude $v_{\mathrm{eff}} \le v_{\mathrm{k}} \le v_{\mathrm{bound}}$. The $v_{\mathrm{bound}}$ constraint is in terms of $\left\{ \phi, \theta_{0}, v_{\mathrm{orb}}, M, M^{\prime} \right\}$ and the $v_{\mathrm{eff}}$ constraint requires specifying the additional parameters $\left\{ e^{\prime}, a^{\prime} \right\}$.

A further restriction on the integration region $R$ of Equation \eqref{eqn:Pi_norm} comes from knowledge of the systemic velocity magnitude $v_{\mathrm{sys}}$. In Equation \eqref{eqn:vsys1}, replacing $\cos\left( \omega \right)$ with Equation \eqref{eqn:cos_omega} gives $v_{\mathrm{sys}}$ in terms of $\left\{ v_{\mathrm{k}}, \phi, \theta_{0}, v_{\mathrm{orb}}, M, M^{\prime} \right\}$,
\begin{equation}
v_{\mathrm{sys}}^{2} = \frac{M_{1}^{\prime 2}}{M^{\prime 2}} v_{\mathrm{k}}^{2} - 2 f \frac{M_{1}^{\prime} M_{2}^{\prime}}{M^{\prime 2}} \frac{\left| \sin\left( \phi \right) \right|}{\tan\left( \theta_{0} \right)} v_{\mathrm{k}} v_{\mathrm{orb}} + f \frac{M_{2}^{\prime}}{M^{\prime 2}} \left( 2 M_{1}^{\prime} + f M_{2}^{\prime} \right) v_{\mathrm{orb}}^{2}. \label{eqn:vsys2}
\end{equation}
Given observational constraints on the $v_{\mathrm{sys}}$ extrema, Equation \eqref{eqn:vsys2} provides quadratic inequalities in $v_{\mathrm{k}}$ that further pare down the integration region $R$ in $( v_{\mathrm{k}}, \phi )$-space for a given $\theta_{0}$.

%===========================================================================
%===========================================================================
\label{lastpage}
\end{document}